\documentclass[fleqn,usenatbib,useAMS]{mnras}

\usepackage{newtxtext,newtxmath}

\usepackage[T1]{fontenc}
\usepackage{ae,aecompl}
\usepackage{cleveref}
\usepackage{gensymb}

\usepackage{xcolor}		

\usepackage{graphicx}	
\usepackage{amsmath}	
\usepackage{amssymb}	
\usepackage{comment}

\newcommand{\todo}{\ifmmode \text{\color{red}\Huge{\(\bullet\)}} \else {\color{red}{\Huge$\bullet$}}\fi}
\newcommand{\tido}{\ifmmode {{\color{red}\bullet}} \else {\color{red}$\bullet$}\fi}

\newcommand{\E        }[1]{\ifmmode 10^{#1} \else $10^{#1}$\fi}
\newcommand{\tE        }[1]{\ifmmode \times10^{#1} \else $\times10^{#1}$\fi}
\newcommand{\til}{\ifmmode \sim \else $\sim$\fi}
\renewcommand{\~} {\ifmmode \sim \else $\sim$\fi}

\newcommand{\pc}	{\ifmmode {\rm pc} \else pc\fi}
\newcommand{\kpc}	{\ifmmode {\rm kpc} \else kpc\fi}
\newcommand{\ld}	{\ifmmode {\rm l.d.} \else l.d.\fi}
\newcommand{\kms}	{\ifmmode {\rm km\,s}^{-1} \else km\,s$^{-1}$\fi}
\newcommand{\cc}	{\ifmmode {\rm cm}^{-3}    \else cm$^{-3}$\fi}
\newcommand{\cmii}	{\ifmmode {\rm cm}^{-2}    \else cm$^{-2}$\fi}
\newcommand{\ergs}	{\ifmmode {\rm erg\,s}^{-1} \else erg s$^{-1}$\fi}
\newcommand{\ergcms}	{\ifmmode {\rm erg\,cm}^{-2}\,{\rm s}^{-1} \else erg\,cm$^{-2}$\,s$^{-1}$\fi}
\newcommand{\ergcmsA}	{\ifmmode {\rm erg\,cm}^{-2}\,{\rm s}^{-1}\,{\rm\AA}^{-1}
\else erg\,cm$^{-2}$\,s$^{-1}$\,\AA$^{-1}$\fi}
\newcommand{  \ergcmsHz  }{\ifmmode{\rm erg\,cm}^{-2}\,{\rm s}^{-1}\,{\rm Hz}^{-1}
                       \else ergs\,cm$^{-2}$\,s$^{-1}$\,Hz$^{-1}$\fi}
\newcommand{\kev}	{\ifmmode {\rm keV} \else keV\fi}

\newcommand{\kB}	{\ifmmode k_{\mbox{\tiny B}} \else $k_{\mbox{\tiny B}}$\fi}

\newcommand{\mic}	{\ifmmode {\rm \mu m} \else $\mu$m\fi}
\newcommand{\vFWHM}	{\ifmmode v_{\mbox{\tiny FWHM}} \else $v_{\mbox{\tiny FWHM}}$\fi}
\newcommand{\vBLR}	{\ifmmode v_{\mbox{\tiny BLR}} \else $v_{\mbox{\tiny BLR}}$\fi}
\newcommand{\sigBLR}	{\ifmmode \sigma_{\mbox{\tiny BLR}} \else $\sigma_{\mbox{\tiny BLR}}$\fi}
\newcommand{\vNLR}	{\ifmmode v_{\mbox{\tiny NLR}} \else $v_{\mbox{\tiny NLR}}$\fi}
\newcommand{\tauBLR}	{\ifmmode \tau_{\mbox{\tiny BLR}} \else $\tau_{\mbox{\tiny BLR}}$\fi}

\newcommand{\Hubble}	{\ifmmode {\rm km\,s}^{-1}\,{\rm Mpc}^{-1} \else km\,s$^{-1}$\,Mpc$^{-1}$\fi}
\newcommand{\NDunit}	{\ifmmode {\rm Mpc}^{-3} \else Mpc$^{-3}$\fi}
\newcommand{\LFunit}	{\ifmmode {\rm Mpc}^{-3}\,{\rm mag}^{-1} \else Mpc$^{-3}$\,mag$^{-1}$\fi}
\newcommand{\MFunit}	{\ifmmode {\rm Mpc}^{-3}\,{\rm dex}^{-1} \else Mpc$^{-3}$\,dex$^{-1}$\fi}

\newcommand{\Msun}{\ifmmode M_{\odot} \else $M_{\odot}$\fi}
\newcommand{\Lsun}{\ifmmode L_{\odot} \else $L_{\odot}$\fi}
\newcommand{\Zsun}{\ifmmode Z_{\odot} \else $Z_{\odot}$\fi}
\newcommand{\Rsun}{\ifmmode R_{\odot} \else $R_{\odot}$\fi}
\newcommand{\mpyr}{\ifmmode \Msun\,{\rm yr}^{-1} \else $\Msun\,{\rm yr}^{-1}$\fi}

\newcommand{\Msol}{\Msun}

\newcommand{\qnote}{\ifmmode q_{0} \else $q_{0}$\fi}
\newcommand{\Hnote}{\ifmmode H_{0} \else $H_{0}$\fi}
\newcommand{\hnote}{\ifmmode h_{0} \else $h_{0}$\fi}
\newcommand{\anote}{\ifmmode a_{0} \else $a_{0}$\fi}
\newcommand{\tnote}{\ifmmode t_{0} \else $t_{0}$\fi}

\def\gsim{\;\rlap{\lower 2.5pt \hbox{$\sim$}}\raise 1.5pt\hbox{$>$}\;}
\def\lsim{\;\rlap{\lower 2.5pt \hbox{$\sim$}}\raise 1.5pt\hbox{$<$}\;}

\newcommand{  \Halpha   }{\ifmmode {\rm H}\alpha \else H$\alpha$\fi}

\newcommand{  \ha       }{\Halpha}
\newcommand{  \Hbeta    }{\ifmmode {\rm H}\beta \else H$\beta$\fi}

\newcommand{  \hb       }{\Hbeta}
\newcommand{  \Hgamma   }{\ifmmode {\rm H}\gamma \else H$\gamma$\fi}
\newcommand{  \Hdelta   }{\ifmmode {\rm H}\delta \else H$\delta$\fi}
\newcommand{  \Lya      }{\ifmmode {\rm Ly}\alpha \else Ly$\alpha$\fi}
\newcommand{  \Lyb      }{\ifmmode {\rm Ly}\beta \else Ly$\beta$\fi}
\newcommand{  \Pa       }{\ifmmode {\rm P}\alpha \else P$\alpha$\fi}
\newcommand{  \Pb       }{\ifmmode {\rm P}\beta \else P$\beta$\fi}
\newcommand{  \Bra      }{\ifmmode {\rm Br}\alpha \else Br$\alpha$\fi}
\newcommand{  \Brg      }{\ifmmode {\rm Br}\gamma \else Br$\gamma$\fi}
\newcommand{  \hii      }{\ifmmode {\rm H}\,\textsc{ii} \else H\,\textsc{ii}\fi}
\newcommand{  \hei      }{\ifmmode {\rm He}\,\textsc{i} \else He\,\textsc{i}\fi}
\newcommand{  \heii     }{\ifmmode {\rm He}\,\textsc{ii} \else He\,\textsc{ii}\fi}
\newcommand{  \HeIIuv   }{\ifmmode {\rm He}\,\textsc{ii}\,\lambda1640 \else He\,\textsc{ii}\,$\lambda1640$\fi}
\newcommand{  \HeIIop   }{\ifmmode {\rm He}\,\textsc{ii}\,\lambda4686 \else He\,\textsc{ii}\,$\lambda4686$\fi}
\newcommand{  \CII	}{\ifmmode \left[{\rm C}\,\textsc{ii}\right]\,\lambda157.74\,\mu{\rm m} \else [C\,{\sc ii}]\ $\lambda157.74\,\mu{\rm m}$\fi}
\newcommand{  \cii	}{\ifmmode \left[{\rm C}\,\textsc{ii}\right] \else [C\,{\sc ii}]\fi}

\newcommand{  \ciii     }{\ifmmode {\rm C}\,\textsc{iii}\right] \else C\,\textsc{iii}]\fi}
\newcommand{  \CIII     }{\ifmmode {\rm C}\,\textsc{iii}\right]\,\lambda1909 \else C\,\textsc{iii}]\,$\lambda1909$\fi}
\newcommand{  \civ      }{\ifmmode {\rm C}\,\textsc{iv}  \else C\,\textsc{iv}\fi}
\newcommand{  \CIV      }{\ifmmode {\rm C}\,\textsc{iv}\,\lambda1549 \else C\,\textsc{iv}\,$\lambda1549$\fi}
\newcommand{  \nii      }{\ifmmode \left[{\rm N}\,\textsc{ii}\right]  \else [N\,\textsc{ii}]\fi}
\newcommand{  \niii     }{\ifmmode {\rm N}\,\textsc{iii} \else N\,\textsc{iii}\fi}
\newcommand{  \NIII     }{\ifmmode {\rm N}\,\textsc{iii}\,\lambda4640 \else N\,\textsc{iii}\,$\lambda4640$\fi}
\newcommand{  \niv      }{\ifmmode {\rm N}\,\textsc{iv}  \else N\,\textsc{iv}\fi}
\newcommand{  \NIVuv    }{\ifmmode {\rm N}\,\textsc{iv}\,\lambda1486 \else N\,\textsc{iv}\,$\lambda1486$\fi}
\newcommand{  \nv       }{\ifmmode {\rm N}\,\textsc{v}   \else N\,\textsc{v}\fi}
\newcommand{\oi}{\ifmmode \left[{\rm O}\,\textsc{i}\right] \else [O\,{\sc i}]\fi}
\newcommand{\OI}{\ifmmode \left[{\rm O}\,\textsc{i}\right]\,\lambda6300 \else [O\,{\sc i}]$\,\lambda6300$\fi}
\newcommand{\oii}{\ifmmode \left[{\rm O}\,\textsc{ii}\right] \else [O\,{\sc ii}]\fi}
\newcommand{\OII}{\ifmmode \left[{\rm O}\,\textsc{ii}\right]\,\lambda3727 \else [O\,{\sc ii}]\,$\lambda3727$\fi}
\newcommand{\oiii}{\ifmmode \left[{\rm O}\,\textsc{iii}\right] \else [O\,{\sc iii}]\fi}
\newcommand{\OIII}{\ifmmode \left[{\rm O}\,\textsc{iii}\right]\,\lambda5007 \else [O\,{\sc iii}]\,$\lambda5007$\fi}
\newcommand{  \OIIIbf   }{\ifmmode {\rm O}\,\textsc{iii}\,\lambda3133 \else O\,\textsc{iii}\,$\lambda3133$\fi}
\newcommand{  \OIIIuv   }{\ifmmode {\rm O}\,\textsc{iii}\,\lambda1663 \else O\,\textsc{iii}\,$\lambda1663$\fi}
\newcommand{  \oiv      }{\ifmmode {\rm O}\,\textsc{iv}  \else O\,\textsc{iv}\fi}
\newcommand{  \OIVuv    }{\ifmmode {\rm O}\,\textsc{iv}\,\lambda1402  \else O\,\textsc{iv}\,$\lambda1402$\fi}
\newcommand{  \OIVIR    }{\ifmmode {\rm O}\,\textsc{iv}\,25.9\,\mu {\rm m} \else O\,\textsc{iv}\,$25.9\,\mu$m\fi}
\newcommand{  \ovi      }{\ifmmode {\rm O}\,\textsc{vi}   \else O\,\textsc{vi}\fi}
\newcommand{  \Ovi      }{\ifmmode {\rm O}\,\textsc{vi}\,\lambda1035 \else O\,\textsc{vi}\,$\lambda1035$\fi}
\newcommand{  \nei      }{\ifmmode {\rm Ne}\,\textsc{i}   \else Ne\,\textsc{i}\fi}
\newcommand{  \neii     }{\ifmmode {\rm Ne}\,\textsc{ii}  \else Ne\,\textsc{ii}\fi}
\newcommand{  \NeiiIR   }{\ifmmode {\rm Ne}\,\textsc{ii}\,12.8\,\mu {\rm m} \else Ne\,\textsc{ii}\,$12.8\,\mu$m\fi}
\newcommand{  \neiii    }{\ifmmode {\rm Ne}\,\textsc{iii} \else Ne\,\textsc{iii}\fi}
\newcommand{  \neiv     }{\ifmmode {\rm Ne}\,\textsc{iv}  \else Ne\,\textsc{iv}\fi}
\newcommand{  \nev      }{\ifmmode {\rm Ne}\,\textsc{v}   \else Ne\,\textsc{v}\fi}
\newcommand{  \NevIR    }{\ifmmode {\rm Ne}\,\textsc{v}\,24.3\,\mu {\rm m} \else Ne\,\textsc{v}\,$24.3\,\mu$m\fi}
\newcommand{  \nevi     }{\ifmmode {\rm Ne}\,\textsc{vi}  \else Ne\,\textsc{vi}\fi}
\newcommand{  \mgi      }{\ifmmode {\rm Mg}\,\textsc{i} \else Mg\,\textsc{i}\fi}
\newcommand{  \mgii     }{\ifmmode {\rm Mg}\,\textsc{ii} \else Mg\,\textsc{ii}\fi}
\newcommand{  \MgII     }{\ifmmode {\rm Mg}\,\textsc{ii}\,\lambda2798 \else Mg\,\textsc{ii}\,$\lambda2798$\fi}
\newcommand{  \sii      }{\ifmmode \left[{\rm S}\,\textsc{ii}\right] \else [S\,\textsc{ii}]\fi}
\newcommand{  \siii     }{\ifmmode {\rm S}\,\textsc{iii} \else S\,\textsc{iii}\fi}
\newcommand{  \siv      }{\ifmmode {\rm S}\,\textsc{iv} \else S\,\textsc{iv}\fi}
\newcommand{  \sili     }{\ifmmode {\rm Si}\,\textsc{i}   \else Si\,\textsc{i}\fi}
\newcommand{  \silii    }{\ifmmode {\rm Si}\,\textsc{ii}  \else Si\,\textsc{ii}\fi}
\newcommand{  \Siliv    }{\ifmmode {\rm Si}\,\textsc{iv}  \else Si\,\textsc{iv}\fi}
\newcommand{  \SilIVuv  }{\ifmmode {\rm Si}\,\textsc{iv}\,\lambda1400  \else Si\,\textsc{iv}\,$\lambda1400$\fi}
\newcommand{  \AlIII   }{\ifmmode {\rm Al}\,\textsc{iii}\,\lambda1857 \else Al\,\textsc{iii}\,$\lambda1857$\fi}
\newcommand{  \Aliii   }{\ifmmode {\rm Al}\,\textsc{iii} \else Al\,\textsc{iii}\fi}
\newcommand{  \caii     }{\ifmmode {\rm Ca}\,\textsc{ii} \else Ca\,\textsc{ii}\fi}
\newcommand{  \feii     }{\ifmmode {\rm Fe}\,\textsc{ii} \else Fe\,\textsc{ii}\fi}
\newcommand{  \feiii    }{\ifmmode {\rm Fe}\,\textsc{iii} \else Fe\,\textsc{iii}\fi}
\newcommand{  \Kalpha   }{\ifmmode {\rm K}\alpha \else K$\alpha$\fi}

\newcommand{ \Lhb   }{\ifmmode L_{\hb} \else $L_{\hb}$\fi}
\newcommand{ \Lha   }{\ifmmode L_{\ha} \else $L_{\ha}$\fi}
\newcommand{ \fwhb  }{\ifmmode {\rm FWHM}\left(\hb\right) \else FWHM(\hb)\fi}
\newcommand{\sighb  }{\ifmmode \sigma\left(\hb\right) \else $\sigma\left(\hb\right)$\fi}
\newcommand{ \ewhb  }{\ifmmode {\rm EW}\left(\hb\right) \else EW(\hb)\fi}
\newcommand{ \fwha  }{\ifmmode {\rm FWHM}\left(\ha\right) \else FWHM(\ha)\fi}
\newcommand{ \ewha  }{\ifmmode {\rm EW}\left(\ha\right) \else EW(\ha)\fi}
\newcommand{ \Lmg   }{\ifmmode L\left(\mgii\right) \else $L\left(\mgii\right)$\fi}
\newcommand{ \fwmg  }{\ifmmode {\rm FWHM}\left(\mgii\right) \else FWHM(\mgii)\fi}
\newcommand{ \Lciv  }{\ifmmode L\left(\civ\right) \else $L\left(\civ\right)$\fi}
\newcommand{ \fwciv }{\ifmmode {\rm FWHM}\left(\civ\right) \else FWHM(\civ)\fi}
\newcommand{ \fwhm  }{\ifmmode {\rm FWHM} \else FWHM\fi} 
\newcommand{ \voff  }{\ifmmode v_{\rm off} \else $v_{\rm off}$\fi} 
\newcommand{ \vmax  }{\ifmmode v_{\rm max} \else $v_{\rm max}$\fi} 

\newcommand{ \mumg  }{\ifmmode \mu\left(\mgii\right) \else $\mu\left(\mgii\right)$\fi}
\newcommand{ \fmg   }{\ifmmode f\left(\mgii\right) \else $f\left(\mgii\right)$\fi}
\newcommand{ \muciv }{\ifmmode \mu\left(\civ\right) \else $\mu\left(\civ\right)$\fi}
\newcommand{ \fciv  }{\ifmmode f\left(\civ\right) \else $f\left(\civ\right)$\fi}

\newcommand{  \auvo     }{\ifmmode \alpha_{\nu,{\rm UVO}} \else $\alpha_{\nu,{\rm UVO}}$\fi}
\newcommand{  \Ledd     }{\ifmmode L_{\rm Edd} \else $L_{\rm Edd}$\fi}
\newcommand{  \lamLlam  }{\ifmmode \lambda L_{\lambda} \else $\lambda L_{\lambda}$\fi}
\newcommand{  \lLl      }{\ifmmode \lambda L_{\lambda} \else $\lambda L_{\lambda}$\fi}
\newcommand{  \nuLnu    }{\ifmmode \nu L_{\nu} \else $\nu L_{\nu}$\fi}
\newcommand{  \nLn      }{\ifmmode \nu L_{\nu} \else $\nu L_{\nu}$\fi}
\newcommand{  \Luv      }{\ifmmode L_{1450} \else $L_{1450}$\fi}
\newcommand{  \Lop      }{\ifmmode L_{5100} \else $L_{5100}$\fi}
\newcommand{  \lLop     }{\ifmmode \log\left(\Lop/\ergs\right) \else $\log\left(\Lop/\ergs\right)$\fi}
\newcommand{  \Lthree   }{\ifmmode L_{3000} \else $L_{3000}$\fi}
\newcommand{  \lLthree  }{\ifmmode \log\left(\Lthree/\ergs\right) \else $\log\left(\Lthree/\ergs\right)$\fi}
\newcommand{  \Lsix      }{\ifmmode L_{6200} \else $L_{6200}$\fi}
\newcommand{  \lLisx     }{\ifmmode \log\left(\Lop/\ergs\right) \else $\log\left(\Lop/\ergs\right)$\fi}
\newcommand{  \Lxray    }{\ifmmode L_{\rm X} \else $L_{\rm X}$\fi}

\newcommand{  \Lhard    }{\ifmmode L_{\rm 2-10} \else $L_{\rm 2-10}$\fi}
\newcommand{  \Lsoft    }{\ifmmode L_{\rm 0.5-2} \else $L_{\rm 0.5-2}$\fi}

\newcommand{\Fthree}{\ifmmode F_{3000} \else $F_{3000}$\fi}
\newcommand{\fuv}{\ifmmode f_{\lambda}\left(1450{\rm \AA}\right) \else $f_{\lambda}\left(1450 {\rm \AA}\right)$\fi}
\newcommand{\fthree}{\ifmmode f_{\lambda}\left(3000{\rm \AA}\right) \else $f_{\lambda}\left(3000{\rm \AA}\right)$\fi}
\newcommand{\fH}{\ifmmode f_{\lambda}\left(1.65\micron\right) \else
$f_{\lambda}\left(1.65\micron\right)$\fi}

\newcommand{\fbol}{\ifmmode f_{\rm bol} \else $f_{\rm bol}$\fi}
\newcommand{\fbolwv}{\ifmmode f_{\rm bol}\left(\lambda\right) \else $f_{\rm bol}\left(\lambda\right)$\fi}
\newcommand{\fbolopt}{\ifmmode f_{\rm bol}\left(5100{\rm \AA}\right) \else $f_{\rm bol}\left(5100{\rm \AA}\right)$\fi}
\newcommand{\fbolthree}{\ifmmode f_{\rm bol}\left(3000{\rm \AA}\right) \else $f_{\rm bol}\left(3000{\rm \AA}\right)$\fi}
\newcommand{\fboluv}{\ifmmode f_{\rm bol}\left(1450{\rm \AA}\right) \else $f_{\rm bol}\left(1450{\rm \AA}\right)$\fi}

\newcommand{\fbolbat}{\ifmmode f_{\rm bol}\left(14-150\,\kev\right) \else $f_{\rm bol}\left(14-150\,\kev\right)$\fi}
\newcommand{\fbolhard}{\ifmmode f_{\rm bol}\left(2-10\,\kev\right) \else $f_{\rm bol}\left(2-10\,\kev\right)$\fi}

\newcommand{\fobs}{\ifmmode f_{\rm obs} \else $f_{\rm obs}$\fi}

\newcommand{  \mbh      }{\ifmmode m_{\rm BH} \else $m_{\rm BH}$\fi}
\newcommand{  \lmbh     }{\ifmmode \log\left(\mbh/\Msun\right) \else $\log\left(\mbh/\Msun\right)$\fi} 
\newcommand{  \lledd    }{\ifmmode L/L_{\rm Edd} \else $L/L_{\rm Edd}$\fi}
\newcommand{  \mmedd    }{\ifmmode \dot{m}/\dot{m}_{\rm \,Edd} \else $\dot{m}/\dot{m}_{\rm \,Edd}$\fi}
\newcommand{  \Lbol     }{\ifmmode L_{\rm bol} \else $L_{\rm bol}$\fi}
\newcommand{  \lbol     }{\ifmmode L_{\rm bol} \else $L_{\rm bol}$\fi}
\newcommand{  \lLbol    }{\ifmmode \log\left(\Lbol/\ergs\right) \else $\log\left(\Lbol/\ergs\right)$\fi} 
\newcommand{  \Lagn     }{\ifmmode L_{\rm AGN} \else $L_{\rm AGN}$\fi}
\newcommand{  \lagn     }{\ifmmode L_{\rm AGN} \else $L_{\rm AGN}$\fi}

\newcommand{  \tgrow     }{\ifmmode t_{\rm growth} \else $t_{\rm growth}$\fi}
\newcommand{  \tAD     }{\ifmmode t_{\rm acc} \else $t_{\rm acc}$\fi}
\newcommand{  \tacc    }{\ifmmode t_{\rm acc} \else $t_{\rm acc}$\fi}
\newcommand{  \tUni      }{\ifmmode t_{\rm Universe} \else $t_{\rm Universe}$\fi}

\newcommand{  \Mdotin	}{\ifmmode \dot{M}_{\rm infall} \else $\dot{M}_{\rm infall}$\fi}
\newcommand{  \Mdotbh	}{\ifmmode \dot{M}_{\rm BH} \else $\dot{M}_{\rm BH}$\fi}
\newcommand{  \Mdotad	}{\ifmmode \dot{M}_{\rm AD} \else $\dot{M}_{\rm AD}$\fi}
\newcommand{  \Mdotacc	}{\ifmmode \dot{M}_{\rm acc} \else $\dot{M}_{\rm acc}$\fi}
\newcommand{  \Mdotthin	}{\ifmmode \dot{M}_{\rm thin} \else $\dot{M}_{\rm thin}$\fi}
\newcommand{  \Mdotdisk	}{\ifmmode \dot{M}_{\rm disk} \else $\dot{M}_{\rm disk}$\fi}

\newcommand{  \Mindot	}{\ifmmode \dot{M}_{\rm infall} \else $\dot{M}_{\rm infall}$\fi}
\newcommand{  \Mbhdot	}{\ifmmode \dot{M}_{\rm BH} \else $\dot{M}_{\rm BH}$\fi}
\newcommand{  \Maddot	}{\ifmmode \dot{M}_{\rm AD} \else $\dot{M}_{\rm AD}$\fi}
\newcommand{  \Maccdot	}{\ifmmode \dot{M}_{\rm acc} \else $\dot{M}_{\rm acc}$\fi}
\newcommand{  \Mthdot	}{\ifmmode \dot{M}_{\rm thin} \else $\dot{M}_{\rm thin}$\fi}
\newcommand{  \Mdsdot	}{\ifmmode \dot{M}_{\rm disk} \else $\dot{M}_{\rm disk}$\fi}

\newcommand{  \as	}{\ifmmode a_{\rm *} \else $a_{\rm *}$\fi}
\newcommand{  \avec	}{\ifmmode \vec{a}_{\rm *} \else $\vec{a}_{\rm *}$\fi}
\newcommand{  \re	}{\ifmmode \eta      	 \else $\eta$\fi}
\newcommand{  \RISCO	}{\ifmmode R_{\rm ISCO}  \else $R_{\rm ISCO}$\fi}

\newcommand{  \mseed    }{\ifmmode M_{\rm seed} \else $M_{\rm seed}$\fi}
\newcommand{  \mbul     }{\ifmmode M_{\rm bulge} \else $M_{\rm bulge}$\fi} 
\newcommand{  \mstar    }{\ifmmode M_{*} \else $M_{*}$\fi} 
\newcommand{  \mgal     }{\ifmmode M_{*} \else $M_{*}$\fi} 
\newcommand{  \mhost    }{\ifmmode M_{\rm host} \else $M_{\rm host}$\fi}
\newcommand{  \mmsmall  }{\ifmmode M_{\rm BH}/M_{*} \else $M_{\rm BH}/M_{*}$\fi}
\newcommand{  \mmlarge  }{\ifmmode M_{*}/M_{\rm BH} \else $M_{*}/M_{\rm BH}$\fi}

\newcommand{  \mmdotlarge}{\ifmmode \dot{M}_*/\Mbhdot \else $\dot{M}_*/\Mbhdot$\fi}
\newcommand{  \mmdotsmall}{\ifmmode \Mbhdot/\dot{M}_* \else $\Mbhdot/\dot{M}_*$\fi}

\newcommand{  \mmwp     }{\ifmmode \left(M_{*}/M_{\rm BH}\right) \else $\left(M_{*}/M_{\rm BH}\right)$\fi}
\newcommand{  \ml       }{\ifmmode M_{*}/L_{*} \else $M_{*}/L_{*}$\fi}
\newcommand{  \mlwp     }{\ifmmode \left(M_{*}/L\right) \else $\left(M_{*}/L\right)$\fi}
\newcommand{  \mlk      }{\ifmmode \left(M_{*}/L_{K}\right) \else $\left(M_{*}/L_{K}\right)$\fi}
\newcommand{  \sigs     }{\ifmmode \sigma_{*} \else $\sigma_{*}$\fi}
\newcommand{  \Reff     }{\ifmmode R_{\rm e} \else $R_{\rm e}$\fi}
\newcommand{  \Rvir     }{\ifmmode R_{\rm vir} \else $R_{\rm vir}$\fi}
\newcommand{  \Rtwo     }{\ifmmode R_{200} \else $R_{200}$\fi}
\newcommand{  \Rfive    }{\ifmmode R_{500} \else $R_{500}$\fi}
\newcommand{  \Rgrp     }{\ifmmode R_{\rm grp} \else $R_{\rm grp}$\fi}
\newcommand{  \nser     }{\ifmmode n_{\rm s} \else $n_{\rm s}$\fi}
\newcommand{  \LSF      }{\ifmmode L_{\rm SF}  \else $L_{\rm SF}$\fi}
\newcommand{  \LFIR     }{\ifmmode L_{\rm FIR} \else $L_{\rm FIR}$\fi}
\newcommand{  \Lfir     }{\ifmmode L_{\rm FIR} \else $L_{\rm FIR}$\fi}
\newcommand{  \LTIR     }{\ifmmode L_{\rm TIR} \else $L_{\rm TIR}$\fi}
\newcommand{  \Ltir     }{\ifmmode L_{\rm TIR} \else $L_{\rm TIR}$\fi}

\newcommand{  \mdyn     }{\ifmmode M_{\rm dyn} \else $M_{\rm dyn}$\fi} 
\newcommand{  \mgas     }{\ifmmode M_{\rm gas} \else $M_{\rm gas}$\fi} 
\newcommand{  \mh       }{\ifmmode M_{\rm h} \else $M_{\rm h}$\fi}
\newcommand{  \mhalo    }{\ifmmode M_{\rm halo} \else $M_{\rm halo}$\fi}
\newcommand{  \sfr      }{\ifmmode {\rm SFR} \else SFR\fi}

\newcommand{ \Lcii     }{\ifmmode L_{\cii} \else $L_{\cii}$\fi}
\newcommand{ \fwcii  }{\ifmmode {\rm FWHM}\cii \else FWHM\cii\fi}

\newcommand{  \spitzer }  {{\it Spitzer}}

\newcommand{  \WISE    }  {{\it WISE}}

\newcommand{\bj}{\ifmmode b_{\rm J} \else $b_{\rm J}$\fi}

\newcommand{\iab}{\ifmmode i_{\rm AB} \else $i_{\rm AB}$\fi}

\newcommand{\jab}{\ifmmode J_{\rm AB} \else $J_{\rm AB}$\fi}
\newcommand{\hab}{\ifmmode H_{\rm AB} \else $H_{\rm AB}$\fi}
\newcommand{\kab}{\ifmmode K_{\rm AB} \else $K_{\rm AB}$\fi}

\newcommand{\jveg}{\ifmmode J_{\rm Vega} \else $J_{\rm Vega}$\fi}
\newcommand{\hveg}{\ifmmode H_{\rm Vega} \else $H_{\rm Vega}$\fi}
\newcommand{\kveg}{\ifmmode K_{\rm Vega} \else $K_{\rm Vega}$\fi}

\newcommand{  \Chisq    }{\ifmmode \chi^{2} \else $\chi^{2}$}
\newcommand{  \nelec    }{\ifmmode n_{e} \else $n_{e}$\fi}     
\newcommand{  \nh       }{\ifmmode n_{\rm H} \else $n_{\rm H}$\fi}     
\newcommand{  \Ncol     }{\ifmmode N_{\rm col} \else $N_{\rm col}$\fi} 
\newcommand{  \NH       }{\ifmmode N_{\rm H} \else $N_{\rm H}$\fi}     

\def\ion#1#2{#1$\;${\small\rm\@Roman{#2}}\relax}

\newcommand{\angstrom}{\textup{\AA}}

\title[Searching for Super-Eddington Quasars]{Searching for Super-Eddington Quasars Using a Photon Trapping Accretion Disc Model}

\author[Q. Pognan et al.]
{Quentin Pognan$^{1,7}$\thanks{E-mail: quentin.pognan@astro.su.se},
Benny Trakhtenbrot$^{2}$\thanks{E-mail: benny@astro.tau.ac.il},
Tullia Sbarrato$^{3,4}$,\newauthor
Kevin Schawinski$^{1,5}$,
and Caroline Bertemes$^{6}$\\
$^{1}$Institute for Particle Physics and Astrophysics, ETH Z\"{u}rich, Wolfgang-Pauli-Str. 27, CH-8093 Z\"{u}rich, Switzerland\\
$^{2}$School of Physics and Astronomy, Tel Aviv University, Tel Aviv 69978, Israel\\
$^{3}$Dipartimento di Fisica ``G. Occhialini'', Universit\`{a} degli Studi di Milano Bicocca, Piazza della Scienza 3, 20126 Milano, Italy\\
$^{4}$INAF-Osservatorio Astronomico di Brera, via E. Bianchi 46, I-23807, Merate, Italy\\
$^{5}$Modulos AG, Technoparkstrasse 1, CH-8005 Z\"{u}rich, Switzerland\\
$^{6}$Department of Physics, University of Bath, Claverton Down, Bath BA2 7AY, UK \\
$^{7}$Institute for Astronomy, Oskar Klein Centre, Stockholm University, Roslagstullsbacken 21, 115 44, Stockholm, Sweden 
}

\date{Accepted XXX. Received YYY; in original form ZZZ}

\pubyear{2019}

\begin{document}
\label{firstpage}
\pagerange{\pageref{firstpage}--\pageref{lastpage}}
\maketitle

\begin{abstract}
Accretion onto black holes at rates above the Eddington limit has long been discussed in the context of supermassive black hole (SMBH) formation and evolution, providing a possible explanation for the presence of massive quasars at high redshifts (z$\gtrsim$7), as well as having implications for SMBH growth at later epochs. 
However, it is currently unclear whether such `super-Eddington' accretion occurs in SMBHs at all, how common it is, or whether every SMBH may experience it. In this work, we investigate the observational consequences of a simplistic model for super-Eddington accretion flows -- an optically thick, geometrically thin accretion disc (AD) where the inner-most parts experience severe photon-trapping, which is enhanced with increased accretion rate. 
The resulting spectral energy distributions (SEDs) show a dramatic lack of rest-frame UV, or even optical, photons. Using a grid of model SEDs spanning a wide range in parameter space (including SMBH mass and accretion rate), we find that 
large optical quasar surveys (such as SDSS) may be missing most of these luminous systems.
We then propose a set of colour selection criteria across optical and infra-red colour spaces designed to select super-Eddington SEDs in both wide-field surveys (e.g., using SDSS, 2MASS and \textit{WISE}) and deep \& narrow-field surveys (e.g., COSMOS). 
The proposed selection criteria are a necessary first step in establishing the relevance of advection-affected super-Eddington accretion onto SMBHs at early cosmic epochs.
\end{abstract}

\begin{keywords}
accretion discs -- black hole physics -- quasars: supermassive black holes -- surveys
\end{keywords}



\section{Introduction}
\label{sec:intro}

Super-massive black holes (SMBHs) with masses ranging from $10^6$ to $10^{10} M_{\odot}$ are believed to inhabit most if not all galaxies in the universe with the majority existing in a quiescent state, and only a small fraction actively accreting matter at a rate large enough to produce sufficient emission to be detected as active galactic nuclei \cite[AGN; e.g.,][]{Soltan:82,Richstone.etal:98}. 
AGN and quasars have been detected at high redshifts up to $z\sim7.5$ \citep[see][and references therein]{Banados.etal:18}, corresponding to the first Gyr after the Big Bang, though their origin is still debated. 
There are several scenarios that may explain the origin of such early, luminous quasars, including `massive seeds' from direct collapse BHs, or lighter seed BHs, e.g. from population III stellar remnants, which undergo extremely fast growth (see, e.g., \citealt{Bromm.Loeb:03,DiMatteo.etal:03,Volonteri.etal:03,Johnson.Bromm:07}, and reviews by \citealt{Natarajan:11} and \citealt{Volonteri:12}).

An alternative mechanism proposed to solve the mystery of super-massive quasars at high redshift is extremely rapid black hole (BH) growth, through super-Eddington accretion -- that is, when the mass accretion rate through the accretion flow exceeds the classical Eddington limit, $\dot{m}\equiv M/M_{\rm Edd} > 1$.\footnote{For super-Eddington accretion, the (normalized) physical accretion rate, $\dot{m}\equiv M/M_{\rm Edd}$, does not necessarily equal to the {\it observed} Eddington ratio, \lledd, as probed by the emergent radiation field. 
Throughout this work $\dot{M}_{\rm Edd}$ is derived through $\dot{M}_{\rm Edd}\equiv L_{\rm Edd}/\eta c^2$, with $\eta=0.083$ being the radiative efficiency appropriate for a moderately-spinning BH \cite[$a\simeq0.5$; see, e.g.,][and references therein]{Volonteri.etal:13}.} 
With super-critical accretion periods, early SMBH could have grown from even stellar-mass BH seeds to become the observed $z \gtrsim 6$ quasar population \citep[e.g.,][]{Madau.etal:14,Volonteri.etal:15,Valiante.etal:16}. 
More generally, super-Eddington accretion can be applied to BHs at any epoch, and so is broadly relevant to SMBH growth and evolution and to AGN physics.
Although super-Eddington accretion has been robustly identified in {\it stellar}-mass compact objects \cite[e.g.,][]{Bachetti.etal:14,Israel.etal:17a,Israel.etal:17b}, and some claims were made about certain AGN \cite[e.g.,][]{Jin.etal:17}, the relevance for super-Eddington accretion for the general AGN population has yet to be established. 

The most common model for accretion onto SMBHs is that of an optically thick, geometrically thin ({\it sub}-Eddington) accretion disc \citep{Shakura.Sunyaev:73,Netzer:13}. 
The intense UV radiation that emerges from the inner-most parts of the disc drives significant (high ionization) line emission, and may provide the seed photons for important  X-ray emission (through Compton up-scattering) -- all of which are commonly used as robust identifiers of AGN, including in large surveys \cite[e.g.,][]{ Baldwin.etal:81,Schmidt.Green:83,Richards.etal:02,Done.etal:12,Brandt.Alexander:15,Arcodia:19}.
Thin accretion discs were shown to account for the broad-band spectral energy distribution (SED) of quasars  \cite[e.g.,][and references therein]{Capellupo.etal:15,Capellupo.etal:16,Jin.etal:16}, particularly once relativistic effects are considered \citep{Jaroszynski.etal:92,Koratkar.Blaes:99,Laor.Davis:11,Davis.Laor:11,Brenneman.etal:13}. 
There are still a number of outstanding issues with the application of thin accretion discs to AGN -- even over the range of accretion rates where they should hold ($0.01 \lesssim \dot{m} \lesssim 0.3$) -- as implied from various probes of the accretion flow scale (see recent commentary by \citealt{Lawrence:18} and \citealt{Antonucci:18}).
 
Other models of accretion flows onto SMBHs have been investigated in order to deal with accretion rate regimes not well described by the classical thin disc. 
Advection dominated accretion flows (ADAFs) attempt to describe very low accretion rates, $\dot{m} \ll 0.01$. 
In this scenario, the flow becomes geometrically thick (and indeed quasi-spherical), and optically thin, leading to extremely low radiative efficiency \citep{Narayan.Yi:94,Narayan.Yi:95,Narayan.etal:98,Yuan.Narayan:14}. 
At high Eddington rates, on the other hand ($\dot{m} \gg 0.3$), the standard thin disc model is expected to fail, as the radiation pressure increases and indeed becomes dominant. 
The hydrostatic equilibrium assumed by a thin disc no longer holds and the disc half-thickness to radius ratio  is expected to increase up to $H/r \sim 1$, motivating instead models of `slim' or `thick'  accretion discs \cite[see, e.g.,][]{Abramowicz.etal:88,Wang.etal:99,Watarai.Fukue:99, Kawaguchi:03}. 

Similarly to ADAF discs, the slim disc model has very low radiative efficiencies, but for physically different reasons. Unlike the ADAF model, the slim disc remains optically thick while becoming geometrically thick \citep{Abramowicz.etal:88,Wang.etal:99,Watarai.Fukue:99}. 
The increased disc thickness greatly diminishes the radiative cooling rate since photons can no longer efficiently escape the disc. Thus, most of the energy is advected towards the BH along with the infalling material, leading to very low radiative efficiencies. Slim discs are thus expected to remain extremely luminous, emitting luminosities that saturate at roughly $L_{\rm Edd}$ \citep{Abramowicz.etal:88,Watarai.Fukue:99,Ohsuga.etal:02,Ohsuga.etal:03,Madau.etal:14}. 
The combination of an increasing physical accretion rate with this saturated, Eddington-limited luminosity naturally leads to progressively low radiative efficiencies, $\eta \equiv (L / \dot{M} c^2) \ll 0.05$.

Analytical models of slim accretion discs have often been suggested as an explanation for observations of AGN apparently accreting at Eddington or super-Eddington rates \cite[e.g.,][]{Collin.etal:02,Kawaguchi:03,Collin.Kawaguchi:04}.
More recent advances in general relativistic radiation magnetohydrodynamical (GRRMHD) codes have allowed to study, numerically, the intricacies of super-Eddington accretion flows onto BHs \cite[see e.g.,][]{Sadowski:09,McKinney.etal:14,Sadowski.etal:14,Sadowski.Narayan:16}. 
Although these studies differ in many specifics, they have provided several consistent insights regarding the nature of such systems.
Most notably, the simulations confirm the saturation of the emergent radiation field at $\sim 1-10 L_{\rm Edd}$, even as the accretion rate are as extreme as $\gtrsim 100\,\dot{M}_{\rm Edd}$ -- thus also confirming the very low radiative efficiencies, of order 0.01 \citep{McKinney.etal:14,Sadowski.etal:14,Sadowski.Narayan:16}. Moreover, the discs are indeed slim (or thick), with $H/r \sim 0.5-1.0$ \citep{McKinney.etal:14,Sadowski.etal:14,Sadowski.Narayan:16}. 
Both types of models were shown to be relevant for the rapid growth of the earliest SMBHs from (pop-III) stellar remnants of $\sim 100 M_{\odot}$ \citep[see, e.g.,][]{Lupi.etal:16,Pezzulli.etal:16}, particularly when certain conditions are met (e.g., high BH spins; \citealt{Madau.etal:14,McKinney.etal:14}). 
The stability of the simulated accretion flows over relatively long timescales further supports this scenario, given that the gas supply from the host galaxy (or beyond) is sufficient. 

In terms of the emission expected from super-Eddington SMBHs, several studies have suggested that the thickened layers of the disc would produce excess UV radiation \cite[see detailed discussion in ][]{Done.etal:12} -- a prediction that seems to be supported by some observational evidence for specific, fast-growing systems (see below).
Many other studies however, have stressed that slim disc models may underestimate the importance of photon trapping effects, which become more significant as the accretion rate increases.   
\citep{Ohsuga.etal:02,Ohsuga.etal:03,Mineshige.Ohsuga:08,Sadowski.etal:14,Sadowski.Narayan:16}.
As such, it is suggested that too many photons are allowed to escape in the regular slim disc model, leading to an overestimation of the disc luminosity \citep{Ohsuga.etal:02,Ohsuga.etal:03,Mineshige.Ohsuga:08,Inayoshi.etal:16,Sakurai.etal:16}. 
Moreover, photon trapping effects are expected to significantly change the shape of the emergent SED, as (UV) photons from the inner parts of the accretion flow are advected onto the BH. 
In either case, the continuum (UV) SEDs expected from super-Eddington SMBHs may differ significantly, from those of normal, sub-Eddington AGN. These differences may then also manifest themselves in significant changes to the X-ray continuum and/or (high-ionization) line emission, as these emission components are broadly thought to be driven by reprocessing of disc UV photons.
All these factors raise the possibility that super-Eddington AGN may look very different from their sub-Eddington counterparts, and hence could in principle be mis-classified, 
or entirely omitted from current databases. 

Observational efforts to determine the accretion rates of (distant) SMBHs focus on unobscured, radiatively efficient AGN.
Large surveys suggest that \lledd\ generally increases with redshift, while still broadly obeying $\lledd\lesssim1$ -- as determined either for SMBHs of given mass \cite[e.g.,][]{McLure.Dunlop:04,Netzer.Trakhtenbrot:07,Trakhtenbrot.Netzer:12}, or from the `break' in the Eddingtron ration distribution function \cite[ERDF; e.g.,][]{Kauffmann.Heckman:09,Schulze.Wisotzki:10,Shen.Kelly:12,Caplar.etal:15,Schulze.etal:15,Weigel.etal:17}.
Indeed, the highest-redshift quasars (at $z\gtrsim5$) are observed to radiate at rates that are consistent with $\lledd\simeq1$ (see, e.g., \citealt{Trakhtenbrot.etal:11,Kelly.Shen:13,Page.etal:14,DeRosa.etal:14,Trakhtenbrot.etal:17b}, but also \citealt{Tang.etal:19}).
At all redshifts, the highest-\lledd\ AGN are the rarest sources, tracing a population with space densities of order $ \lesssim 10^{-9}\, {\rm Mpc}^{-3}$.
Given the expectation that super-Eddington AGN would saturate at roughly $\lledd \lesssim 10$ (and the large systematic uncertainties on \lledd\ measurements in quasars), it is entirely possible that a fraction of the highest-\lledd\ quasars are indeed accreting at super-Eddington rates, making them a (yet to be determined) fraction of the brightest and rarest AGN at all redshifts.
Some recent observational studies have suggested that the X-ray through optical SEDs of a few AGN are consistent with slim disc models \cite[e.g.,][and references therein]{Done.Jin:16,Jin.etal:16}. Such analysis, however, requires exquisite multi-wavelength data, which currently cannot be obtained for large (high-redshift) AGN samples (but see \citealt{Tang.etal:19}), and relies heavily on certain slim disc models which can account for significant emission components in the extreme-UV and soft X-ray regimes.

The observational effort to identify (populations of) super-Eddington AGN and to understand their role in SMBH evolution thus faces several different challenges: 
(1) understanding the unique emission features that differ super-Eddington AGN from their sub-Eddington counterparts; 
(2) designing the selection criteria that will robustly identify super-Eddington AGN in large (likely wide-field) multi-wavelength surveys;
(3) confirming the high physical accretion rates of the sources under study (i.e., measuring, or at least constraining, $\dot{m}$ rather than \lledd); 
and, ultimately, (4) forming a representative census of super-Eddington systems (preferably, at several redshifts), to constrain the typical timescales SMBHs spend in this extreme regime.

In this paper, we wish to contribute to the first two steps in this road-map: we investigate the SEDs expected to emerge from a rather simplistic model of a thin disc affected by severe photon trapping, and how these relate to large extra-galactic surveys.
We test the SDSS quasar selection algorithm on the models both with and without added host emission. 
We discuss the processes involved in Section \ref{sec:methods}, 
and present our results in terms of selection completeness in Section \ref{sec:results}. 
We then propose new super-Eddington AGN colour selection criteria for both wide and deep field surveys in Section \ref{sec:selection}. 
We summarise our main findings and potential future directions in Section \ref{sec:conclusion}. 
Throughout this work, a cosmology with $\Omega_{\rm \Lambda} = 0.7$, $\Omega_{\rm M} = 0.3$ and $H_0 = 70\, \kms\,{\rm Mpc}^{-1}$ is assumed.

\section{Model SEDs and Colour Selection Algorithms}
\label{sec:methods}

The ultimate goal of this work is to suggest observational selection criteria to successfully identify potentially super-Eddington SMBHs at significant redshifts ($z \geq 0.5$), to be applied in both wide-field and deep-drill multi-wavelength surveys. 
We approach this by generating a grid of super-critical SEDs using a simple photon trapping model covering a wide range in basic SMBH parameters, with Eddington ratio as our main focus. 
We then test whether the model super-Eddington SEDs would be detected by current methods, by applying the quasar spectroscopic follow-up target selection algorithm of the Sloan Digital Sky Survey \citep[SDSS;][]{Newberg.Yanny:97,York.etal:00,Richards.etal:02}. 
Following the results of the SDSS selection algorithm, we then suggest new selection criteria across optical and infra-red colour spaces to specifically target super-Eddington AGN which would follow our AD model. 
We present such criteria for wide field surveys such as SDSS, the 2 Micron All Sky Survey \cite[2MASS;][]{Skrutskie.etal:06} and the {\it Wide-field Infrared Survey Explorer} \cite[\textit{WISE};][]{Wright.etal:10}, as well as for deep multi-wavelength surveys such as the Cosmic Evolution Survey \cite[COSMOS;][]{Scoville.etal:07}. 

\subsection{General Considerations}
\label{sec:gencon}

For high Eddington ratios at or above the Eddington limit, the radiation pressure from the innermost region of the AD is expected to render the disc geometrically thick \citep{Abramowicz.etal:88,Wang.etal:99,Watarai.Fukue:99}. 
This has for effect to increase the radiative diffusion timescale to a point where it is longer than the accretion timescale. 
Thus photons within a certain radius are `trapped' and advected inwards towards the black hole, instead of escaping outwards  \citep{Wang.Zhou:99,Ohsuga.etal:02,Ohsuga.etal:03}. 
Although the exact form of the typical `photon trapping radius' depends on various assumptions made such as the exact shape of the disc and the dominant heating mechanism, its effect can be generally described regardless of model specifics. 

As previously noted in Section \ref{sec:intro}, the various regimes of AGN accretion are often described in terms of the Eddington ratio $\lambda_{\rm Edd} = \lledd$, which can be rewritten directly in terms of the mass accretion rate such that $\dot{m} \equiv \dot{M}/\dot{M}_{\rm Edd}$.
Though the two definitions are equivalent for sub-Eddington accretion regimes where the radiative efficiency linking luminosity to accretion rate is constant, this ceases to be the case for super-Eddington accretion (i.e. with slim discs), where the luminosity is expected to saturate at about $(1-10) \times L_{\rm Edd}$, even as $\dot{m}$ continues to increase \citep{Ohsuga.etal:02,McKinney.etal:14,Sadowski.etal:14,Sadowski.Narayan:16}. 
As this work deals with super-Eddington accretion focusing on the mass accretion rate, we will hereafter refer to the Eddington ratio as the latter definition, which focuses on the normalized physical accretion rate, $\dot{m}$.

The photon trapping effects of the slim disc are also expected to modify the total luminosity of the disc, as well as its effective temperature profile. 
Notably, as $\dot{m}$ increases, photon trapping is expected to be more important since radiative pressure progressively thickens the disc. Thus, we expect more photons to be advected onto the BH, leading to a reduced luminosity compared to a classical thin disc with a naively scaled-up accretion rate.
A standard thin disc has an overall effective temperature profile of $T_{\rm eff} \sim r^{-3/4}$ \citep{Shakura.Sunyaev:73}.  
In general, photons emitted closer to the innermost stable circular orbit (ISCO) are more likely to be trapped and advected into the BH, resulting in a potentially flatter temperature profile than for a classic disc. 
This would effectively reduce higher-energy emission such as far UV in the emergent SED. 
A lack of UV photons may also affect the soft X-ray emission, which is believed to arise from inverse-Compton scattering of UV photons.
Furthermore, emission lines originating from ionized circumnuclear gas, and commonly found for sub-Eddington AGN (e.g., [O\,{\sc iii}]\,$\lambda\lambda4959,5007$, [N\,{\sc ii}]\,$\lambda\lambda6549,6583$, and [S\,{\sc ii}]\,$\lambda\lambda6718,6732$), may also be weak or missing, as they are driven by the UV emission from the central engine.

The magnitude of all these effects, as well as the temperature profile of the disc, will depend heavily on the dominant heating mechanism that is assumed. Notably, a viscous heating mechanism only effective in the equatorial regions of the disc will lead to the vast majority of photons within the trapping radius being advected onto the BH, while assuming uniform heating through the disc will allow some photons emitted closer to the disc surface to escape. In the latter case, the photon trapping effects are mitigated, and the disc luminosity is expected to continue increasing with $\dot{m}$, as opposed to remaining in the $1-10 L_{\rm Edd}$ range \citep{Ohsuga.etal:02}. 
The realistic scenario may be even more complicated than currently considered, though many studies agree that some form of photon trapping yielding the effects described above is required in order to successfully account of super-Eddington accretion onto SMBHs  \citep{Ohsuga.etal:02,Ohsuga.etal:03,Mineshige.Ohsuga:08,McKinney.etal:14,Sadowski.etal:14,Inayoshi.etal:16,Sakurai.etal:16,Sadowski.Narayan:16}.
In what follows, we thus proceed by investigating a rather simplistic model for photon trapping, and show that this can already result in intriguing insights regarding the identification of super-Eddington AGN.

A simple expression for the photon trapping radius can be obtained by equating the radiative diffusion and accretion timescales as described in \citet{Ohsuga.etal:02,Ohsuga.etal:03}:
\begin{equation}
r_{\rm trap} = \frac{3}{2}\, \dot{m}\, h \, r_{\rm s} \,\, , 
\label{eq:rtrap}
\end{equation}
where $r_{\rm s}$ is the Scwharzschild radius, $h$ is the ratio of half-disc thickness to radius, and $\dot{m}$ is the accretion rate, normalized to the critical value, as defined above. 
In the radiation pressure dominated inner region of the thickened accretion disc, $h$ is expected to be of order unity \citep{Ohsuga.etal:02}. 
Thus photon trapping effects become important at ${\sim}(3/2) r_{\rm s}$ when $\dot{m}=1$, and increasingly further out as accretion rate increases.
Specifically, when $\dot{m}>10$, the corresponding trapping radius, $r_{\rm trap} > 15\,r_{\rm s}$, becomes larger than $R_{\rm ISCO}$ (for any BH spin), and significant changes to the emergent SED are thus expected.
Depending on the model of heating assumed with this trapping radius (equatorial plane only vs. uniform heating), the accretion efficiency is expected to scale between $\eta \propto \dot{m}^{-1.0}$ and $\eta \propto \dot{m}^{-0.5}$ \citep{Ohsuga.etal:02,Ohsuga.etal:03}. 
This implies that the radiative efficiency of a highly super-Eddington BH is extremely low, as is consistent with other theoretical and numerical studies (see Section \ref{sec:intro}). 

\begin{table}
\caption{Parameters of model SEDs. Eddington ratios are 1, 3, 10, 30, 50, 100}
\label{tab:models}
\resizebox{\linewidth}{!}{
\begin{tabular}[width=0.9\linewidth]{l c c c}
\hline
Parameter & Min. value & Max. value & Step size \\
\hline
BH mass, $log(M_{\rm BH}/M_{\odot})$ & 7 & 11 & 0.5 \\
Eddington ratio, $\dot{M}/\dot{M}_{\rm Edd}$ & 1 & 100 & variable \\
Redshift, $z$ & 0.5 & 2.0 & 0.1 \\
Inclination angle, $i$ & 10\degree & 45\degree & variable\\
\hline
\end{tabular}
}
\end{table}

\begin{figure}
\includegraphics[width=1.10\linewidth]{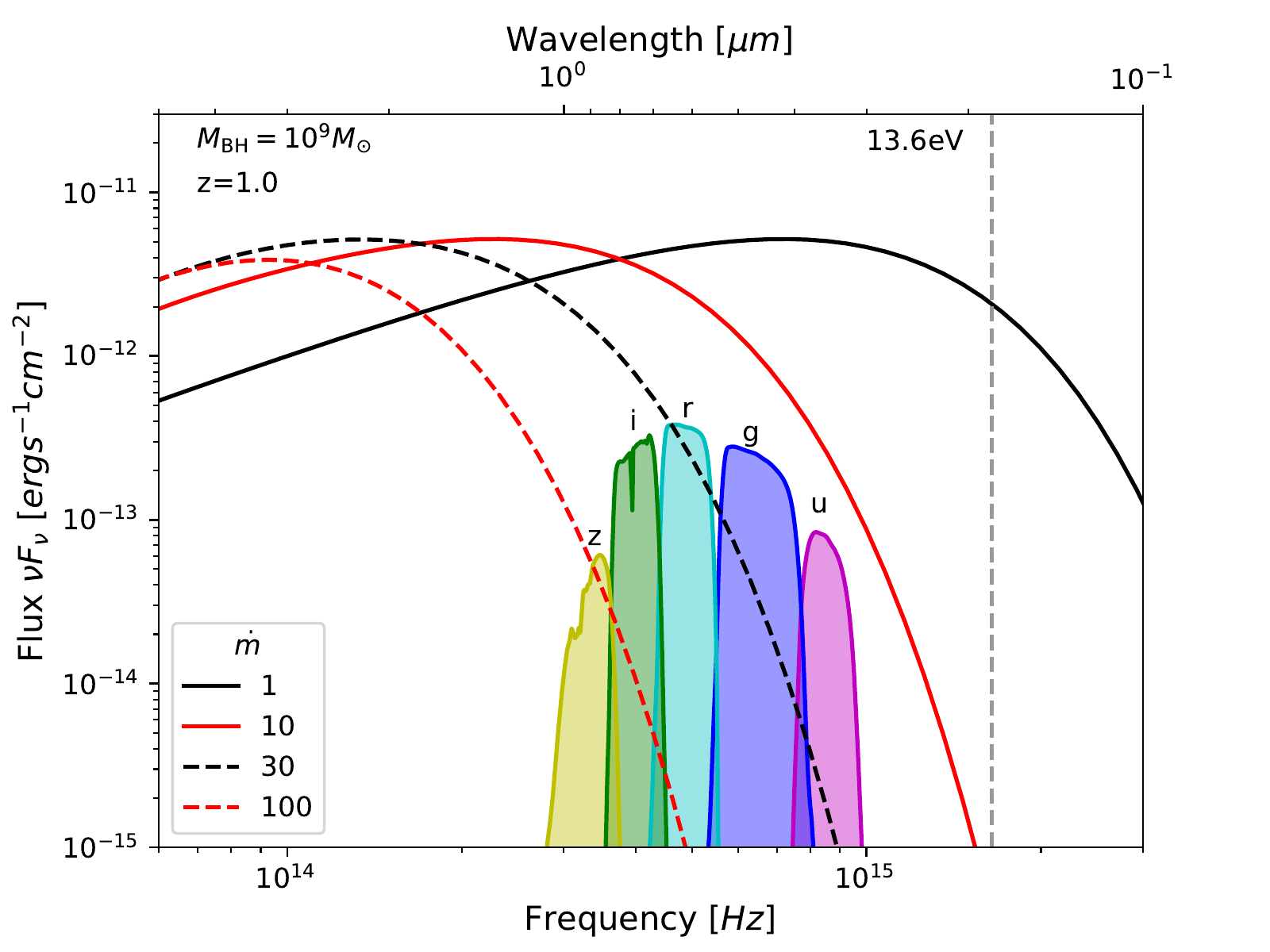}
\caption{Illustrating the effects of increasing Eddington ratio and photon trapping in the context of SDSS filters. 
Our model SEDs become progressively redder with increasing Eddington ratio. There will be few ionising photons as most of the SED is below the 13.6eV line. The SEDs all have $\rm log(M_{\rm BH}/M_{\odot}) = 9$ at redshift 1.0 with inclination angle 30\degree and are plotted in the observed frame.}
\label{fig:SED_filters}
\end{figure}

\subsection{Model SEDs}
\label{sec:model}

We proceed with generating a large grid of model SEDs that combine standard AGN accretion flows with the simplistic realization of photon trapping, to gain quantitative insights on the effects this would have on the observational nature of highly super-Eddington AGN.

Our model starts with a standard Shakura-Sunyaev thin disc \citep{Shakura.Sunyaev:73}. 
We then adopt the expression for the photon trapping radius in Equation \ref{eq:rtrap}, and consider the simple case where every photon within this radius is advected onto the BH. 
In the heating mechanisms described above, this most closely corresponds to assuming purely equatorial heating where the majority of photons within the trapping radius are advected. Thus we remove any emission from the region of the AD within the photon trapping radius, essentially truncating the effective temperature profile for small radii. 
Outside the photon trapping radius, our model remains the standard  thin disc. 
Although a real {\it slim} disc would have a smoother transition from thick to thin regimes, a step function from a half-height to radius ratio $h \sim 1$ to $h \sim 0$ is sufficient for our purposes as it captures, and indeed highlights, the effects of photon trapping on the emergent SED. 

\begin{figure*}
\center
\includegraphics[trim={0.2cm 0 1.5cm 0},clip,width = 0.48\textwidth]{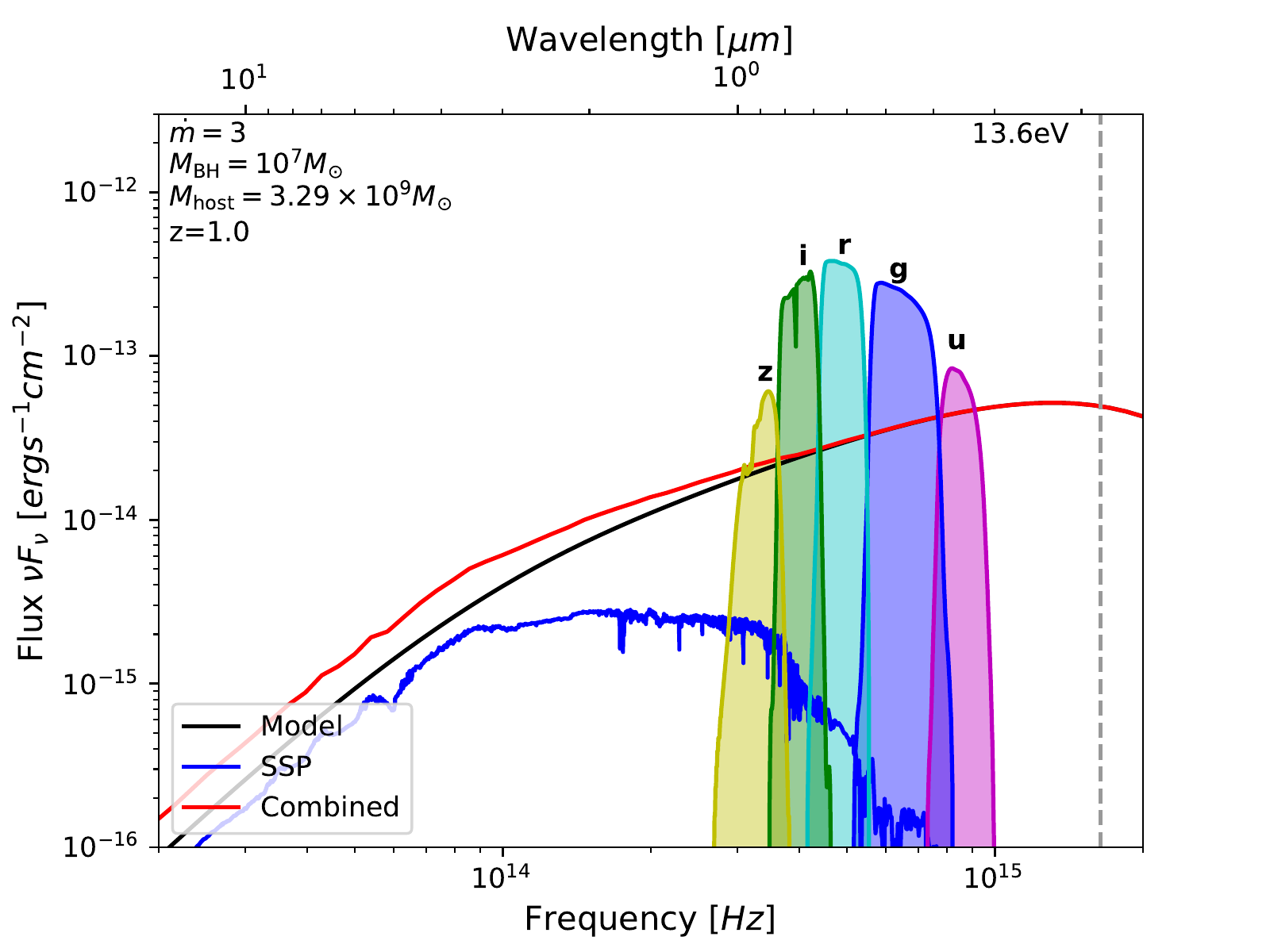}
\includegraphics[trim={0.2cm 0 1.5cm 0},clip,width = 0.48\textwidth]{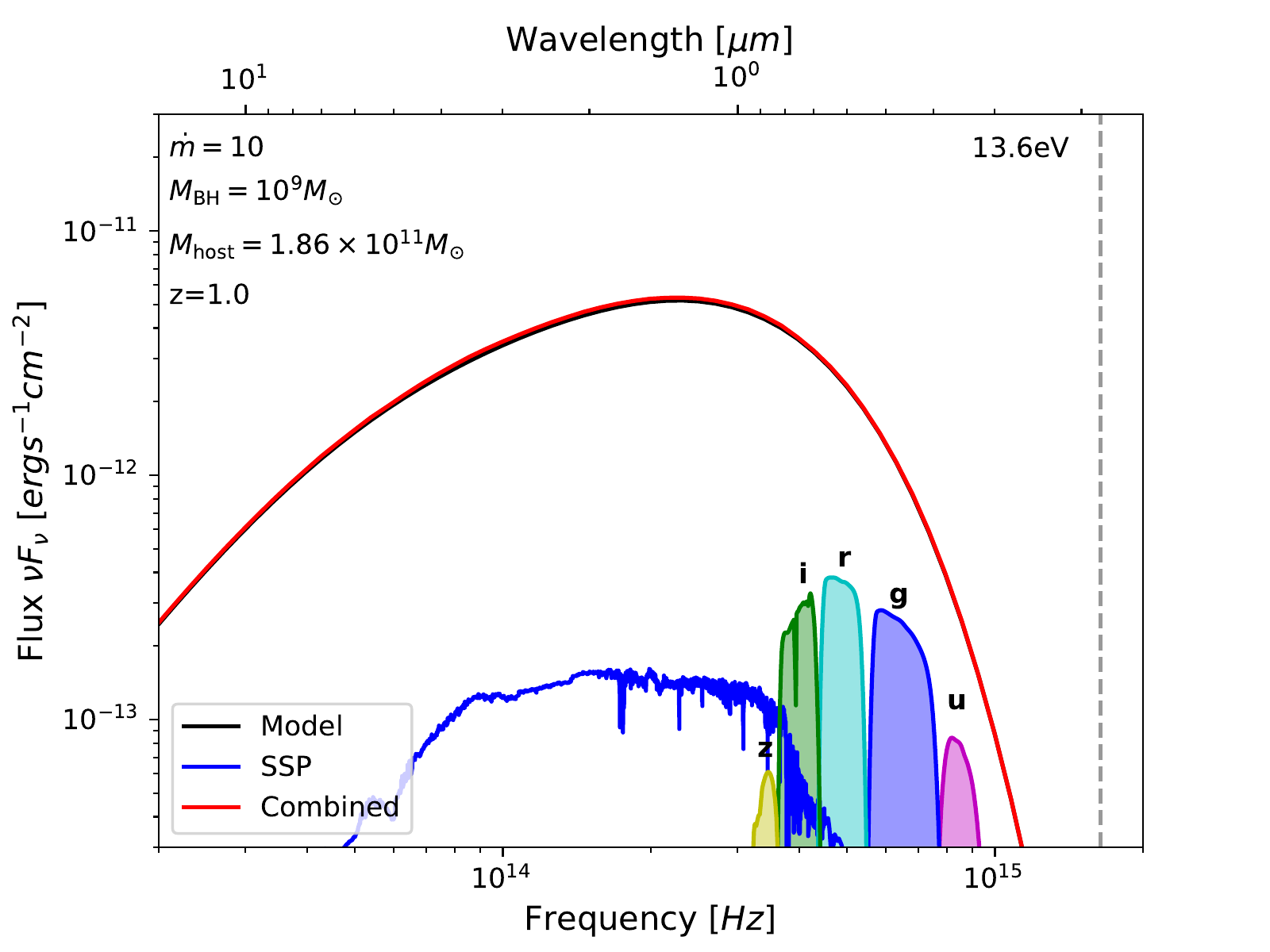}
\includegraphics[trim={0.2cm 0 1.5cm 0},clip,width = 0.48\textwidth]{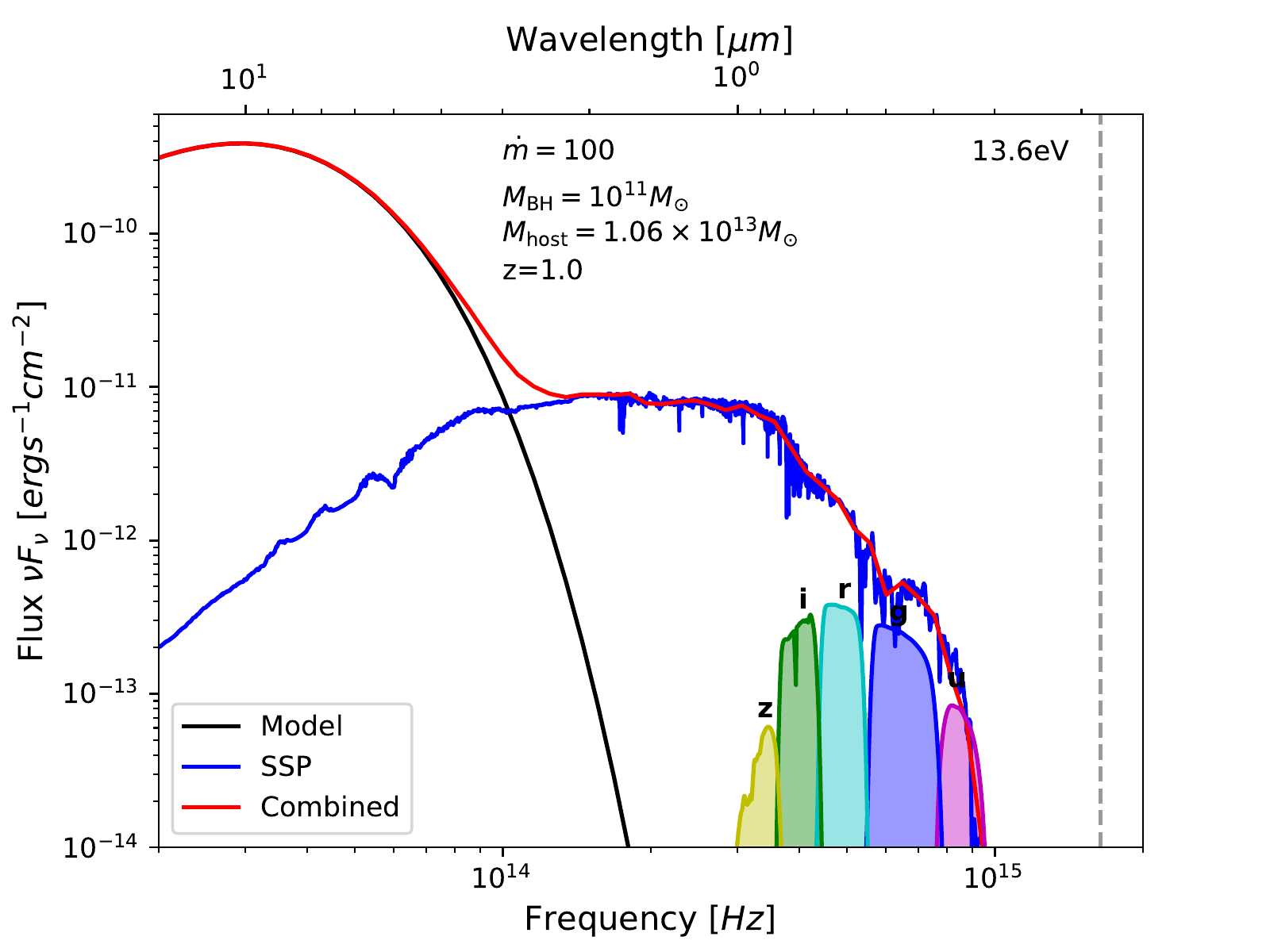}
\caption{The effect of adding host galaxy emission to our super-Eddington SEDs.
In each panel, we show the total calculated emission (red), which is composed of the super-Eddington SEDs (black) and a 1 Gyr old SSP (blue). 
The latter is scaled to a total mass that corresponds to the BH mass (see text for details).
The left top panel is chosen to represent a typically bluer model with low BH mass and Eddington ratio. The right top panel represents the `average' SED model with an intermediate BH mass and mildly super-Eddington ratio. Finally the bottom panel represents the reddest models with high BH mass and Eddington ratios. The models are all at $z=1$ with an inclination angle of 30\degree.}
\label{fig:SSP_SED}
\end{figure*}

We stress that we choose to omit more complicated physical features that may further affect the SED, such as outflows from the inner parts of the accretion flow, dust, and/or line emission and absorption, and only consider non-spinning SMBHs. 
The choice of taking a non-spinning BH for the AD model is motivated by the fact that the photon trapping radius extends to $r_{trap} \gg 10 \,r_{\rm s}$ for highly super-Eddington discs (see Eq.~\ref{eq:rtrap}). Thus, the SED changes expected for spinning SMBHs, which are ultimately linked to the inner-most parts of the disc, are expected to be heavily suppressed. Furthermore, in the interest of keeping the model relatively simple, we prefer to avoid taking into account potentially complex physics occurring for highly spinning BHs. 
Emission lines are not expected to significantly affect the broad-band SED, and moreover we expect that the ionizing radiation that drives line emission, will be suppressed as the UV photons from the inner disc are advected into the BH. 
This also supports ignoring dust effects, as the dust will have little UV emission to absorb and reprocess as infra-red emission. 
Other features such as outflows and jets are ignored as they are difficult to model and out of scope for the simple AD model used here, although their potential role in real super-Eddington systems cannot be ignored \cite[e.g.,][]{Dotan.Shaviv:11}.

We choose a grid of parameters spanning a range of BH mass, accretion rate, inclination angle and redshift. 
Our models cover BH masses and accretion rates in the range $\mbh = 10^7 - 10^{11}\, M_{\odot}$ and $\dot{m} = 1 - 100$, respectively, and we use three inclination angles, $10$, $30$, and $45\degree$.
Finally, each model SED is redshifted and scaled down to correspond to a range of redshifts $0.5 \leq z \leq 2.0$. 
The ranges and step sizes in all four parameters are listed in Table~\ref{tab:models}. 
Our grid thus yields 2592 individual super-Eddington model SEDs.
The parameter space choices are motivated mostly by observational considerations.
The mass range consistent with studies of large quasar samples for $z < 2$ \citep{McLure.Dunlop:04,Fine.etal:08,Trakhtenbrot.Netzer:12,Schulze.etal:15}.
The inclination angles of $10, 30, 45$ degrees are motivated by our choice to pursue mostly unobscured AGN. 
With a maximum redshift of $z=2$, we avoid the need to deal with significant absorption by the inter-galactic medium (IGM), which would affect the (rest-frame) UV part of the SEDs at higher redshifts, regardless of the intrinsic SED. This would have a degenerate reddening effect to that of the photon trapping, and thus would be counter-productive to the goals of this study. 

Figure \ref{fig:SED_filters} exemplifies the effect of increasing $\dot{m}$ for a specific choice of SED input parameters ($\mbh=10^9\,M_{\odot}$ at $z=1$), in the context of the SDSS flux limit for quasar spectroscopy.
The SDSS optical filters curves are scaled such that the top of the $i$-band filter curve is at $i_{\rm AB}=20.2$ -- the flux limit for spectroscopic follow-up observations in the legacy SDSS effort (see Section \ref{sec:SDSS} below). 
Thus, a model SED with $i$-band flux below this limit would be too faint to be targeted for SDSS spectroscopy, regardless of the SED shape (i.e., colours). 
The 13.6~eV line is plotted to illustrate how most of the model SEDs have very little ionizing emission. As predicted from the general considerations of photon trapping effects applied to an otherwise standard accretion disc, the SEDs are quasi-blackbody with little ionising UV emission, and increasing $\dot{m}$ shifts the peak SED emission to redder wavelengths. 
This suggests that quasar selection criteria that are originally designed to focus on the excess emission in the blue (or indeed near-UV) regime may fail to select SED shapes similar to our models. 
We investigate, quantitatively, the consequences of our simple photon trapping model on both flux- and colour-based aspects of the SDSS quasar selection algorithm in Section~\ref{sec:results}.

In addition to our grid of pure AD SEDs, we consider the effect of host galaxy emission by adding a simple stellar population (SSP) SED to the AD models. 
We obtain the SSP SEDs from the \texttt{GALAXEV} code presented in \citet{Bruzual.Charlot:03}\footnote{Available at \url{http://www.bruzual.org}}. The SSPs have a Chabrier IMF~\citep{Chabrier:03} with solar metallicities. 
We select three SSP ages of 0.2, 1, and 4 Gyr, in order to sample a broad range of ages and the resulting significant differences in the (rest-frame) blue regime, where our super-Eddington SEDs are suppresed and the host emission is expected to become significant. 
The SSP SEDs are initially scaled assuming that the {\it total} stellar mass follows the relation between BH and bulge mass observed in the local universe, following \citet[their Eqn.~11]{Kormendy.Ho:13}, and then redshifted and scaled to match the model SED redshift. 
These choices are meant to provide a rough estimation of the total host mass, and to provide an indication of how much host emission could affect the observed (total) colours of the super-Eddington model SEDs, with the age of the stellar population being they key factor. 
They are by no means extensive or intended to represent a detailed investigation of the intricacies of SMBH-host relations or their evolution. 

Figure \ref{fig:SSP_SED} shows the effect of adding a 1 Gyr old SSP to several of our super-Eddington model SEDs, all at $z=1$ but with different combinations of \mbh\ and $\dot{m}$, ranging from the bluest (left panel) to reddest (right panel) models. 
Appendix Figures~\ref{fig:200MYrSSP_SED} and \ref{fig:4GYrSSP_SED} demonstrate the 0.2 and 4 Gyr old SSPs, respectively, to the same super-Eddington models. 
Clearly, adding host galaxy emission will have little effect on the colours of mildly super-Eddington models ($\dot{m} \sim 10$) and masses that are comparable with the break in the black hole mass function (BHMF), of roughly $\mbh \sim 10^9\, M_{\odot}$ \cite[ e.g.,][]{McLure.Dunlop:04,Kauffmann.Heckman:09,Vestergaard.etal:08,Schulze.Wisotzki:10,Shen.Kelly:12,Schulze.etal:15,Weigel.etal:17}, since the SMBH-related SED is broad enough to dominate over the entire spectral range where the stellar contribution is relevant. 
The same stellar population is expected to slighly enhance the red emission for the bluest SMBH-related models -- those with low BH masses and Eddington ratios ($\mbh \sim 10^7 - 10^8 \,\Msun$, $\dot{m} = 1-3$). 
However, the most striking effect is for the reddest models: highly super-Eddington systems with high BH masses ($\mbh \gtrsim 10^{10}\,\Msun$, $\dot{m} \gtrsim 50$): 
as shown in the rightmost panel of Figure \ref{fig:SSP_SED}, in such cases the host emission dominates the optical regime, while the SMBH-related emission is essentially limited to the IR regime, at (observed) wavelengths $\lambda_{\rm obs} > 4\, \mic$. 
More generally, we may thus expect that the optical colours for such extremely high-\mbh\ and $\dot{m}$ systems will solely depend on the host galaxy properties (i.e., the specifics of its stellar and dusty gas content), while showing little or no evidence for the otherwise luminous, fast-growing SMBHs at their centres.

\subsection{SDSS Quasar Selection Algorithm}
\label{sec:SDSS}

The SDSS quasar spectroscopy selection algorithm is described in extensive detail in \citet{Newberg.Yanny:97} and \citet{Richards.etal:02}. In what follows, we recall its most basic and relevant features.
The algorithm was designed to provide a balance between high sample completeness ($> 90\%$) and purity ($>65\%$), mainly focusing on unobscured, (rest-frame) UV bright quasars. The algorithm uses the magnitudes from the five SDSS optical bands, $ugriz$ \citep{York.etal:00,Doi.etal:10}, to define a four-dimensional colour space, consisting of $u-g$, $g-r$, $r-i$ and $i-z$. These are then divided again into two subspaces, relying either on the $ugri$ or on the $griz$ bands, and used primarily for identifying low- and high-redshift targets respectively. The colour-colour spaces are then populated with multi-dimensional exclusion and inclusion zones, in which targets are automatically rejected or selected respectively. Exclusion zones are areas of colour space especially prone to contamination from non-AGN, point-like sources (i.e., stars), while inclusion zones are areas of colour space which are expected to be predominantly inhabited by quasars of all redshifts. It should be noted that the UV excess inclusion zone in the $ugr$ colour-colour space\footnote{Namely, $u-g<0.6$ in the AB system, which is used hereafter for all SDSS-related magnitudes and colours.} is the only 2-dimensional special zone. If zones overlap, exclusion zones are taken as priority over inclusion zones. 

\begin{figure*}
\center
\includegraphics[trim={0.6cm 0 1.5cm 0},clip,width = 0.48\textwidth]{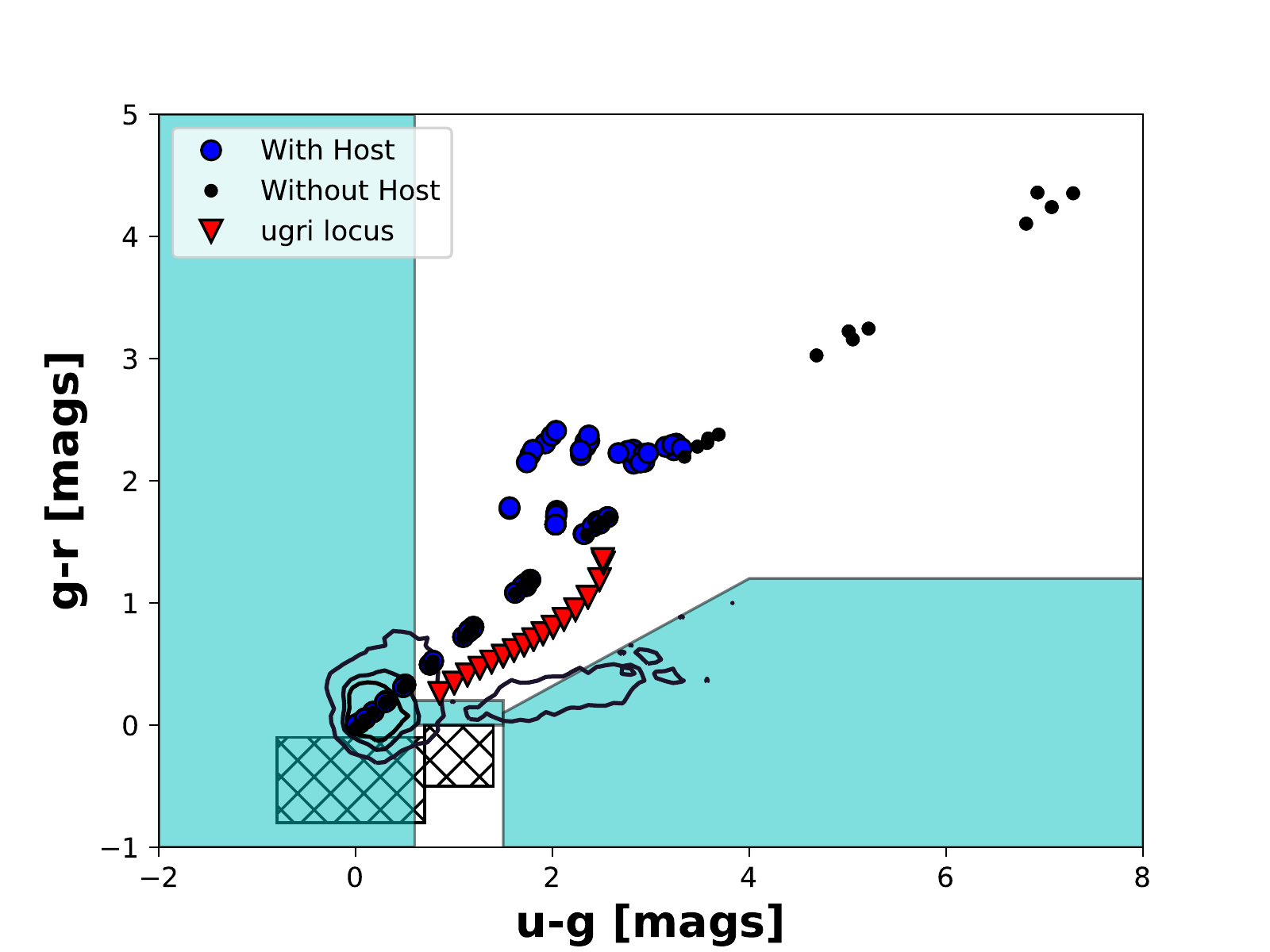}
\includegraphics[trim={0.6cm 0 1.5cm 0},clip,width = 0.48\textwidth]{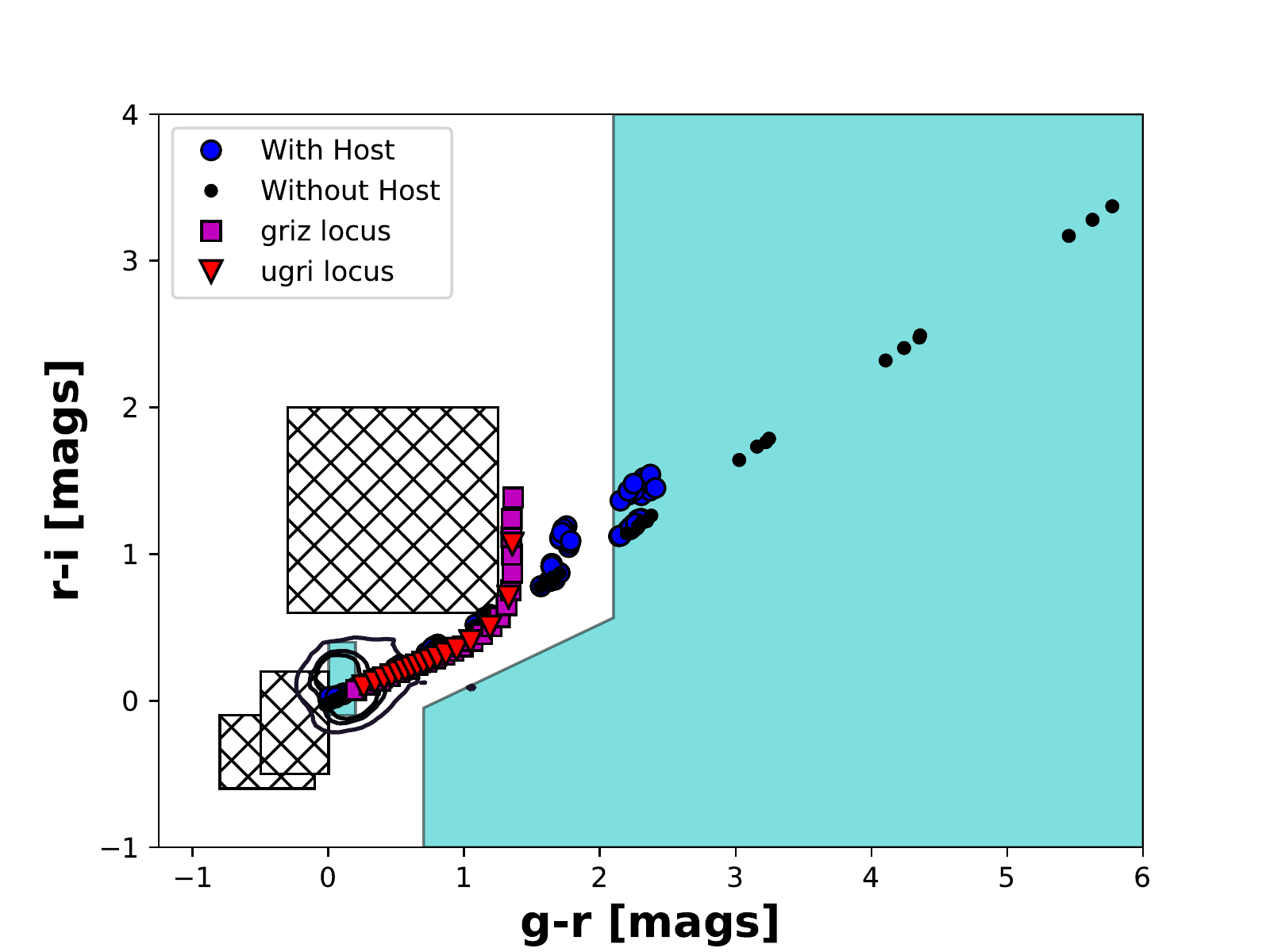}
\includegraphics[trim={0.6cm 0 1.5cm 0},clip,width = 0.48\textwidth]{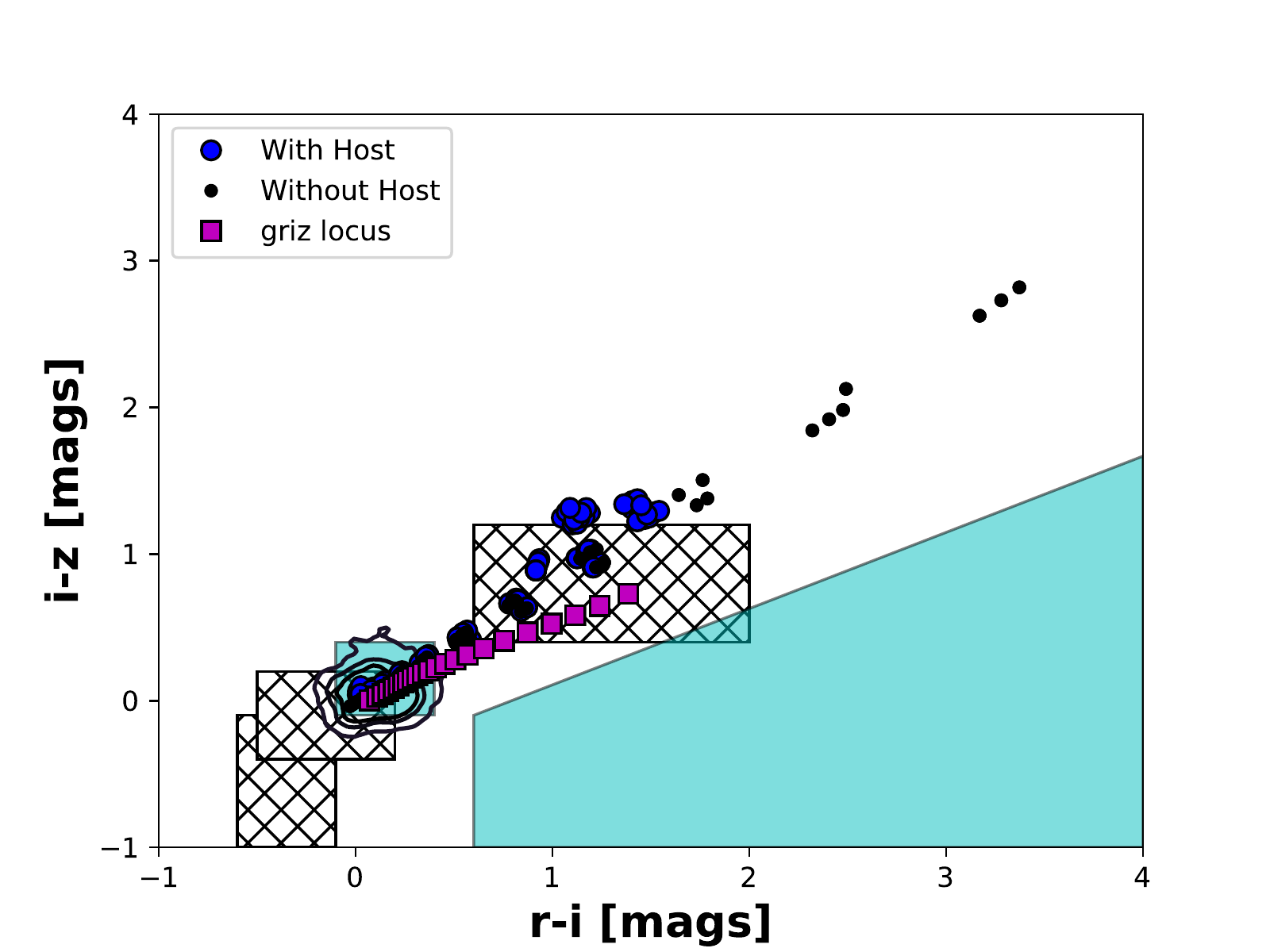}
\caption{Super-Eddington Quasar models at z=1.0 in \textit{ugriz} colour spaces. The cyan regions are SDSS inclusion zones and the crossed regions are SDSS exclusion zones, while the contours represent 60\%, 75\% and 90\% of SDSS DR7 quasars that have been selected using the colour selection algorithm. The red downwards triangles are the low redshift stellar locus centres, and the purple squares are the high redshift stellar locus. Note that \textit{gri} colour space has both stellar locii represented. Finally, the black points are the Super-Eddington models, and the blue points are the models with a 1GYr old SSP added. Note that these extend into very red colours (A-B > 10) and thus are truncated here. The blue points are only visible when the SSP makes a substantial colour difference, and are otherwise covered by the black points.}
\label{fig:colours}
\end{figure*}

In order to deal with possible contamination from normal galaxies at high redshift, and foreground stars, the quasar spectroscopic selection algorithm further uses the `stellar locus' as an additional set of exclusion rules \cite[see][]{Newberg.Yanny:97}. 
The construction of this locus is complex and multidimensional with $ugri$ and $griz$ projections, and is described fully in appendix A of \citet{Richards.etal:02}. 
It should be noted that as quasars are expected to be point sources, low redshift galaxies classified as extended objects are automatically rejected by the algorithm. 
A point source that is not in any special zone, and is sufficiently distant from the stellar locus in either $ugri$ or $griz$ colour space is then selected for follow up spectroscopy. 
Along with the elaborate colour criteria, targets must be within certain flux limits for spectroscopy with an $i$-band AB-like magnitude between 15 and 19.1 for the  $ugri$ space, which is designed to target low-redshfit quasars, and between 15 and 20.2 for the $griz$ colour space, which is designed to target fainter, high-redshift quasars. 
We finally note that, in addition to these purely-optical selection criteria, the SDSS also used a separate selection process for radio sources, based on cross-matching the optical imaging data with the Faint Image of the Radio Sky at Twenty cm survey \cite[FIRST;][]{Helfand.etal:15}. However, only a small percentage of the quasar candidates ($\sim 5\%$) \citep{Schneider.etal:10} are found relying \emph{solely} on this method. 
Thus we do not consider radio-based selection in the present study and focus instead on the optical selection criteria. 
%

\section{Selection of Super-Eddington SEDs in SDSS}
\label{sec:results}

In this section we examine the position of the super-Eddington SEDs in SDSS colour-colour space and discuss the effects of various parameters on the optical colours. We then examine how this affects the completeness of SDSS according to the quasar selection algorithm with regards to the super-Eddington SEDs.
For this, we obtain synthetic photometry for our model SEDs using the SDSS filter response functions (as seen in Figures \ref{fig:SED_filters} and \ref{fig:SSP_SED}).

\subsection{Magnitudes and Colours}
\label{subsec:colour_results}

The position of our $z=1$ model SEDs in the SDSS colour-colour space are shown in Figure~\ref{fig:colours}. In order to compare the model colours to those of real, though sub-Eddington, quasars, we also show the SDSS DR7 colour selected quasar sample \cite[black contours;][]{Schneider.etal:10}. We note that the colour selected DR7 quasars are not all \textit{solely} selected using the colour algorithm, and may also have radio components identified in cross-matching with FIRST. The blue and hatched regions in Fig.~\ref{fig:colours} represent the quasar colour algorithm special inclusion and exclusion zones (respectively; as described in Section \ref{sec:SDSS}). 
These regions are all 4-dimensional apart from the UV excess zone in $ugr$ colour space with $u-g < 0.6$. 
The black points in Figure \ref{fig:colours} represent the pure AD super-Eddington models, while the blue points also include host emission. We note that many of the colours do not change with host emission, especially for the bluer models (see Fig.~\ref{fig:SSP_SED}), and the black super-Eddington points lie exactly over the blue points with host emission. However, the redder super-Eddington SEDs end up having relatively \textit{bluer} colours that are driven by host (stellar) emission, consistent with our expectations (see bottom panel of Figure \ref{fig:SSP_SED}).

Figure~\ref{fig:colours} clearly shows that, regardless of host emission, only a small fraction of the super-Eddington model SEDs overlap with the actual observed SDSS quasar population. The models, even at relatively low (super-)Eddington ratios of $\dot{m} \sim 1-3$, do not span the SDSS DR7 colour selected quasar contours, though they do lie firmly within the contours. The subset of quasars with redder $u-g$ colours in the $ugr$ colour space arises from $z > 2$ quasars whose \textit{u}-band emission has been distinguished by IGM absorption. Since we limit ourselves to models with $z \leq 2$ and do not account for IGM effects, it is then not surprising that our models omit this area of colour space. 
We have further verified that simple sub-Eddington versions of our model SEDs, where there is little or no photon trapping, {\it do} correspond to the observed population of colour-selected, sub-Eddington SDSS quasars (see Appendix \ref{sec:subEdd} and Figure~\ref{fig:SDSS_subEdd} therein). 

Other factors are also expected to play a role in the difference between the model and observed colours. 
Since at the low-$\dot{m}$ regime ($\dot{m} \leq 3$) the effects of photon trapping are not yet significant (see Eq.~\ref{eq:rtrap}), colour differences may also be due to observed quasars having some contamination from dust \cite[e.g.,][]{Collinson.etal:15,Baron.etal:16}; effects of AGN outflows \cite[e.g.,][]{Slone.Netzer:12}; or other considerable contributions from broad emission lines \citep[see e.g.][]{Hao.etal:13}, which are ignored in our model. 
Furthermore, the observed SDSS quasar population is mostly sub-Eddington \cite[e.g.,][]{Trakhtenbrot.Netzer:12,Shen.Kelly:12,Kelly.Shen:13,Netzer.Trakhtenbrot:14}, and so some measure of difference between our model SEDs and real quasars is expected, as even the standard, sub-Eddington thin disc SED shape depends on $\dot{m}$. 
A non-negligible number of model SEDs are also close to, or even overlapping, with the stellar locus. This is not entirely surprising, as a significant part of the SDSS quasar contours overlap with the stellar locus as well, and little colour difference was expected between the bluer models and standard quasars.

Most importantly, many of our super-Eddington models are much redder than the standard quasar contours or the stellar locus. We note that Figures \ref{fig:colours} and \ref{fig:SDSS_subEdd} only cover the colour range considered by SDSS, while our entire grid of model SEDs, including those at $z > 1$, extends much further into redder colours, formally reaching $i-z \simeq 12$ and/or having essentially no emission in the $u$ band. 
As noted previously, the addition of host emission to these redder models often yields bluer optical colours than for the pure AD emission. In the most extreme case for $z=1$ as, plotted in Figure \ref{fig:colours}, the reddest model goes from $u-g \sim 25$ for pure AD emission to $u-g \sim 2$ once our simple host emission model is added. As host emission will dominate the optical emission for these extremely red AD models (see Figure \ref{fig:SSP_SED}), it follows that such combined SEDs would have galaxy-like colours and, at lower redshifts, may overlap with the stellar locus. 
These considerations suggest that our super-Eddington models may be most robustly identified in the IR regime -- a direction we investigate further in Sections~\ref{sec:NIR_sel} and \ref{sec:MIR} below.

Strikingly, none of the models fall within any of the multi-dimensional SDSS spectroscopy inclusion zones, though some do fall in the UV excess inclusion zone in the $ugr$ space. These are the bluest models, with $\dot{m} \leq 3$, and which indeed still have some UV emission. Similarly, most exclusion zones are avoided, though a very small number of models fall in the White Dwarf + A star pair exclusion zone in \textit{griz} space (middle and rightmost panels in Figure \ref{fig:colours}). 
Thus, we expect that most rejected models will be either due to stellar locus proximity or not falling within the required flux limits (see Section \ref{subsec:completeness}). 

\begin{figure*}
\center
\includegraphics[trim={0.6cm 0 1.5cm 0},clip,width = 0.48\textwidth]{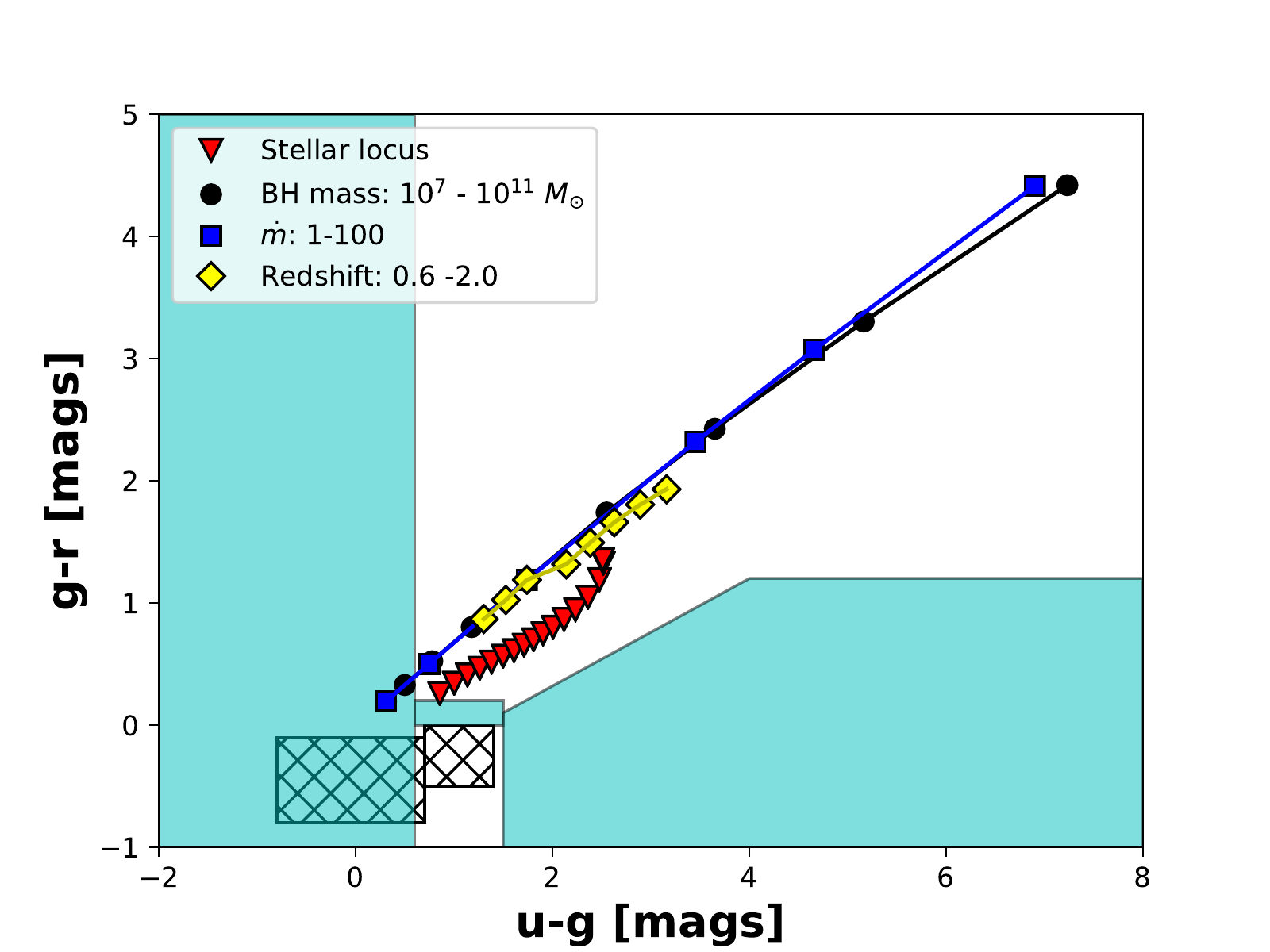}
\includegraphics[trim={0.6cm 0 1.5cm 0},clip,width = 0.48\textwidth]{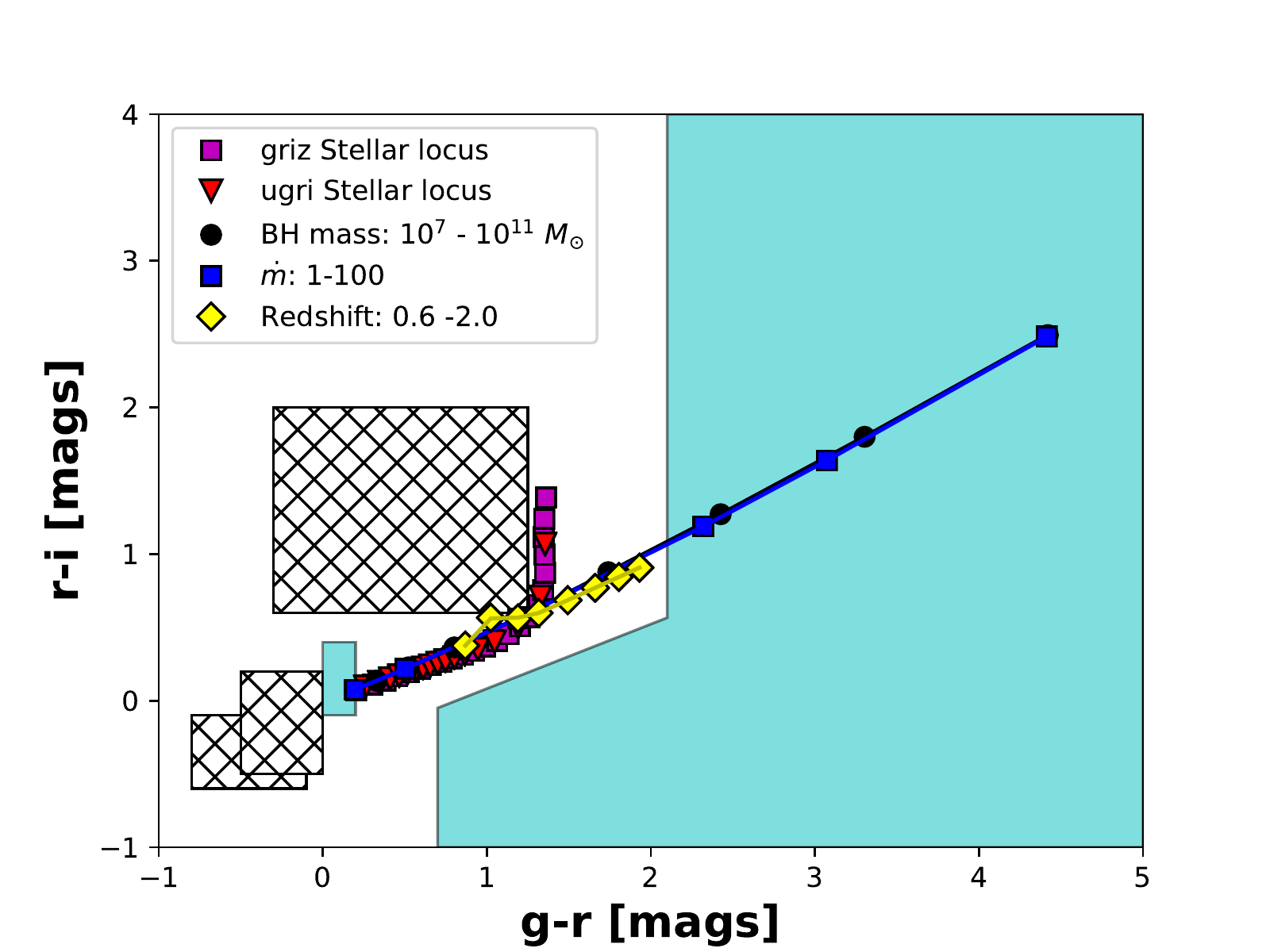}
\includegraphics[trim={0.3cm 0 1.5cm 0},clip,width = 0.48\textwidth]{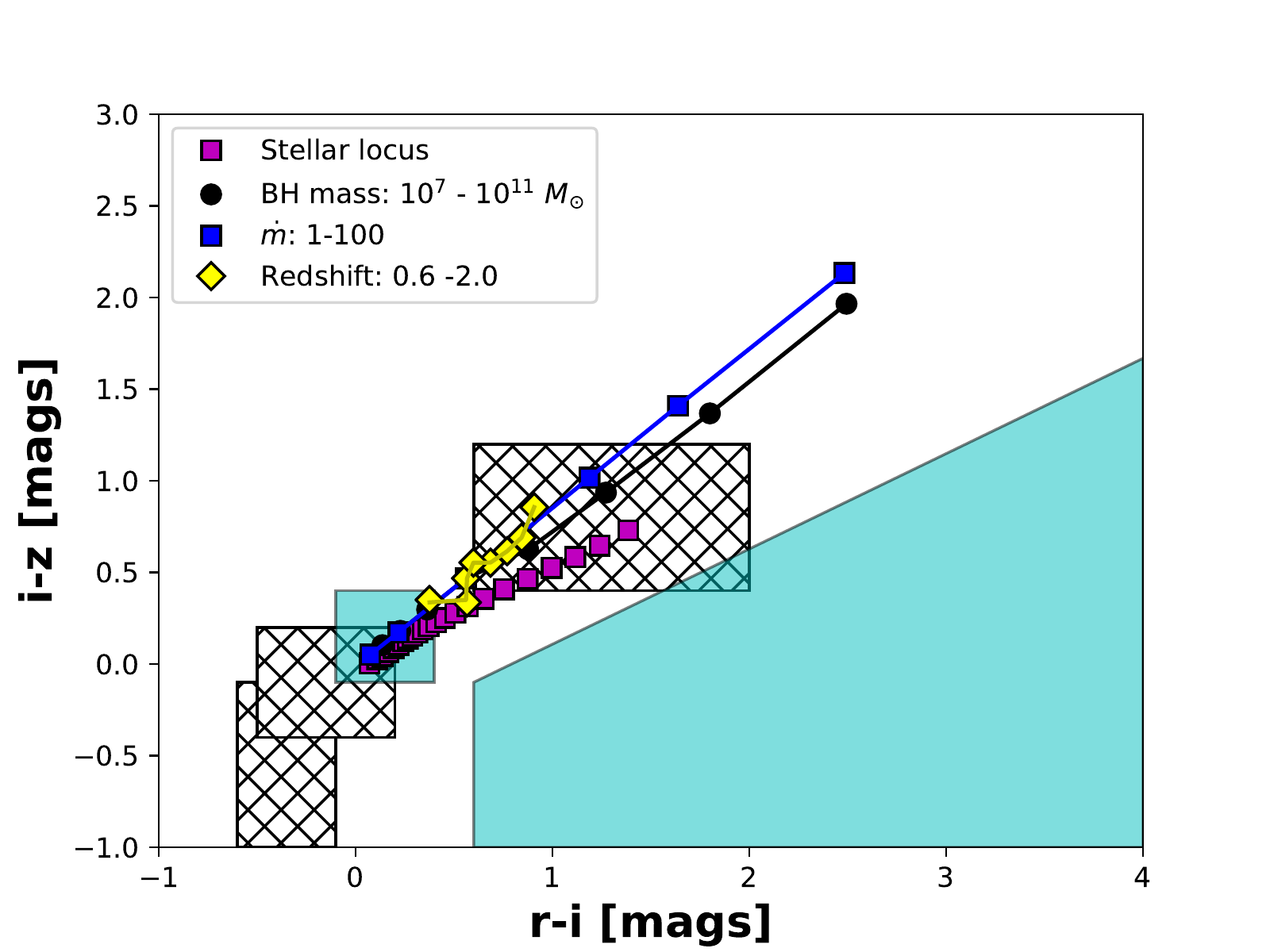}
\caption{ The effect of BH mass, Eddington ratio and redshift on our models. Increasing these values are relatively degenerate in optical colour space, effectively reddening the colours. The standard model here has a BH mass of $10^9 M_{\odot}$, an Eddington ratio of 10 and a redshift of 1.0. All models have an inclination angle of 30 degrees. Note that the redshift steps here are $\Delta z = 0.2$.}
\label{fig:SDSS_tracks}
\end{figure*}

Figure \ref{fig:SDSS_tracks} shows how the colours evolve when varying model parameters independently. We see a somewhat degenerate effect between increasing $\dot{m}$, \mbh, and $z$, where higher values in any of these parameters lead to redder colours. 
These trends follow naturally from the nature of both the standard thin disc and our simplistic photon-trapping model. 
Increasing $\dot{m}$ leads to large photon trapping radii, thus removing progressively redder photons from the SED (see Figure \ref{fig:SED_filters}). 
Increasing \mbh\ reddens the SED by moving the ISCO outwards (i.e. $r_{\rm ISCO} = 3 r_{\rm s} \propto M_{\rm BH}$), and thus reducing the temperature and emissivity of the inner AD regions.
We note that we do not vary the inclination angle for the tracks shown in Figure \ref{fig:SDSS_tracks}, as it has very little effect on colours relative to the other parameters, and thus we expect it to be negligible when applied to the SDSS quasar selection algorithm. 

\begin{figure}
\includegraphics[width=1.1\linewidth]{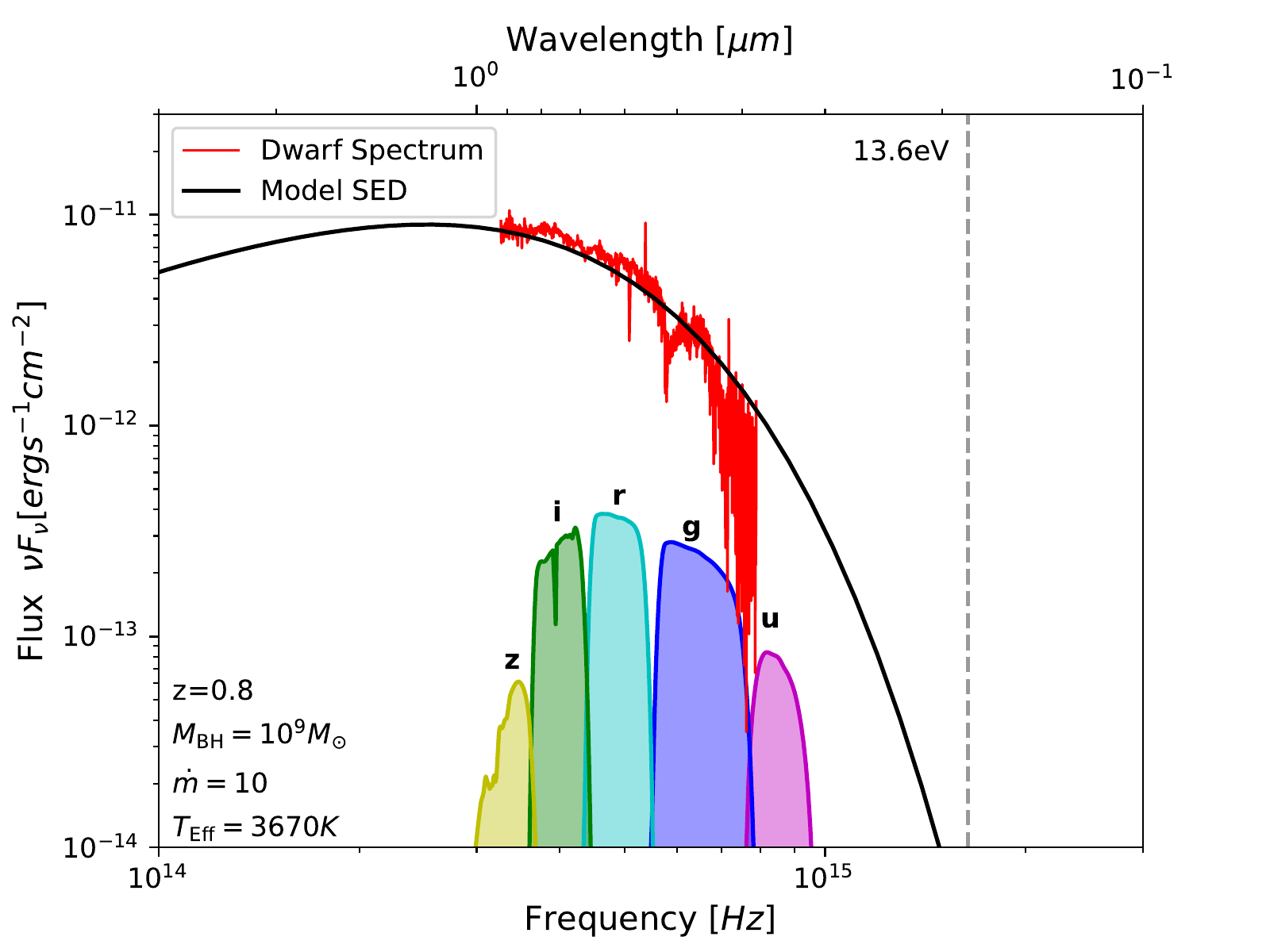}
\caption{Plot of a model SED in black with the spectrum of an M1 dwarf star in red over the SDSS optical filters. The optical spectrum of the dwarf star is scaled to fit the flux of the model SED. Given the close fit of the dwarf star spectrum to the model SED, it is likely that foreground stars will be a major source of contaminants in optical colour space.}
\label{fig:SED_dwarf}
\end{figure}

The quasi-blackbody shape of our (high-$\dot{m}$) super-Eddington model SEDs, and the need to distinguish them from foreground stellar sources motivates us to calculate the (rough) effective temperature, $T_{\rm Eff}$, for each model, using Wien's displacement law, and further associate it with a corresponding stellar type. 
We demonstrate this visually in Figure \ref{fig:SED_dwarf}, where the optical spectrum of an M1 dwarf taken from SDSS DR12 is compared with one of the model SEDs -- a hypothetical system with $\mbh=10^9\,\Msun$ and $\dot{m}=10$, placed at $z=0.8$. 
These model parameters make it representative of an `average' model for the range of parameters in this study (particularly, its mass is close to the break in the BHMF; see above). 
The dwarf star blackbody spectrum is remarkably similar to our model super-Eddington AGN SED, apart from the prominent absorption feature around $5000 \angstrom$ (from TiO). 
This supports the results shown in Figure \ref{fig:colours}, where many of the model SEDs are shown to have optical colours consistent with the stellar locus. 
Moreover, the close similarity between the M-dwarf spectrum and the super-Eddington SED may provide a hint for why super-Eddington AGN with a `truncated' thin (or slim) disc have not yet been identified: if a population of (rare) super-Eddington AGN photometrically resembles stellar sources, it is possible that super-Eddington quasars may have been dismissed from any further follow up. 
Unfortunately, as these stars are both long-lived and extremely common in the Milky Way, managing to distinguish a rare super-Eddington quasar from a dwarf star without using costly spectroscopic, or multi-wavelength resources remains challenging.

\subsection{SDSS Completeness}
\label{subsec:completeness}

We next quantify just how incomplete the SDSS may be to the sort of super-Eddington AGN depicted by our model SEDs. 
We apply the SDSS quasar spectroscopy selection algorithm to our synthetic flux measurements (magnitudes) and can thus determine, for each super-Eddington model SED, whether it would have been included in the legacy spectroscopic SDSS quasar data set. 
We further bin our models into three redshift bins of size $\Delta z = 0.5$, ranging over the entire redshift range of $z = 0.5 - 2.0$, which we hereafter refer to as the low-, mid- and high-redshift bins, respectively. Each bin has 972 individual SEDs, where the models located at bin boundaries are shared among the bins.\footnote{For example, the set of model SEDs with $z=1.0$ are found both in the low- and mid-redshift bins.} 

The results of the SDSS selection algorithm applied to the super-Eddington model SEDs are tabulated in Table \ref{tab:selection}. From the entire set of super-Eddington SEDs, many appear to be too faint to be selected for follow-up spectroscopy, that is they do {\it not} fall within the $i$-band magnitude range of $[15,20.2]$ for the $griz$ selection, or if not chosen by any of the $griz$ criteria, they fall outside of the $ugri$ space $i$-band magnitude range of $[15,19.1]$. 
We call these models `unobservable' hereafter (and in Table \ref{tab:selection}), while the models within the flux limits are called `observable'. 
It is immediately clear that increasing the redshift has a drastic effect on the number of models that fall within the spectroscopic flux limits. The number of observable models going from the low to high $z$ bins falls from 626 to 418 for the pure AD SEDs, and from 765 to 521 for the host-added SEDs. 
The number of models selected by the algorithm, and accordingly the total selection percentage, also drop with redshift, from $\sim 30$ to $\sim 10\,\%$ (for both cases with and without host emission). 
Focusing instead on the percentage of observable models selected for SDSS spectroscopy, we find that these, too, drop with increasing redshift. However, it is perhaps more interesting to note that the addition of host emission lowers the observable selection percentage for the low- and mid-$z$ bins (from $43.5\%$ to $25.5\%$ and from $33.6\%$ to $23.9\%$, respectively), yet raises it slightly for the high $z$ bin (from $25.4\%$ to $29.2\%$). 
In the case of the first two redshift bins, the number of observable models significantly increases when host emission is added (from 626 to 765 and from 494 to 724, respectively). However, it is likely that the newly observable models then fall within the stellar locus or into an exclusion zone (see Figure \ref{fig:colours}), thus leading them to not being selected for spectroscopy, and thus reducing the percentage of observable selected (mock) sources.
In the high-$z$ bin, the number of observable models also increases from 418 to 521, and contrary to the other redshift bins, the number of selected models also increases from 106 to 151, which corresponds to a selection percentage increase from $10.9\%$ to $15.6\%$. 
We conclude that the completeness of SDSS for the super-Eddington model SEDs remains quite low, regardless of host emission: the percentage of mock systems that would have been selected for follow-up spectroscopic observations never exceeds $30\%$. 

\begin{figure*}
\center
\includegraphics[trim={0.3cm 0 1.6cm 0.6cm},clip,width = 0.48\textwidth]{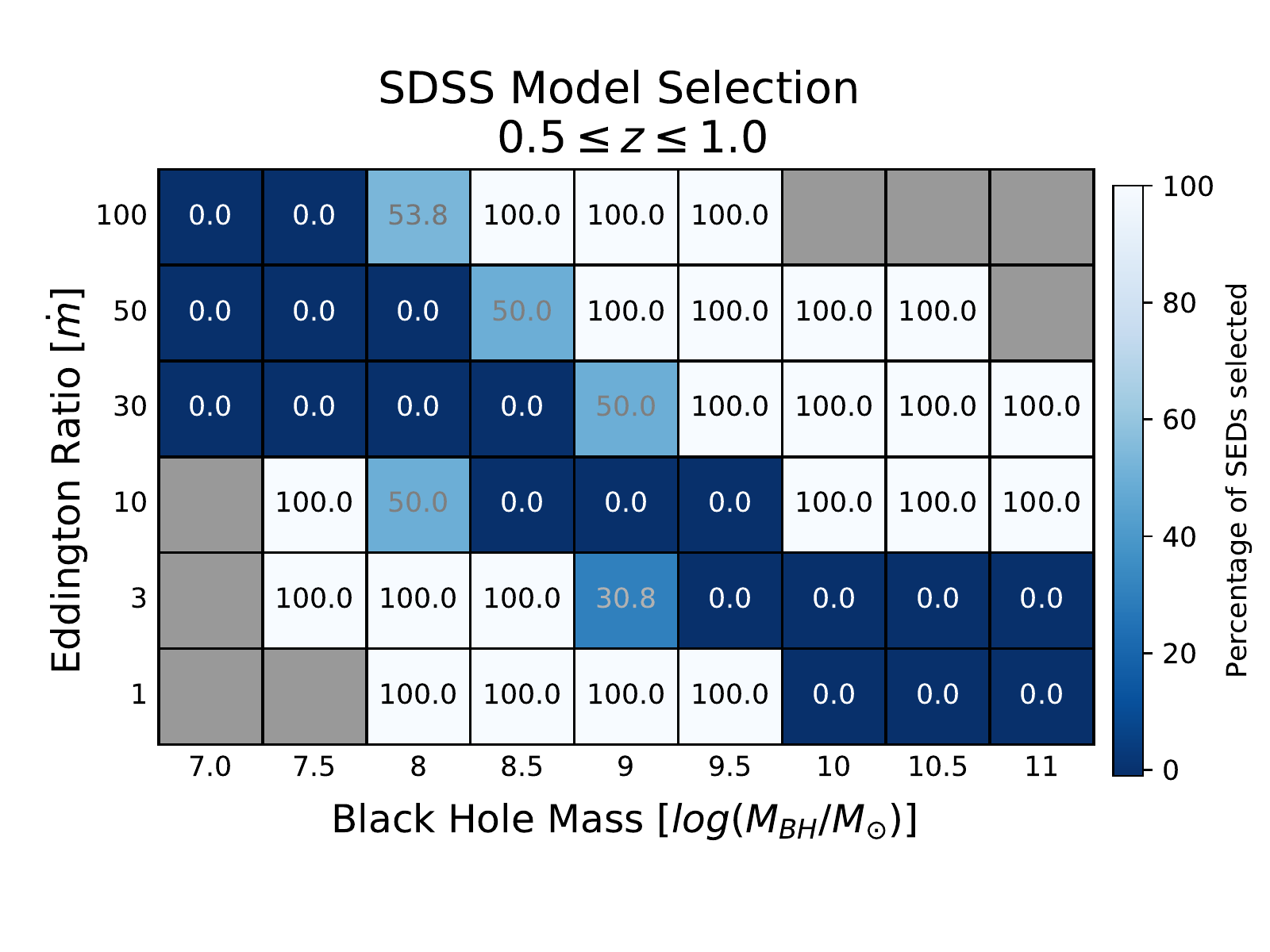}
\includegraphics[trim={0.3cm 0 1.6cm 0.6cm},clip,width = 0.48\textwidth]{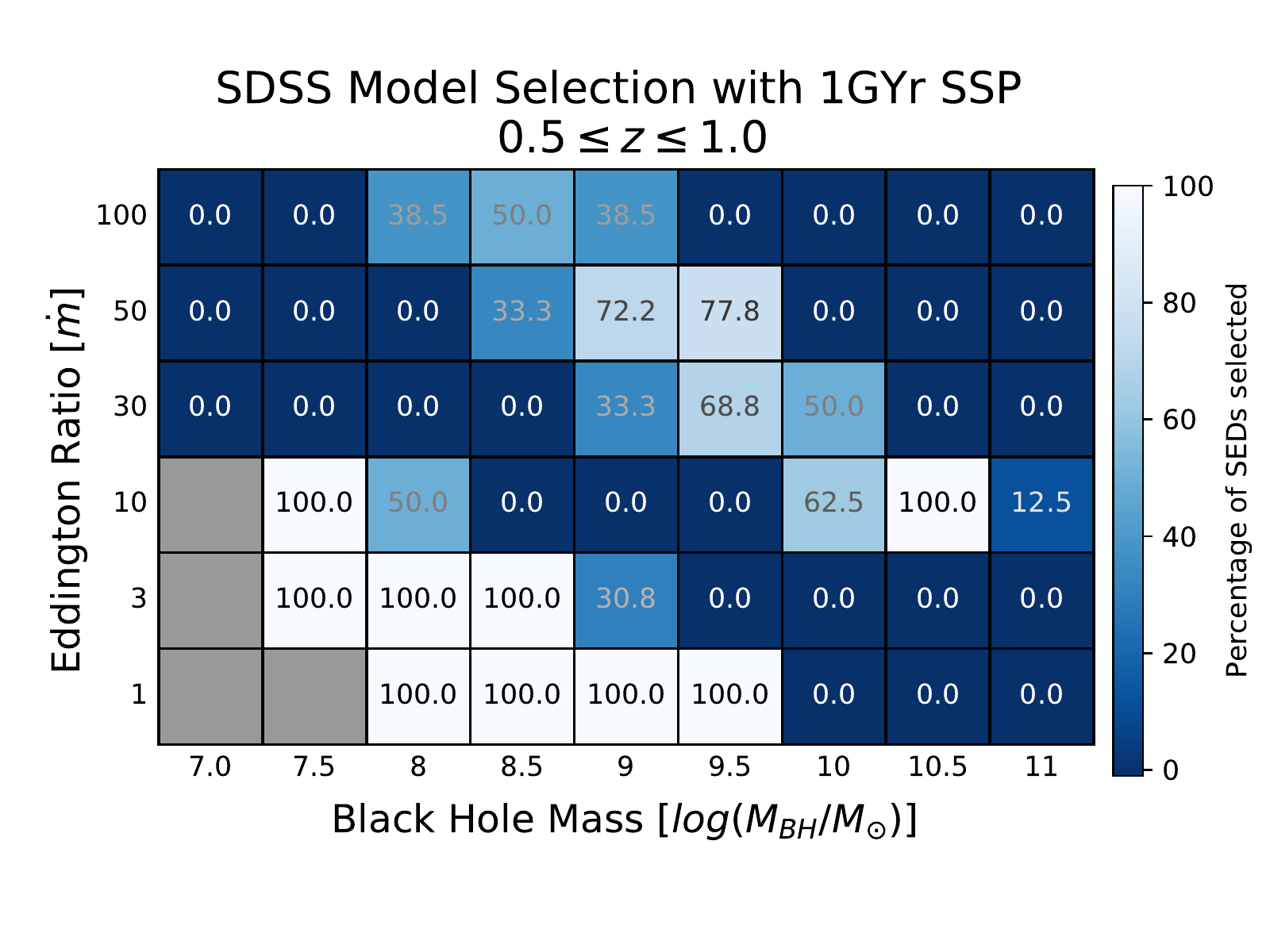}
\includegraphics[trim={0.3cm 0 1.6cm 0.6cm},clip,width = 0.48\textwidth]{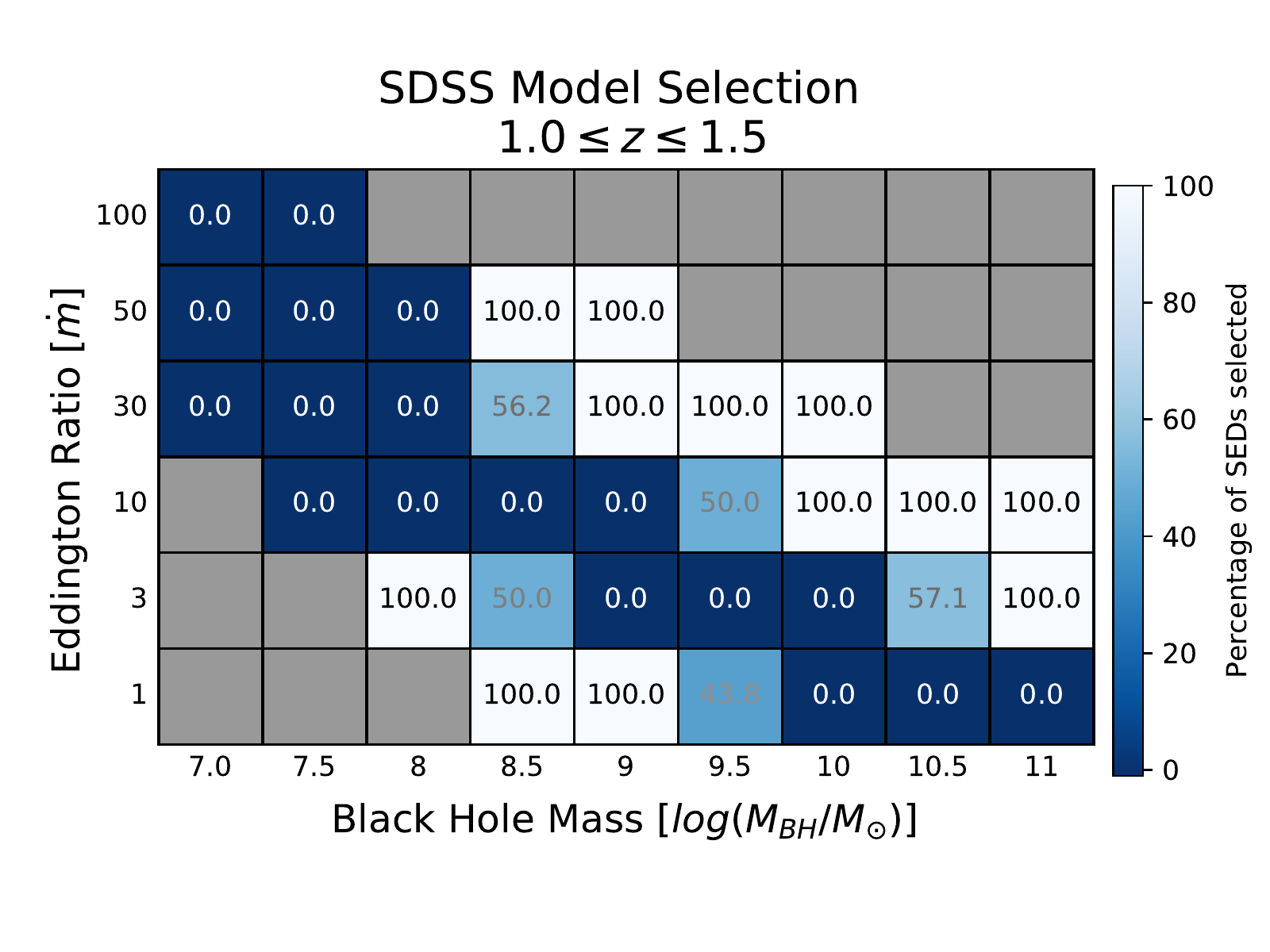}
\includegraphics[trim={0.3cm 0 1.6cm 0.6cm},clip,width = 0.48\textwidth]{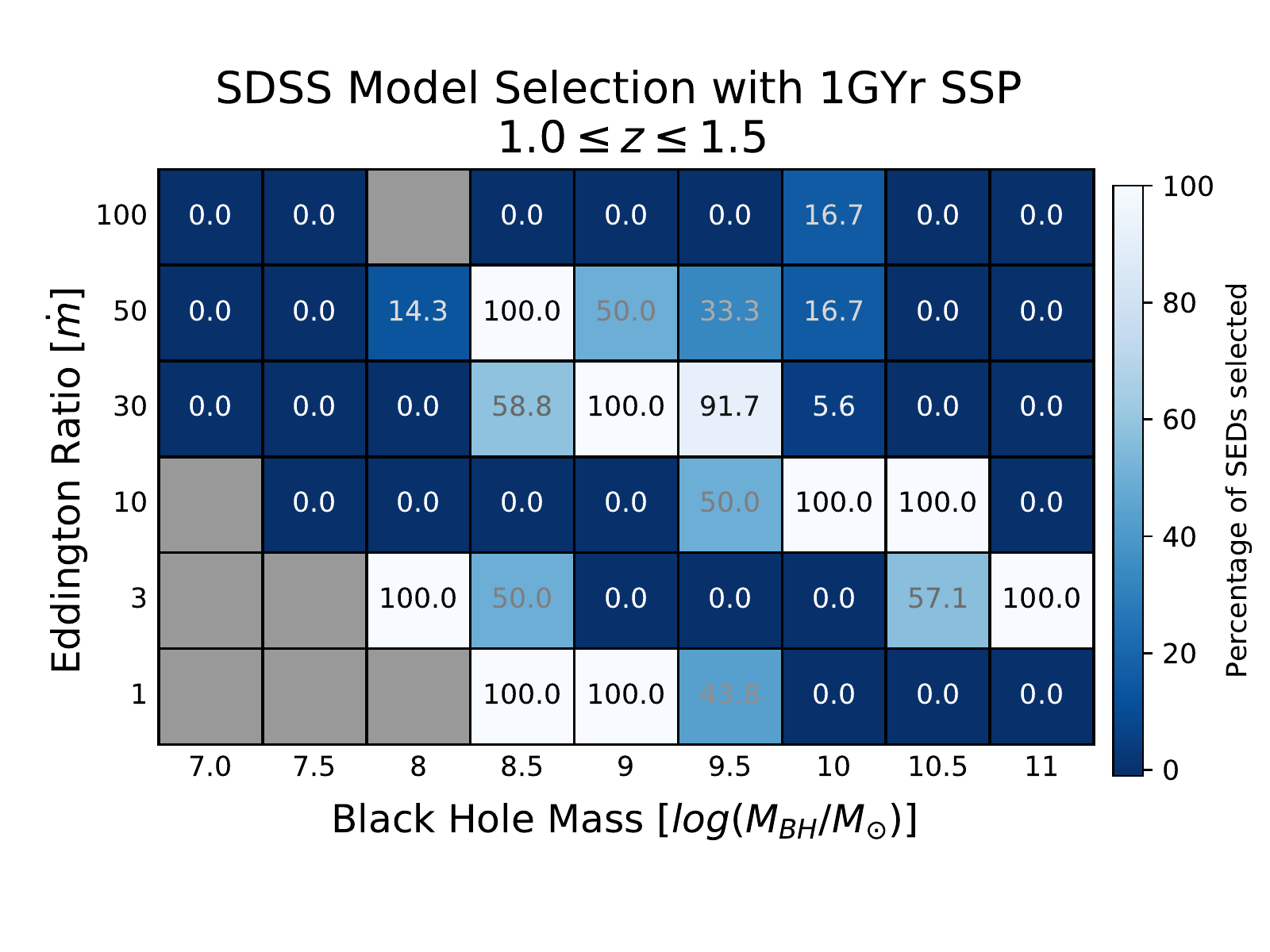}
\includegraphics[trim={0.3cm 0 1.6cm 0.6cm},clip,width = 0.48\textwidth]{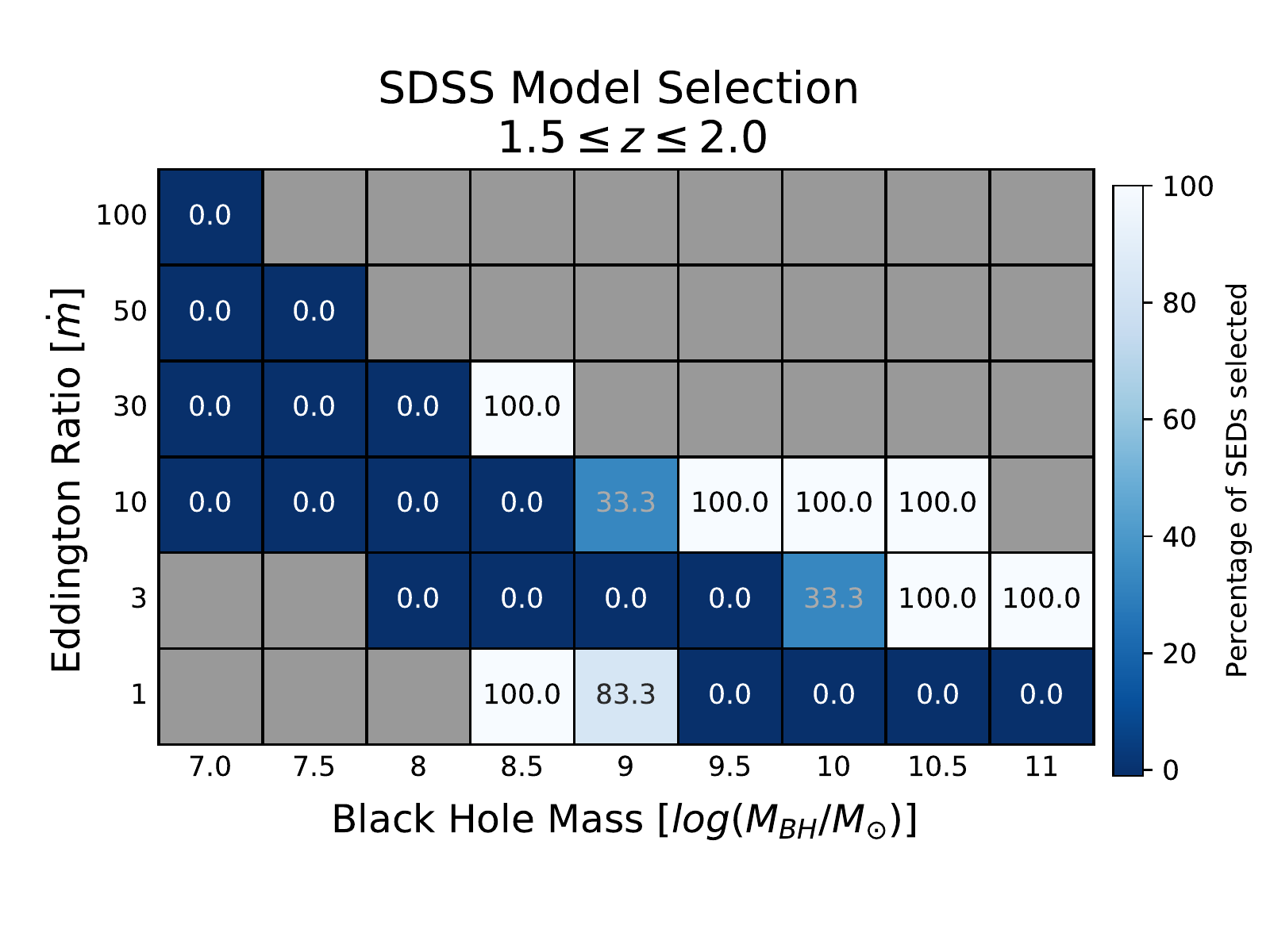}
\includegraphics[trim={0.3cm 0 1.6cm 0.6cm},clip,width = 0.48\textwidth]{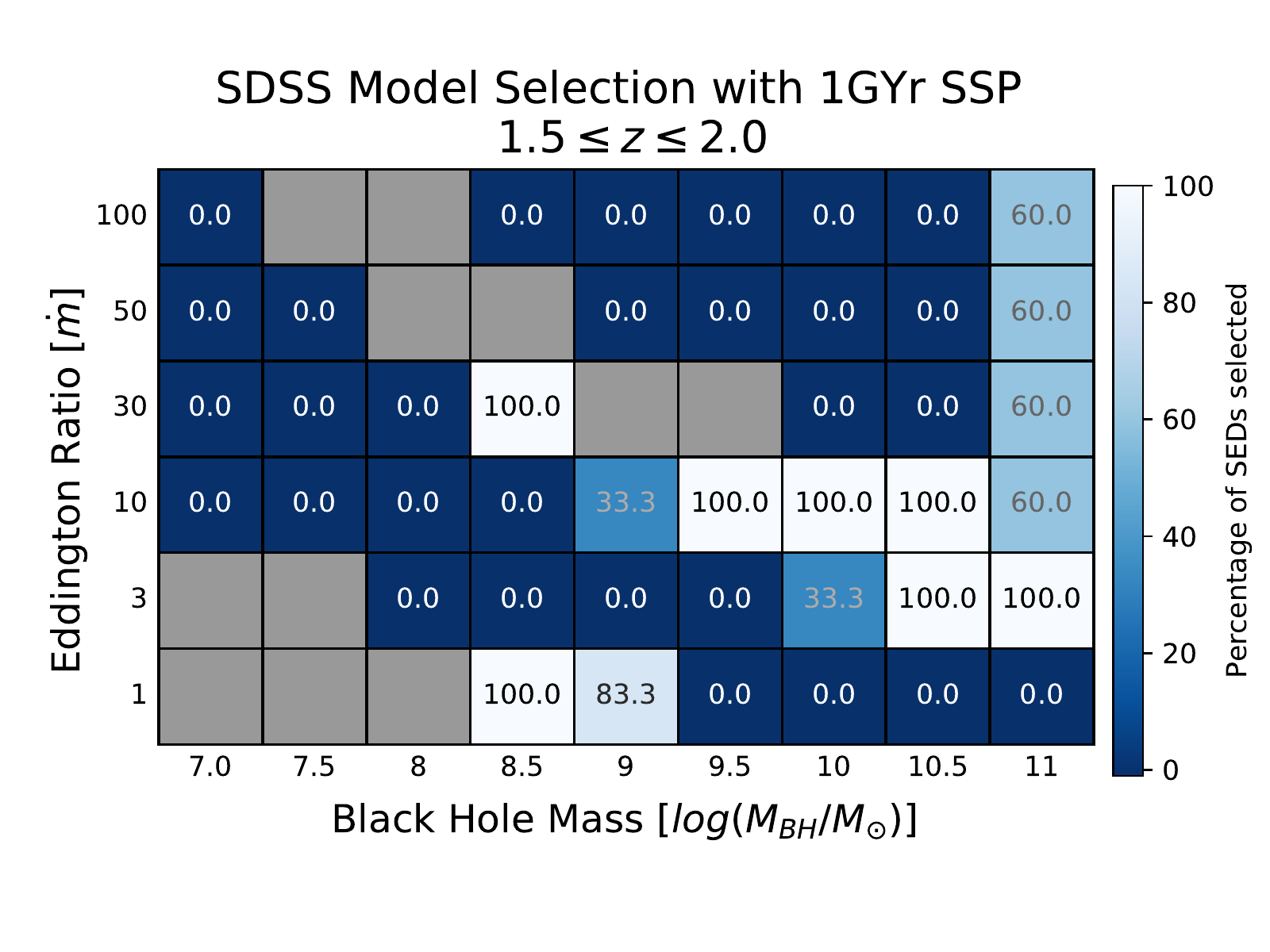}
\caption{Selection grids for three redshift bins in terms of observable models (i.e. within SDSS magnitude limits). The grayed out squares represent parameter spaces with no observable models. The percentage of SEDs selected are with regards to the number of observable models in that part of parameter space, as opposed to the total number of models in that space.}
\label{fig:sel_grids_flux}
\end{figure*}

Figure \ref{fig:sel_grids_flux} further illustrates the percentage of model SEDs that would be considered both `observable' and selected for SDSS spectroscopy, across a grid of $\dot{m}$ and \mbh, for the three redshift bins we consider, and for both pure AD emission and for AD with host galaxy emission. 
This allows the effects of varying input parameters to be studied more closely, as well as visualising the effect of redshift on particular subsets of models.
In what follows we discuss the effects of varying redshift, BH mass, and accretion rate, on the possibility that the corresponding super-Eddington model SEDs would be selected for SDSS spectroscopy.

We emphasize that the percentages presented here are relative to the number of {\it observable} models, not total number of models with a certain set of input parameters, where we recall that a model SED is considered `observable' if it is brighter than the $i$-band flux limit for SDSS spectroscopic follow-up. 
This means that the percentages shown in Fig.~\ref{fig:sel_grids_flux} reflect only the colour-based selection criteria, but on the other hand not all the squares have an identical number of models included in the corresponding percentage calculation. 
This is an important nuance, especially for the bins in parameter space which have 100\% selection rate but also border other bins with no observable models (i.e., squares with 100\% neighbouring squares with 0\%). 
These parts of parameter space may have many models that are outside the flux limits (`unobservable') and thus not included in the calculation of the selection percentage. 
Parts of parameter space that do not have any models within the flux limits are shaded in gray.

As shown numerically in Table \ref{tab:selection}, the parameter space of observable models in Figure \ref{fig:sel_grids_flux} is reduced when moving from the low to high redshift bins. We first consider the left side of Figure~\ref{fig:sel_grids_flux}, representing pure AD emission models. The redder models with high accretion rates and masses ($\dot{m} \geq 30$ and $\mbh \geq 10^9\,\Msun$) quickly become too faint to be observable as redshift is increased. This is due to the peak of the SED shifting to wavelengths greater than the $i$-band which is used for the flux limit, and thus insufficient flux is present in the $i$-band for follow up spectroscopy.

\begin{table*}
\caption[SDSS Super-Eddington SED Selection Statistics]{SDSS Super-Eddington SED Selection Statistics}
\label{tab:selection}
\resizebox{\linewidth}{!}{
\def\arraystretch{1.3}%
\begin{tabular}[]{c c c c c c c c}
\hline
 & Redshift & Observable & Unobservable & Selected & Total Percentage & Observable Percentage\\
 & bin & SEDs & SEDs & & Selected & Selected \\
\hline
 & 0.5$\leq$z$\leq$1.0 & 626 & 346 & 272 & 28.0 & 43.5 \\
No host & 1.0$\leq$z$\leq$1.5 & 494 & 478 & 166 & 17.0 & 33.6 \\
 & 1.5$\leq$z$\leq$2.0 & 418 & 554 & 106 & 10.9 & 25.4 \\
\hline
 & 0.5$\leq$z$\leq$1.0 & 765 & 207 & 195 & 20.1 & 25.5 \\
With host & 1.0$\leq$z$\leq$1.5 & 724 & 248 & 173 & 17.8 & 23.9 \\
 & 1.5$\leq$z$\leq$2.0 & 521 & 451 & 151 & 15.6 & 29.2 \\
\hline
\end{tabular}
}
\end{table*}

The bluest, pure AD models with both low $\dot{m}$ and low \mbh\ are almost always unobservable, regardless of redshift. Such models are often just slightly fainter than the $i=19.1$ limit of the low-$z$ ($ugri$) criteria of the selection algorithm. 
Increasing the model redshift then adds flux to the $i$-band, while also dimming the overall flux of the SED (due to the larger corresponding distance). 
In the case of the bluest models, the latter effect is more significant than the former. 
Thus, as these models reach the higher redshift bin, their $i$-band flux drops and they fall below the the $i=20.2$ magnitude limit of the high-$z$ ($griz$) selection criteria, and remain unobservable. 

The models with `intermediate colours', that are neither on the blue nor the red ends of the range covered by our model SEDs, form an anti-diagonal band across the parameter grid in Figure~\ref{fig:sel_grids_flux}, and remain mostly observable regardless of redshift. 
However, many of these models are never selected for spectroscopy by the colour-based algorithm, leading to the low selection percentages seen along the corresponding anti-diagonals (i.e., 0\%).
This is most likely due to them being in close proximity to the stellar locus and thus being rejected (see \cref{fig:colours,fig:SDSS_tracks}). Increasing the redshift is expected to redden some of the models and move them away from the stellar locus, but can also lead to them falling below the flux limits for spectroscopy (see, e.g., the behaviour of the square with $\dot{m} = 50$ and $\mbh = 10^8\,\Msun$ across the redshift sequence, as an example of this behaviour).

The effect of host emission can be clearly visualised in the right panels of Figure \ref{fig:sel_grids_flux}. 
For the reddest models that were previously unobservable, adding host emission allows them to exceed the $i$-band flux limit and thus remain observable even in the higher redshift bin.  However, their colours prevent them from being selected for spectroscopy. 
This can be seen when comparing the highest $\dot{m}$ and \mbh\ regions in adjacent panels of Fig.~\ref{fig:sel_grids_flux} (i.e., with or without host emission): squares which were gray when considering pure AD models (left panels) turn to have 0\% selection when host emission is added (right panels).
This is consistent with the optical emission being dominated by host emission, yielding galaxy-like colours which would not be marked as potential AGN candidates (see Figures \cref{fig:SSP_SED,fig:colours}). 
Surprisingly, some of the most extremely red, highest-mass models ($\dot{m} \geq 10$ and $\mbh = 10^{11}\,\Msun$) in the high redshift bin with added host emission appear to be selected by the algorithm. This is unexpected as these models should have optical emission entirely dominated by the host galaxy, and thus one would expect that they would not be selected by the {\it quasar} selection algorithm. 
However, it should be noted that these model SEDs have extreme BH and host galaxy masses (i.e., of order of $M_{\rm host}\simeq 10^{13}\,\Msun$), and are thus expected to be extremely rare, if they exist at all at $z\gtrsim1.5$ \cite[e.g.,][]{Ilbert.etal:13,Davidzon.etal:17}. 
As such, the number or fraction of high-mass, high-redshift super-Eddington model SEDs potentially selected for SDSS spectroscopy should be interpreted with caution.

Figure \ref{fig:sel_grids_flux} also allows us to visualise the effect of changing the individual $\dot{m}$ and \mbh\ parameters of our SEDs. 
From Figure \ref{fig:SDSS_tracks}, it is clear that increasing these two parameters has the somewhat degenerate effect of reddening the expected optical colours. 
Starting from the bluest models with $\dot{m} \leq 10$ and $\mbh \leq 10^8\,\Msun$ which are initially unobservable, we see that increasing the BH mass and/or Eddington ratio will increase $i$-band flux (see, e.g., Figure \ref{fig:SED_filters}), eventually exceeding the flux limit and resulting in observable SEDs. 
However, purely increasing $\dot{m}$ while keeping $\mbh \leq 10^8 \Msun$ often does not yield SEDs that are selected by the spectroscopic selection algorithm. 
As increasing $\dot{m}$ significantly shifts the SEDs to redder wavelengths, the steps in $\dot{m}$ may be too large to place these SEDs into a selection area (i.e. observable values would lie in the range ($10 < \dot{m} < 30$ for $\mbh \leq 10^8 \Msun$, depending on redshift). Above $\dot{m} = 30$, these low mass models are observable, but most likely with colours in close proximity to the stellar locus, and thus rejected by the selection algorithm. Conversely, increasing \mbh\ alone has a smaller reddening effect per step size, allowing models with $\dot{m} \leq 10$ to be selected when increasing the BH mass, until the high mass $\mbh \geq 30$ models reach stellar-like colours and are thus rejected. 

The vast majority of selected models lie in the area of parameter space that yields colours redder than the stellar locus, but still blue enough to have the required $i$-band flux to be deemed observable. Naturally, increasing $\dot{m}$ and/or \mbh\ for these models will shift the peak of their emission to yet redder wavelengths, decreasing their $i$-band flux so that it drops below the flux limit for spectroscopy, and resulting in these SEDs being unobservable. For the SEDs with host emission, the AD emission will be reddened, leading to the host dominating optical emission. Such models will remain observable thanks to the host's $i$-band flux, but will also have galaxy-like optical colours, and are not selected by the quasar algorithm.

\begin{figure*}
\center
\includegraphics[trim={0.2cm 0 1.5cm 0},clip,width = 0.48\textwidth]{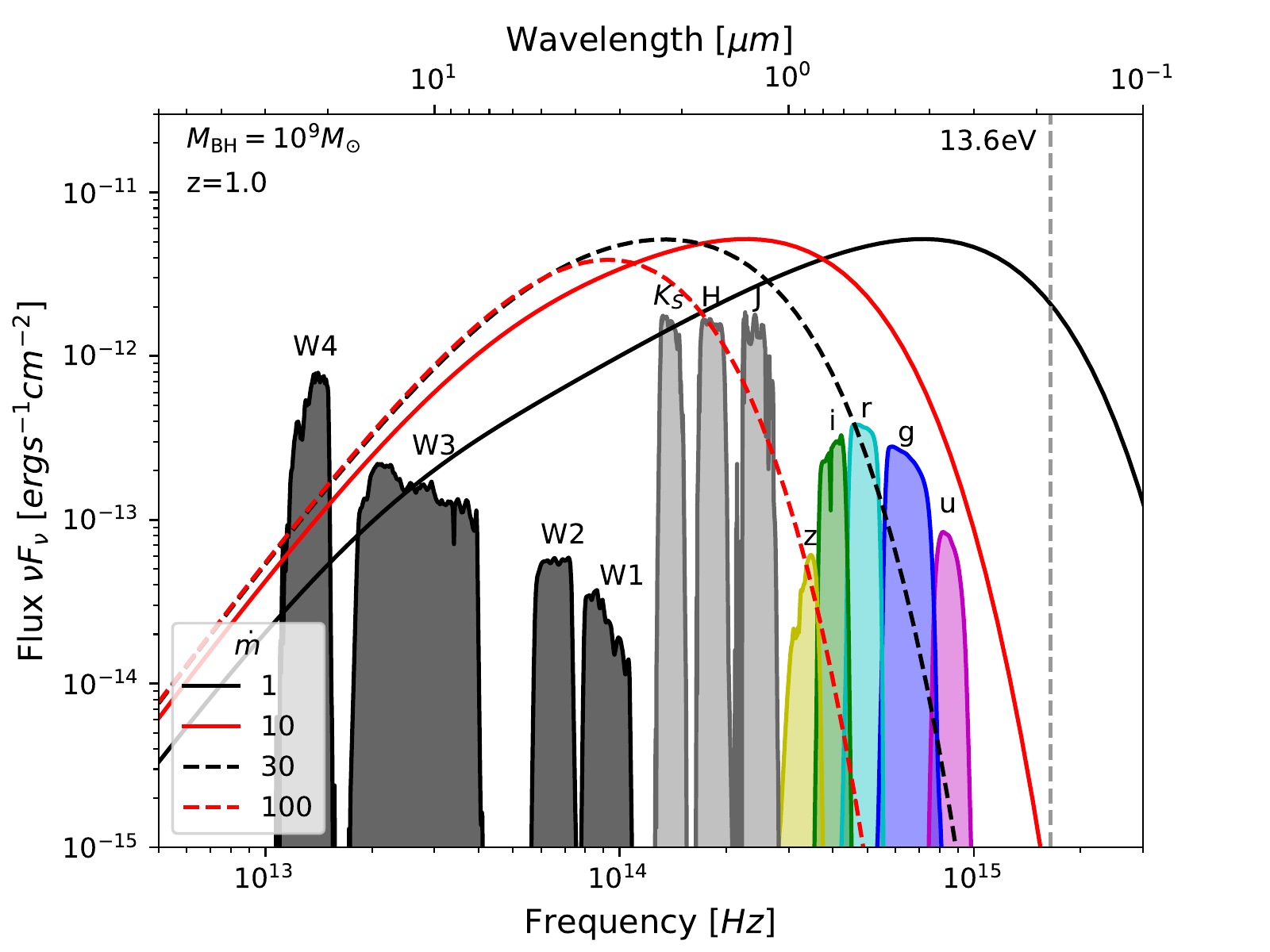}
\includegraphics[trim={0.2cm 0 1.5cm 0},clip,width = 0.48\textwidth]{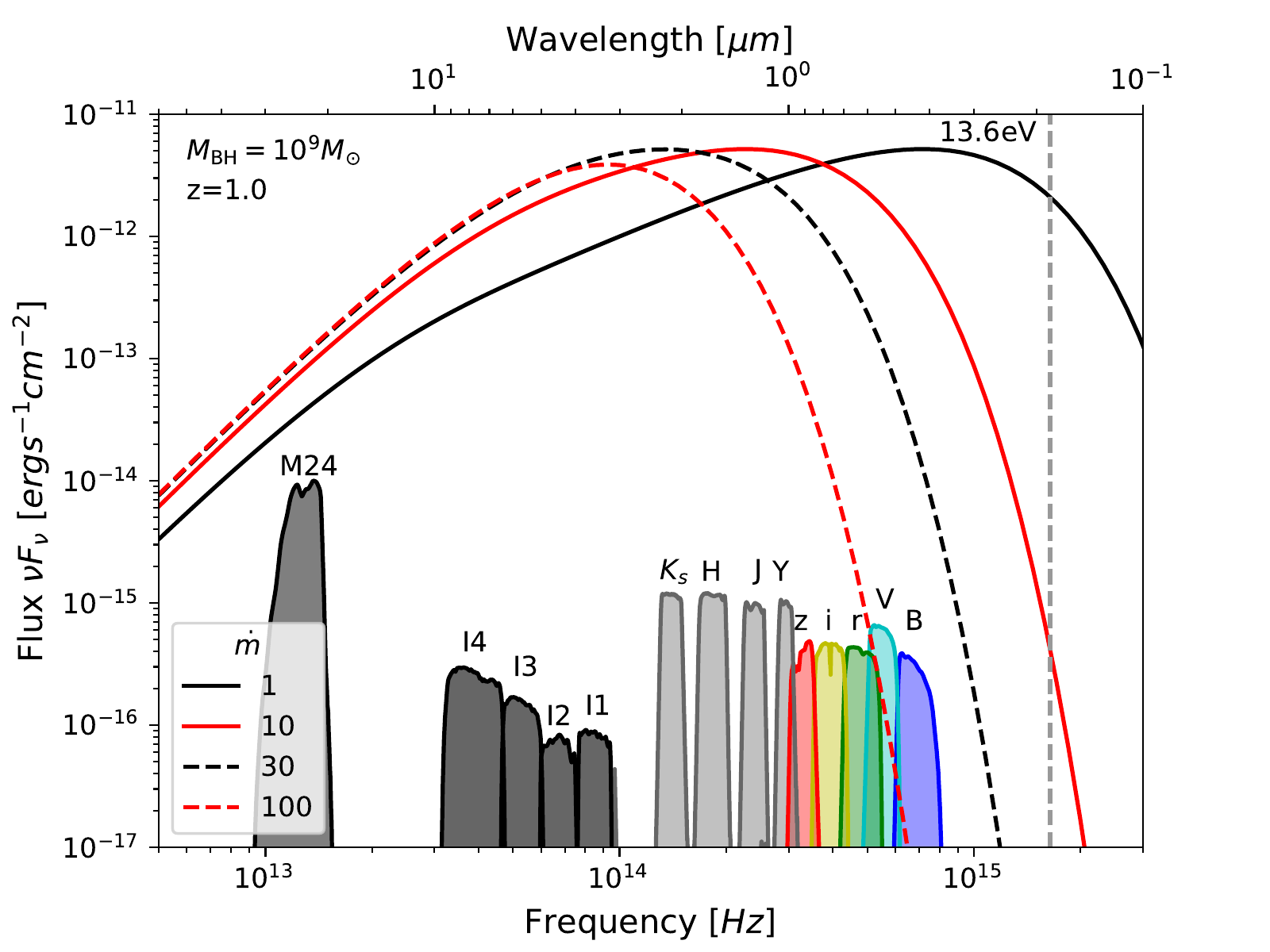}
\caption{Plot of selected SEDs with SDSS, 2MASS and \textit{WISE} filters (first panel) and COSMOS deep field survey filters (second panel) showing the effect of increasing Eddington ratios. The SEDs all have $\mbh = 10^9\,M_{\odot}$ at $z= 1.0$ with inclination angle 30\degree and are plotted in the observed frame. The filters are all scaled such that their peaks are at the flux limit of their respective surveys. The exceptions are the SDSS filters, for which only the \textit{i}-band is scaled to the 20.2 AB magnitude spectroscopic limit, and the rest of the filters are scaled in relation to this maximum.}
\label{fig:all_filters}
\end{figure*}

Table~\ref{tab:selection} and Figure~\ref{fig:sel_grids_flux} show that should super-Eddington AGN that resemble our models exist, large scale surveys using optical colour selection criteria such as SDSS would miss a large fraction of the population, especially at higher redshifts. Furthermore, our model predicts SEDs with little to no emission lines as well as potentially lacking an important X-ray component. This implies that even if a super-Eddington quasar were successfully targeted, follow up spectroscopy may incorrectly identify it as not being an AGN, and/or make redshift determination challenging (if host emission is sub-dominant). 
It must be noted that the model is purely for the accretion disc emission, whereas real AGN will have many more SED components such as outflows \citep[e.g.][]{Slone.Netzer:12}, dust \citep[e.g.][]{Collinson.etal:15}, while other AGN parameters such as BH spin can also play an important role in the resultant AD SED \citep[e.g.][]{Capellupo.etal:15,Bertemes.etal:16}. 
If we now also take into account the realistic difficulties of observing AGN such as dust obscuration and contamination from other sources, selection rates are expected to further decrease. 
Notably, effects such as dust obscuration and AGN outflow would complicate a colour based selection algorithm, as these would reduce the rest-frame UV content of the SEDs. 
In conjunction with other effects such as photon trapping and high black hole masses pushing the ISCO out to larger radii, an important loss of UV and soft X-ray photons would be expected, potentially making common AGN classification schemes -- such as significant X-ray emission \citep{Brandt.Alexander:15}, excess mid-IR emission \citep{Stern.etal:12}, and/or line ratio diagrams \citep{Baldwin.etal:81,Kewley.etal:06,Schawinski.etal:07}, unreliable. 

We thus we conclude that a realistic survey of quasars would yield an extremely low selection fraction of super-Eddington AGN, if these are affected by photon trapping in the inner parts of their accretion flows, and that those sources that {\it are} selected for follow-up study may be misidentified still.

\section{In Search of New Selection Criteria}
\label{sec:selection}

Given our findings in the previous Section, we now proceed to suggest new colour-based selection criteria designed specifically for our super-Eddington AGN SEDs. 
We present these selection criteria both in the context of wide-field, relatively shallow surveys, and of much deeper but narrow extragalactic surveys, which cover a broad range of wavelengths from optical to mid infra-red (MIR). 
As mentioned in Section \ref{sec:intro}, super-Eddington quasars are expected to be very rare given the general shapes, and current determinations of AGN luminosity functions and Eddington ration distribution functions.\footnote{Keeping in mind these may be incomplete for high-$\dot{m}$ sources, given the selection effects outlined in section \ref{subsec:completeness}.} 
As such, wide field surveys providing ample sky coverage maximize the probability of finding such rare sources. 
Conversely, deep, narrow-field and multi-wavelength surveys can provide the richer data required to identify extremely red, high-redshift sources, or otherwise to allow one to avoid well-understood types of faint sources (i.e., high-redshift galaxies). %
In order to benefit from the advantages of both types of surveys, in what follows we focus on the sub-set of models set at $z=1$ to investigate new selection criteria for our super-Eddington SED models.

We rely on the SDSS, the 2 Micron All Sky Survey \citep[2MASS;][]{Skrutskie.etal:06} and the \textit{Wide Field Infrared Survey Explorer} \citep[\WISE;][]{Wright.etal:10} as wide-field surveys, and use the Cosmic Evolution Survey \citep[COSMOS;][]{Scoville.etal:07} as the benchmark of deep, narrow-field surveys. 
In both cases, we use photometric bands and data that range from the blue end of the optical regime, through the near-IR (NIR) to the mid-IR (MIR), keeping in mind that our model SEDs yield very red optical colours and thus may be better suited to being observed in the IR. 

Figure \ref{fig:all_filters} illustrates the photometric bands (filters) we use, with several model SEDs with varying $\dot{m}$ over-plotted. 
Importantly, the response curve of each filter in Figure \ref{fig:all_filters} is scaled according to the respective depth in the relevant survey.
Specifically, 
the SDSS $i$-band is scaled to the maximal 20.2 AB magnitude limit for follow-up spectroscopy \citep{Newberg.Yanny:97,York.etal:00}, and the other SDSS bands are scaled to according to their transmission relative to the $i$-band; 
the depths of the 2MASS bands are taken from \citet{Skrutskie.etal:06} as $[J,H,K_S] = [15.8,15.1,14.3]$ Vega magnitudes; 
and the depths of the \textit{WISE} bands depths are taken from \citet{Wright.etal:10} as $[W1,W2,W3,W4] = [17.3,15.8,11.6,8.0]$ Vega magnitudes. 
The COSMOS filter depths from optical to the \spitzer/IRAC I4 (8.6\,\mic) band are taken from the COSMOS 2015 catalogue \citep{Laigle.etal:16} as $[B,V,r^{+},i^{+},z^{++}] =  [27.0,26.2,26.5,26.2,25.9]$ AB magnitudes for the optical bands, $[Y,J,H,K_S] = [24.8,24.7,24.3,24.0]$ AB magnitudes for the NIR bands, and $[I1,I2,I3,I4] = [25.5,25.5,23.0,22.9]$ AB magnitudes for the MIR bands. The MIPS $24\,\mic$ band depth is taken from \citet{LeFloch:09} as $80\mu J$ which corresponds to an AB magnitude of 19.1.

\begin{table*}
\caption{Super-Eddington AGN Colour Selection Criteria}
\label{tab:colour_cuts}
\begin{tabular}[]{l c c}
\hline
Survey & Colour Space & Colour Cuts\\
\hline
SDSS        & $u g r$         & $(u-g > 2.5)  \wedge\  (g-r > 3.0)$ \\
SDSS        & $g r i$         & $(g-r > 3.0)  \wedge\  (r-i > 2.0)$ \\
SDSS        & $r i z$         & $(r-i > 2.0)  \wedge\  (i-z > 2.0)$ \\
VISTA       & $J H K_{\rm s}$ & $(J-H > 1.0)  \wedge\  (H-K_{\rm s} > 0.7)$\\
\WISE       & $W(1-3)$        & $(W1-W2 > 0.4) \wedge\ (W2-W3 < 2.0)$ \\
\WISE       & $W(2-4)$        & $(W2-W3 < 2.0) \wedge\	(W3-W4 > 0.7)$ \\
COSMOS      & $B V r$         & $(B-V > 2.0)   \wedge\ (V-r^{+} > 1.0)$ \\ 
COSMOS      & $V r i$         & $(V-r^{+} > 1.0) \wedge\ (r^{+}-i^{+} > 1.5)$ \\ 
COSMOS      & $r i z$         & $(r^{+}-i^{+} > 1.5) \vee\ (i^{+}-z^{++} > 1.5)$ \\ 
Ultra-VISTA & $Y J H$         & $(Y-J < -1.0)    \vee\	 (J-H > 2.0)$ \\
Ultra-VISTA & $J H K_{\rm s}$ & $(J-H > 2.0)    \vee\ (H-Ks > 1.5)$\\
\spitzer/IRAC & $I(1-3)$ & $([3.6\mic]-[4.5\mic] > 1.0)$ \\ 
\spitzer/IRAC & $I(2-4)$ & no cut possible\\ 
\spitzer/IRAC+MIPS & $I(3-4),\,M(2,4)$ & $([5.7\mic]-[7.8\mic] > -0.25) \wedge\ ([7.8\mic]-[24.0\mic] < -0.5) $ \\ 
\hline
\end{tabular}
\end{table*}

\begin{figure*}
\center
\includegraphics[trim={0.3cm 0 1.5cm 0.6cm},clip,width = 0.39\textwidth]{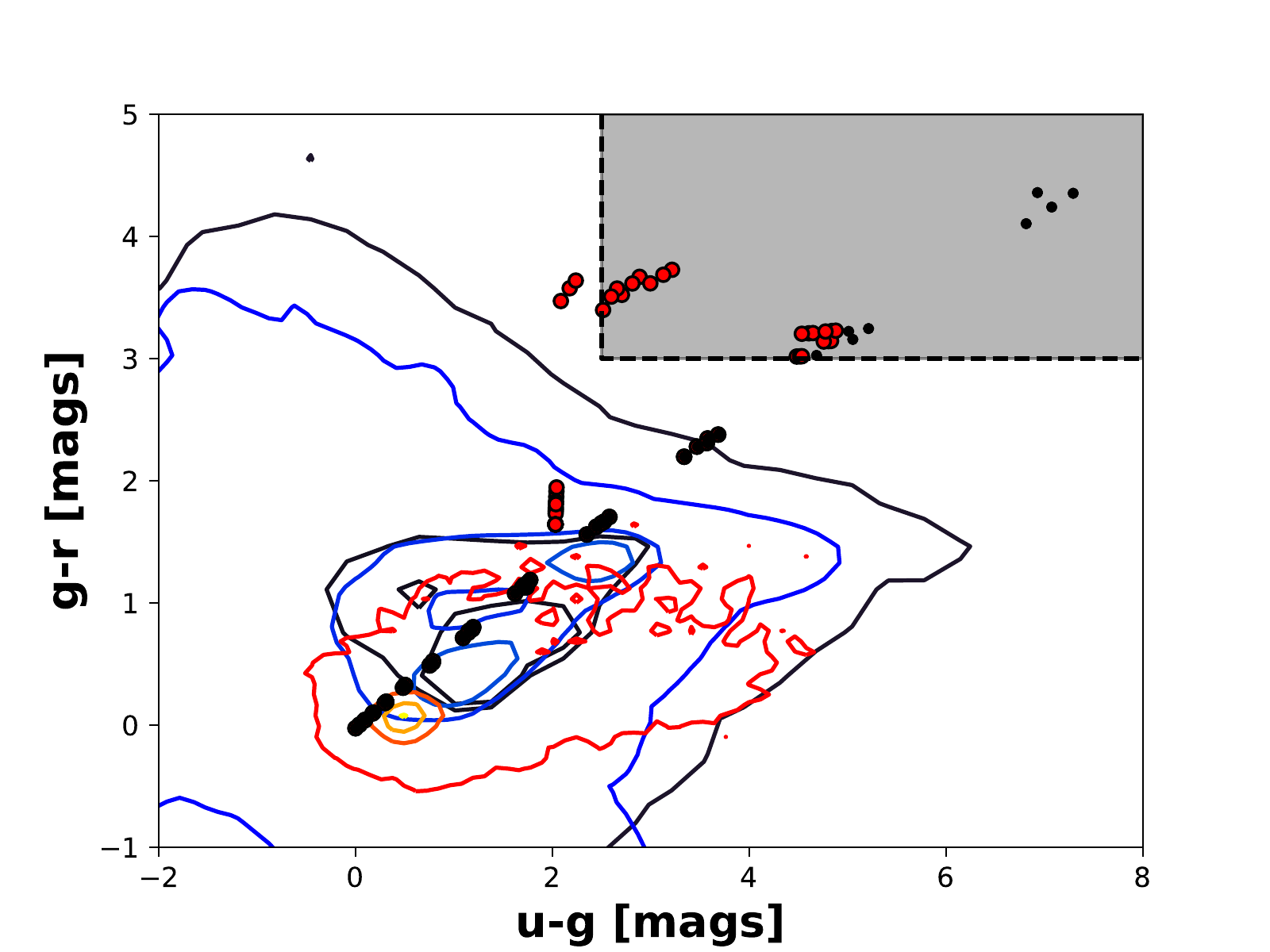}
\includegraphics[trim={0.3cm 0 1.5cm 0.6cm},clip,width = 0.39\textwidth]{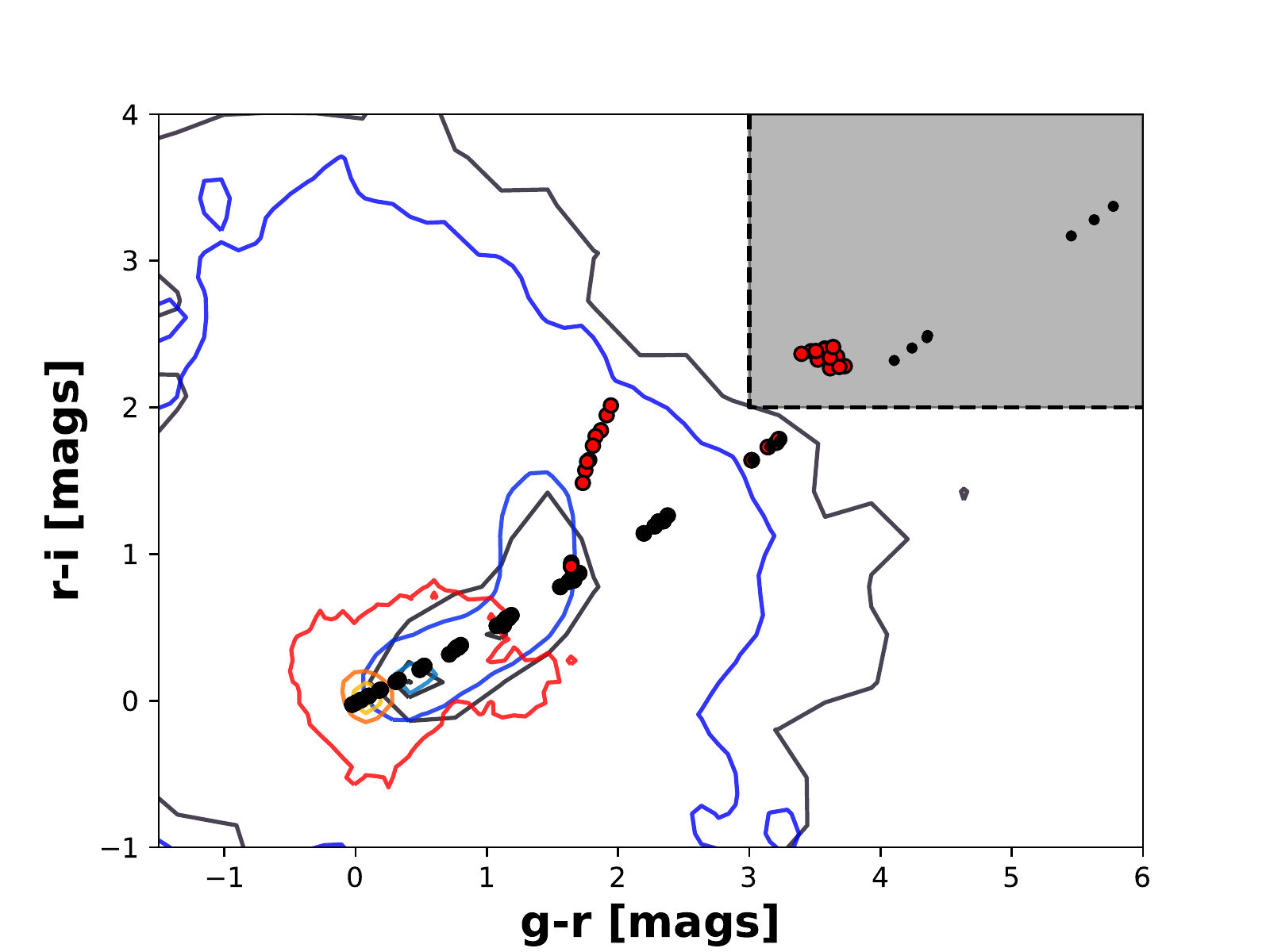}
\includegraphics[trim={0.3cm 0 1.5cm 0.6cm},clip,width = 0.39\textwidth]{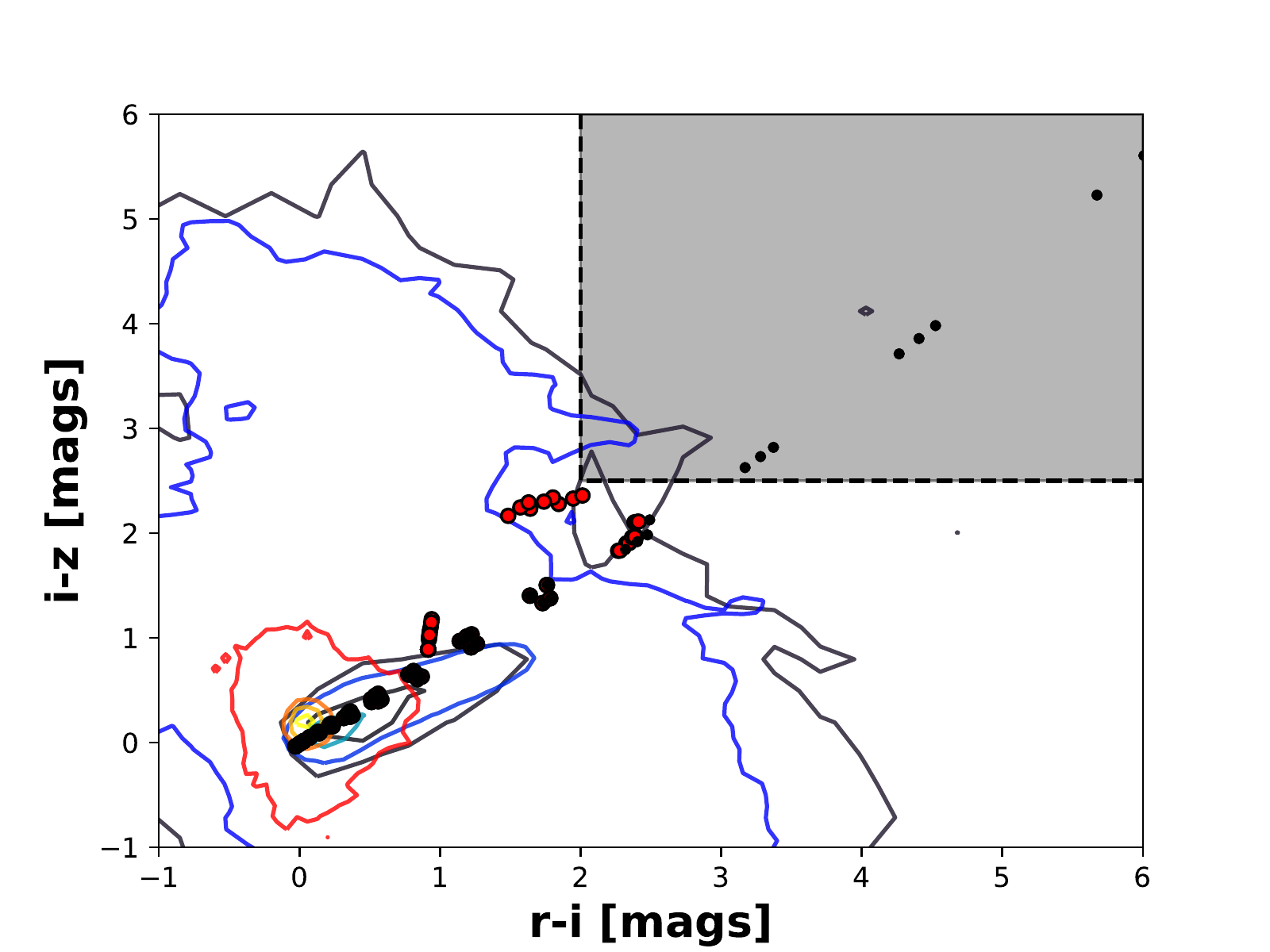} \\
\includegraphics[trim={0.3cm 0 1.6cm 0.6cm},clip,width = 0.39\textwidth]{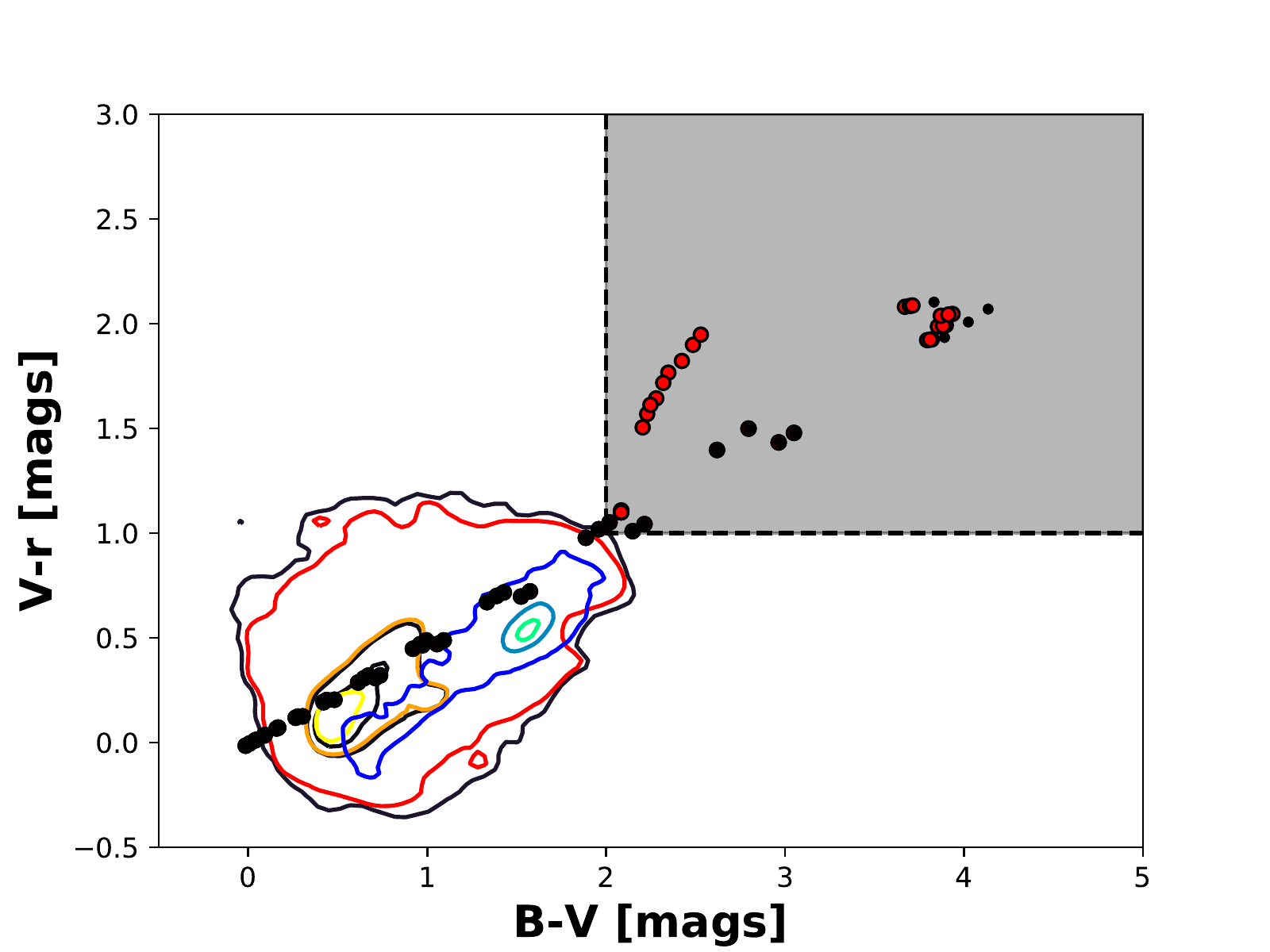}
\includegraphics[trim={0.3cm 0 1.6cm 0.6cm},clip,width = 0.39\textwidth]{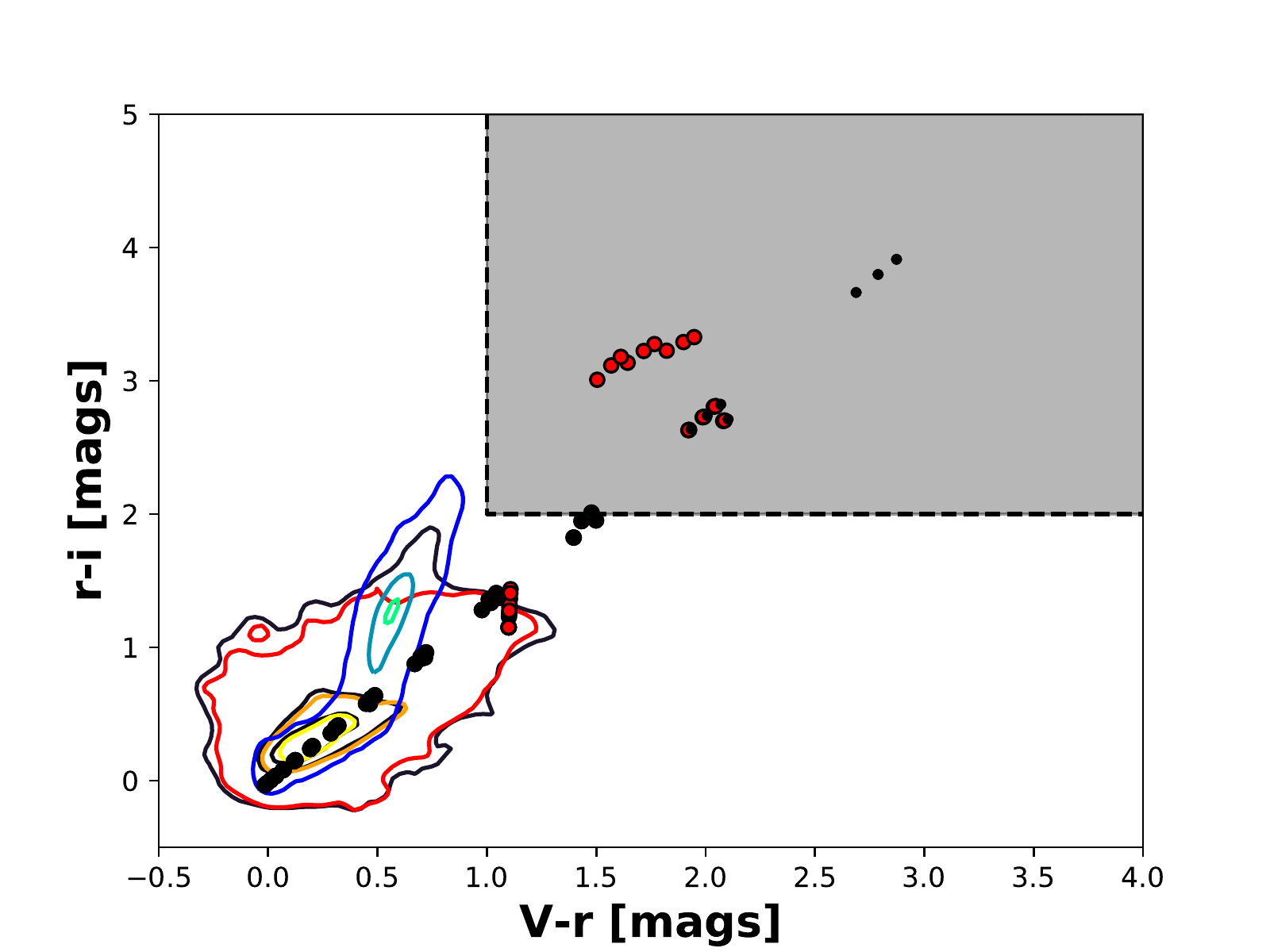}
\includegraphics[trim={0.3cm 0 1.6cm 0.6cm},clip,width = 0.39\textwidth]{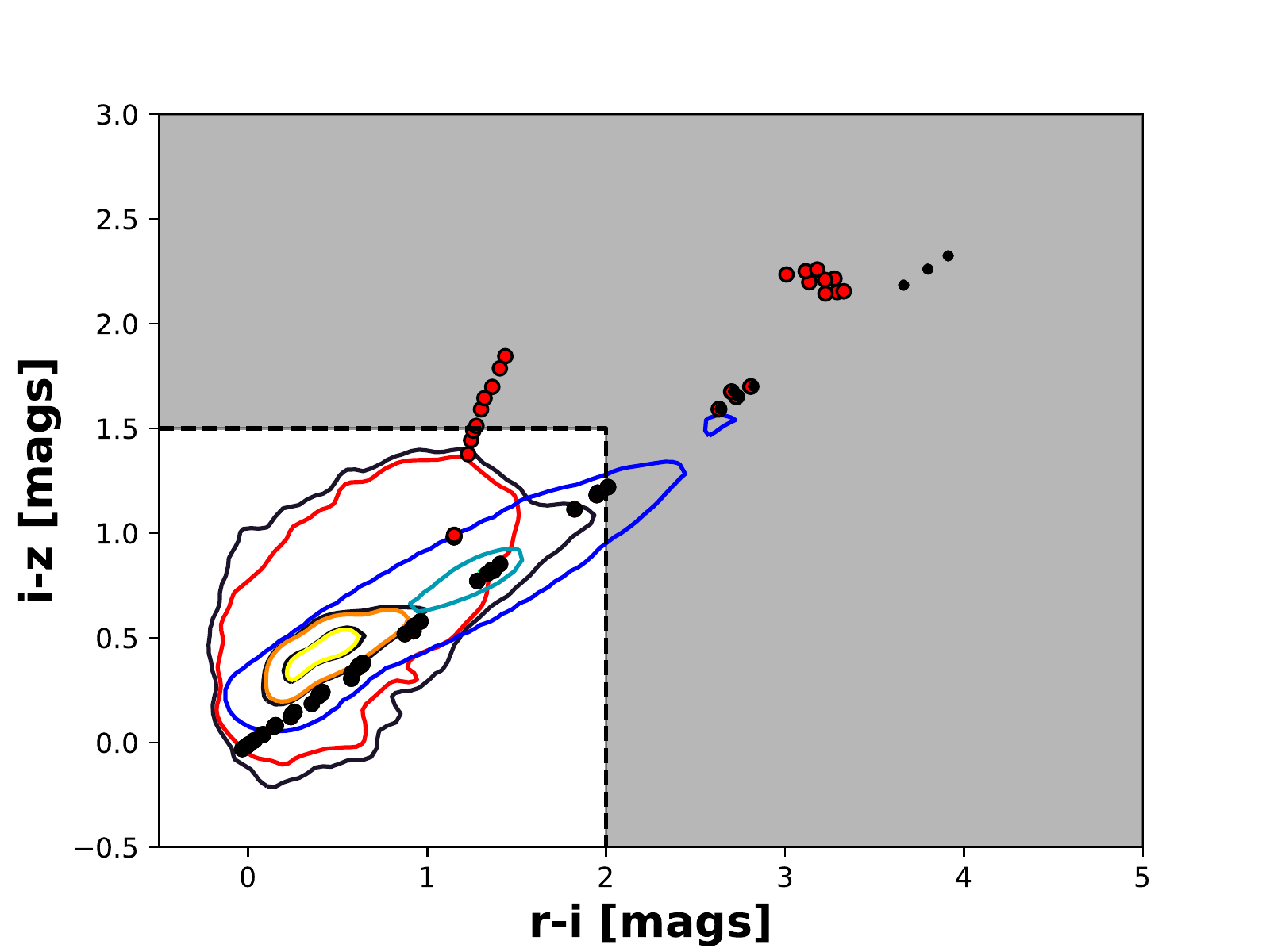}
\caption{Suggested optical selection criteria, represented by the grey areas, for SDSS (top three panels) and COSMOS (bottom three panels). For SDSS, the red contours are for spectroscopically confirmed AGN, the blue contours are stars and the black contours are all photometric point sources. For COSMOS, the red contours are galaxies, the blue contours are stars (i.e. point sources) and the black contours are all sources. The contour values are logarithmically spaced from 35\% to 98\% inclusion of sources.
The black points represent our  $z=1$ model SEDs, while the red points represent  the model SEDs with a 1 Gyr old stellar population added.}
\label{fig:opt_sel}
\end{figure*}

\begin{figure*}
\center
\includegraphics[trim={0.2cm 0 1.5cm 0},clip,width = 0.48\textwidth]{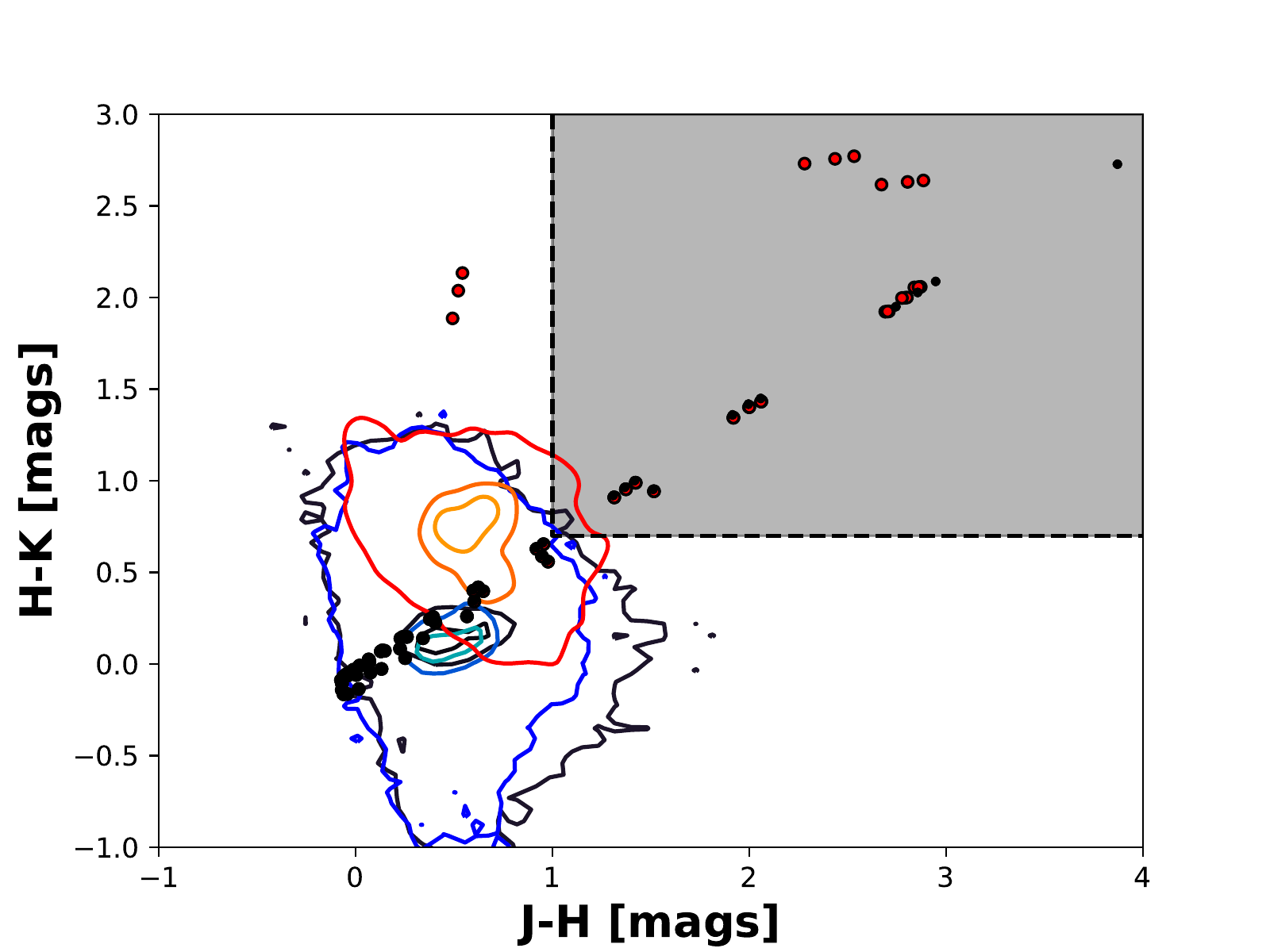} \\
\includegraphics[trim={0.2cm 0 1.5cm 0},clip,width = 0.48\textwidth]{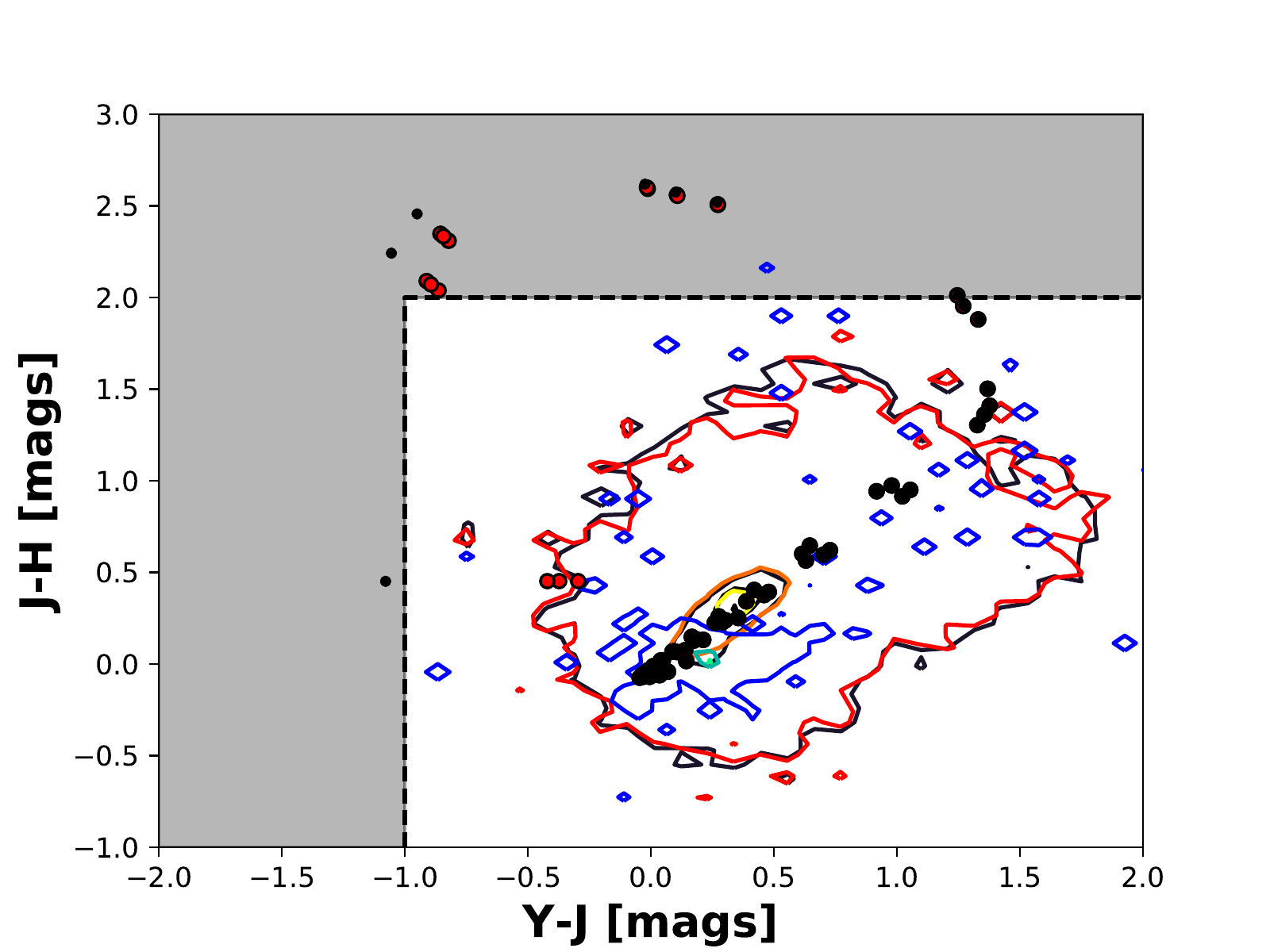}
\includegraphics[trim={0.2cm 0 1.5cm 0},clip,width = 0.48\textwidth]{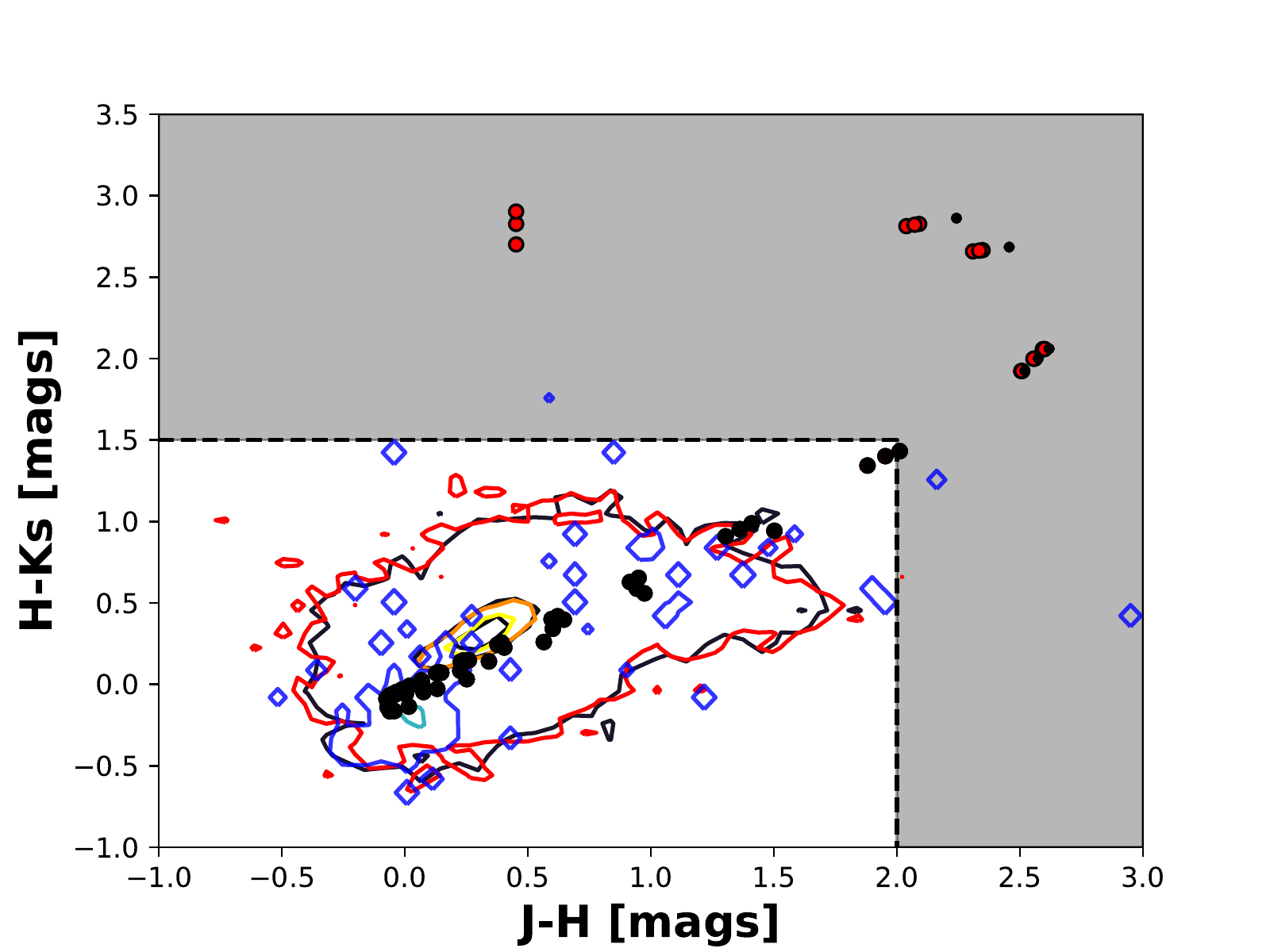}
\caption{Suggested NIR selection criteria, represented by the grey areas, for 2MASS (top panel) and COSMOS with Ultra VISTA (bottom panels). For 2MASS, the sources are cross-matched with SDSS. The red contours are for spectroscopically confirmed AGN, the blue contours are stars and the black contours are all photometric point sources. %
For COSMOS, the red contours are galaxies, the blue contours are stars (i.e. point sources) and the black contours are all sources. The contour values are logarithmically spaced from 35\% to 98\% inclusion of sources. 
The black points represent our $z=1$ model SEDs alone, while the red points are the model SEDs with a 1 Gyr old stellar population added.}
\label{fig:NIR_sel}
\end{figure*}

Figures \ref{fig:opt_sel}, \ref{fig:NIR_sel}, and \ref{fig:MIR_sel} show the optical, NIR, and MIR colour-colour spaces (respectively) we use to investigate new selection criteria for our super-Eddington SED models. 
These figures also show the synthetic colours of our $z=1$ model SEDs, as point-like symbols, overlaid on contours that trace the distribution of certain photometric sources, identified from the aforementioned surveys.
Specifically, the wide-field survey sources are first taken from the SDSS DR12 point source catalogue \citep{Alam.etal:15}, and then cross-matched to 2MASS and \WISE\ for NIR and MIR colours, respectively. We note that we now use the SDSS DR12 catalogue as opposed to DR7, as the colour selection algorithm is no longer in question, but rather the location of all known quasars in colour space. As such, the DR12 catalogue offers a much more complete quasar sample, especially at $z > 2$, than the DR7 sample.
All the COSMOS sources come from the 2015 COSMOS catalogue \citep{Laigle.etal:16}. 

In what follows, we seek to identify regions in colour-colour space that are dominated by our model super-Eddington AGN SEDs, and are relatively free of robustly identified sets of other types of sources (i.e., spectroscopically confirmed galaxies, stars, sub-Eddington AGN) from the aforementioned survey catalogues. 
Naturally, not every photometric source can be spectroscopically classified, and yet a large majority of sources were still labelled as stars, galaxies or sub-Eddington AGN via (template-based) modeling of their multi-band photometry. 
Some of these less robustly classified sources will be found in regions of colour-colour space where our super-Eddington models also exist. 
Given the similarity between some of our models and certain (cool) stellar sources, such as what is presented in Figure \ref{fig:SED_dwarf}, we expect that photometrically-classified `stars' could in fact be `hidden' super-Eddington AGN. 
Conversely, some stars may contaminate any suggested selection criteria that aim for high completeness for super-Eddington AGN. 
We therefore try to clearly distinguish between (1) those sources that are highly unlikely to be misidentified as super-Eddington AGN (i.e., either by being spectroscopically confirmed as stars or galaxies, or simply having colours that greatly differ from our models), and (2) potential `contaminants', which are real, unclassified photometric sources whose colours overlap with those of our super-Eddington AGN model SEDs.

The colour cuts in each set of optical, NIR, and MIR wavebands appear as grey-shaded regions in \cref{fig:opt_sel,fig:NIR_sel,fig:MIR_sel}. These are discussed in in detail in subsections \ref{sec:opt_sel}, \ref{sec:NIR_sel}, and \ref{sec:MIR} below (respectively), and listed in Table \ref{tab:colour_cuts}.

We stress that, although the criteria presented here use colours specific to the chosen surveys, the ideas are generally applicable to any set of optical-NIR-MIR (extragalactic) observations. 
In order to apply these colour selection criteria, we suggest using criteria in similar wavelengths simultaneously, i.e. using all optical colour, NIR or MIR criteria simultaneously. %
This is similar to the way the SDSS quasars selection algorithm uses a 4-dimensional colour space to select photometric candidates for follow up spectroscopy. It is insufficient to select a photometric object based on a single colour-colour space, as this object may be a clear contaminant when viewed in a different colour-colour space. The colour cuts are designed to ideally yield a sample of super-Eddington quasars that resemble our models, with little contamination, and so an object that falls within the cuts in the majority of the colour-colour spaces has a higher chance of being a target of interest. 

Furthermore, the wide and deep field survey criteria can be used either separately or be combined, depending on data available. Notably, surveys of different depths can be susceptible to different contaminants and sources. 
As a simple example, a very red galaxy at intermediate redshift may appear to be an interesting point-like source in a lower-resolution, wide-field survey. 
However, cross checking with the selection criteria we define for deeper surveys (e.g., COSMOS) may reveal that this potentially interesting source is indeed consistent with a red, inactive galaxy.
On the other hand, surveys like COSMOS which focus on galaxies may dismiss super-Eddington AGN as (foreground) stellar sources, based on spectral template fitting (see, e.g., Fig.~\ref{fig:SED_dwarf} and section 4.5 in \citealt{Laigle.etal:16}). 
Here, cross-checking with the selection criteria we define for the wide-field surveys, which include a significant population of stellar sources, could assist in characterizing the source under question.

\subsection{Optical Selection Criteria}
\label{sec:opt_sel}

Figure \ref{fig:opt_sel} shows the colour selection cuts for the optical colours for both SDSS and COSMOS. For SDSS, we also consider the DR12 photometric point source catalogue \citep{Alam.etal:15} as a general indication of where non super-Eddington sources lie in colour-colour space. From the results of the SDSS quasar selection presented in Section \ref{sec:selection}, we expect most contaminants for our models to be stars, thus successful colour criteria must avoid the bulk of these objects. To this end, the stellar population of the SDSS point source catalogue is also plotted. Finally, we also plot the SDSS DR12 quasar catalogue \citep{Paris.etal:17} to clearly indicate the area populated by standard sub-Eddington AGN. 

For the COSMOS sources, we similarly consider the entire photometric catalogue as a general indication of the populated area of colour space in COSMOS and then also plot the galaxy and stellar contours. The stars in COSMOS are identified through multi-band spectral fitting (see Section 4.5 of \cite{Laigle.etal:16} for details). The COSMOS stellar contours give a very important indication of what kind of foreground contaminants one can expect from such a deep, narrow-field survey.

In both cases, the bluer models with lower \mbh\ and $\dot{m}$ are in heavily populated regions of colour space. 
The sources populating these regions are mostly stars and `standard' quasars in SDSS, with the bluest models falling near the peak of the quasar contours. This is somewhat understandable, as the model SEDs with low $\dot{m}$ are not expected to appear so different from sub-Eddington quasars with $\dot{m} \lesssim 1$. 
For COSMOS, the colour distribution corresponding to different sources overlap significantly, although even these overlapping regions are dominated by galaxies.

Contrary to the blue models, the redder models are found in regions of the three optical colour-colour spaces that are mostly empty of known photometric sources. 
We stress that the optical colour spaces we show, and particularly the bluer $ugr$ and $BVr$ spaces, are truncated, as the reddest models without added host emission extend to extremely red colours, up to $u-g \approx 30$ for $z=1$. 
Thus, we suggest optical colour cuts to select these redder models that lie away from the well-studied photometric sources, with $(u-g > 2.5)~\wedge~(g-r>3.0)~ \wedge ~(r-i >2.0)~ \wedge ~(i-z > 2.0)$ for SDSS,  and $(B-V > 2.0)~ \wedge ~(V-r^{+}>1.0)~ \wedge ~(r^{+}-i^{+} >1.5)~ \wedge ~(i^{+}-z^{++} > 1.5)$ for COSMOS. 

As they have quite red colours, the models selected by these colour cuts would most likely have high $\dot{m}$ and \mbh. These are expected to be very rare even in the case of elusive super-critical BHs, as the BH masses are much higher than the $10^9 M_{\odot}$ break in the BHMF \citep{McLure.Dunlop:04,Kauffmann.Heckman:09,Shen.Kelly:12,Weigel.etal:17}, and such truly gigantic SMBH are already rare even in the sub-Eddington regime. Furthermore, we draw attention to the fact that the colour cuts here do not take into account the limited depths of the respective surveys. As shown in Figure \ref{fig:sel_grids_flux}, many models are potentially unobservable in the optical as they lie outside of flux limits (i.e., the SDSS $i$-band magnitude limit). 
While non-detection may be less of an issue for deep surveys, the reddest models which reach colours of $u-g \gtrsim 30$ are still likely to be beyond the deeper flux limits, and so completely undetectable in optical wavebands. 
We also note that the redder models with high \mbh\ are found to be much bluer when host emission is added, as is consistent with the example SED shown in the rightmost panel of Figure \ref{fig:SSP_SED}. As the host emission can dominate the optical emission for these extreme super-Eddington models, it is not surprising that these models are then found well within known sources' contours when host emission is taken into account (e.g. red points in top panels of Figure \ref{fig:opt_sel}). 

Thus while optical colour cuts may favour very red models by default, consideration of host emission, survey flux limits, as well as other complicated physical features not included in the model may in fact render this selection more difficult than implied. Notably, many of the red models that lie away from the known sources' contours in optical space are optically faint, if at all detected. This factor along with host emission leading to super-Eddington AGN appearing galaxy-like in terms of optical colours, strongly supports also investigating these objects in other wavebands such as the NIR and MIR. 

\subsection{Near Infra-red Selection Criteria}
\label{sec:NIR_sel}

The NIR colour cuts are shown in Figure \ref{fig:NIR_sel}, with wide-field (2MASS) and deep, narrow-field (COSMOS) surveys in the top and bottom panels, respectively. 
The known sources contours are the same as for the optical colour-colour spaces, where the SDSS point sources have been cross-matched with the 2MASS all sky Point Source Catalogue \citep[PSC,][]{Skrutskie.etal:06}, while the entire PSC is used for the photometric point source contours. The COSMOS NIR contours come from the Ultra VISTA DR3 Deep Field survey \citep{McCracken.etal:12}. As for optical colour spaces, the bluer models are in areas where all the known sources are present, while the redder models are further away from the SDSS contours. 
For COSMOS, on the other hand, the vast majority of our models overlap with the contours of known sources. 
Where we are able to define NIR selection cuts, they are more robust against the effect of host emission, especially for the COSMOS survey (bottom two panels of Figure \ref{fig:NIR_sel}). 
These cuts are given by $(J-H > 1.0)~ \wedge ~(H-K_{s} >0.7)$ for 2MASS and $(Y-J < -1.0)~ \wedge ~(J_H > 2.0)~ \wedge ~(H-K_{s} >1.5)$ for COSMOS. 
Although added host emission still shifts the reddest super-Eddington SEDs towards bluer colours, they nevertheless remain in areas mostly free of other sources, with the exception of the few outlying contaminants from the photometric point source catalogues.

The true benefit of using NIR colours in conjunction with optical and/or MIR colours, in addition to the robustness against host contamination, lies in the implicit flux limits of the surveys. Unlike the optical colour spaces, the NIR colours are all within $\rm |colour| \lesssim 4$, and NIR magnitudes are often bright, as many of the model SEDs have strong emission or even peak in the NIR bands (see Figure \ref{fig:all_filters}). Thus, model SEDs that would be potentially undetected in optical or MIR bands are most likely detectable in at least one NIR band. 
The NIR cuts can then act as a complementary selection method to the optical colour cuts which also target very red models, but run the risk of not being able to reliably detect those models. 
The two main advantages of the NIR cuts -- higher detection probability and robustness to (blue) host contamination -- mark them as favorable starting points for any real search for the kind of super-Eddington sources that our models mimic.

\begin{figure*}
\center
\includegraphics[trim={0.3cm 0 1.6cm 0.6cm},clip,width = 0.48\textwidth]{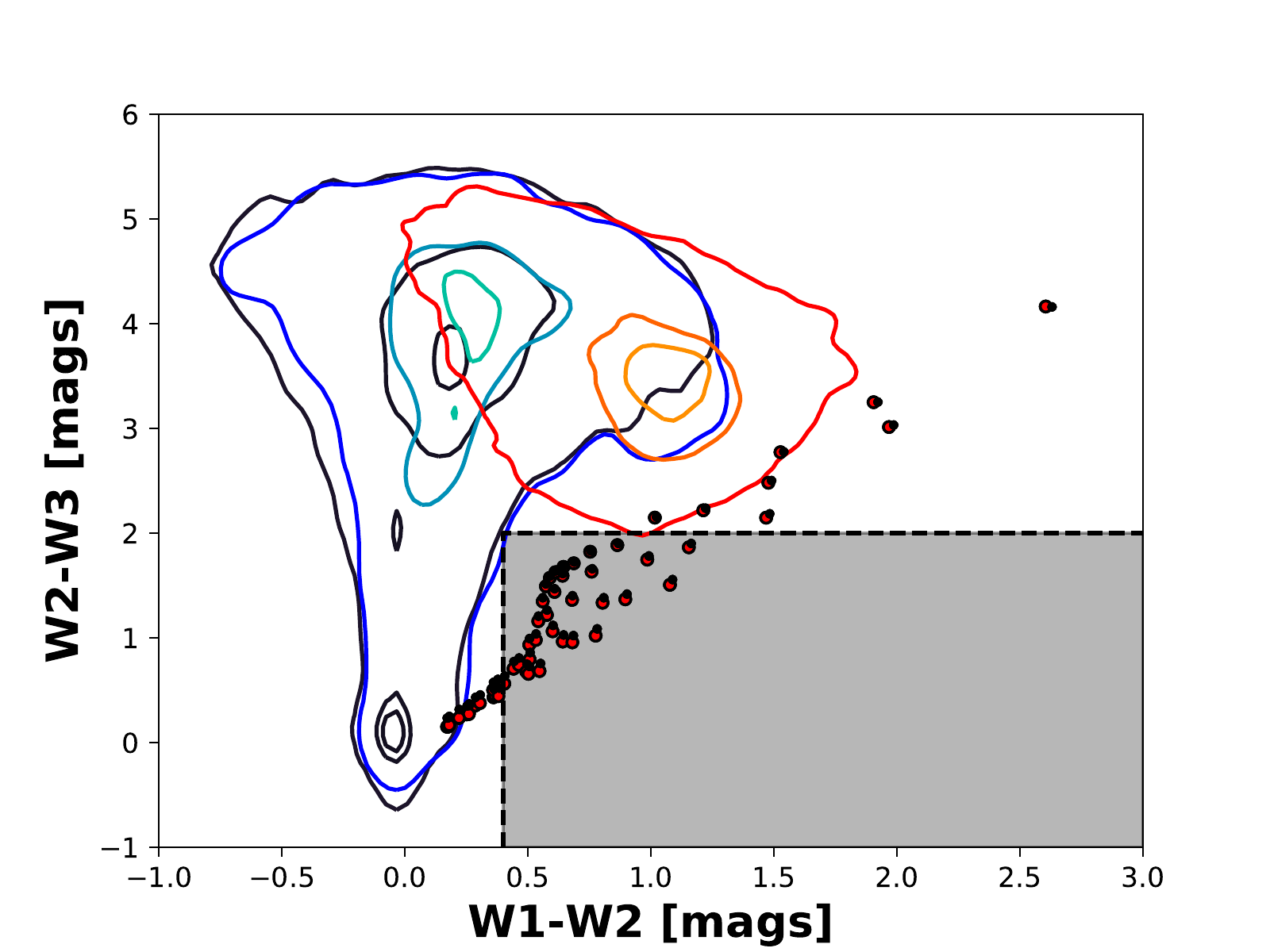}
\includegraphics[trim={0.3cm 0 1.6cm 0.6cm},clip,width = 0.48\textwidth]{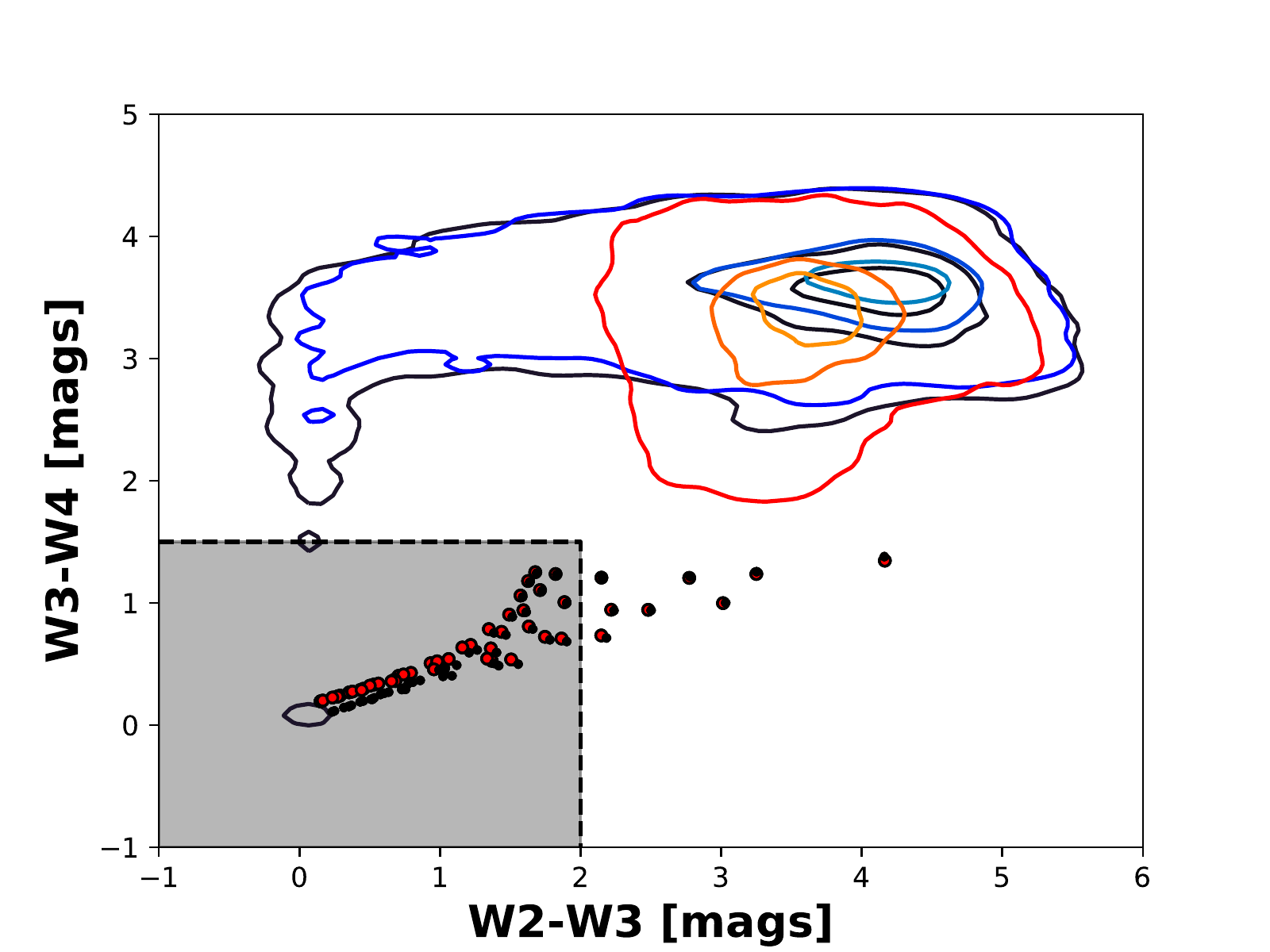}
\includegraphics[trim={0.3cm 0 1.6cm 0.6cm},clip,width = 0.48\textwidth]{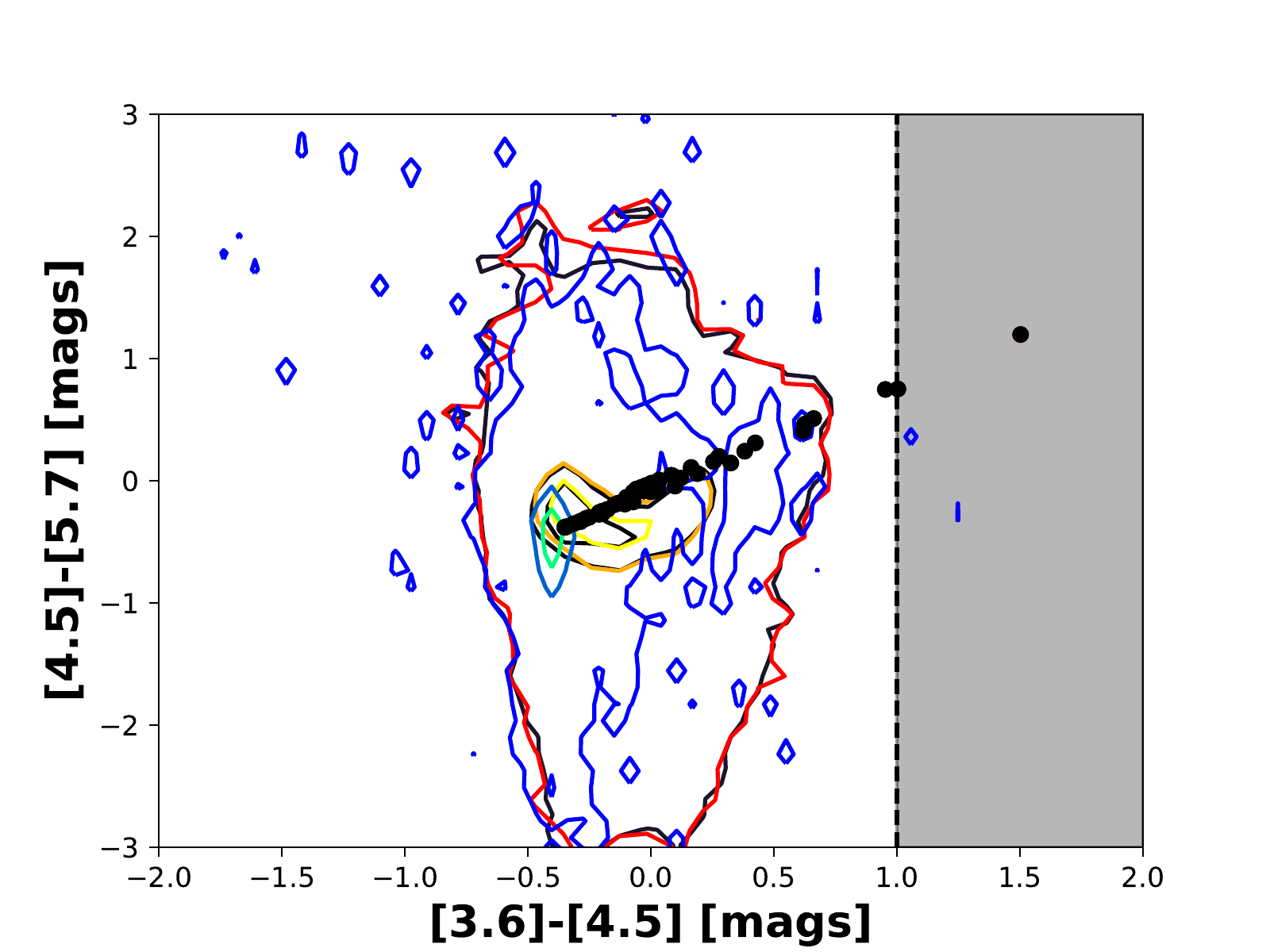}
\includegraphics[trim={0.3cm 0 1.6cm 0.6cm},clip,width = 0.48\textwidth]{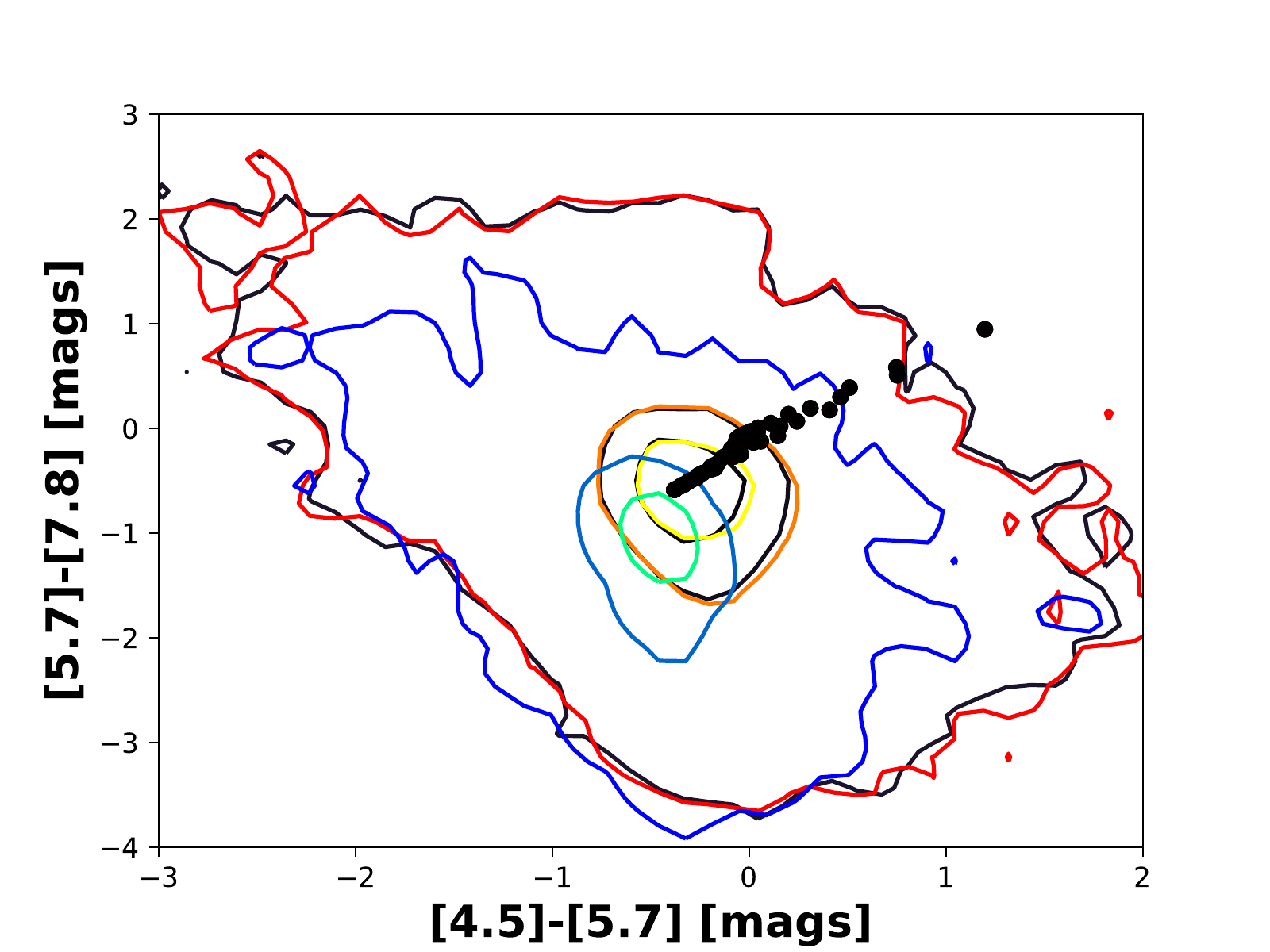}
\includegraphics[trim={0.3cm 0 1.6cm 0.6cm},clip,width = 0.48\textwidth]{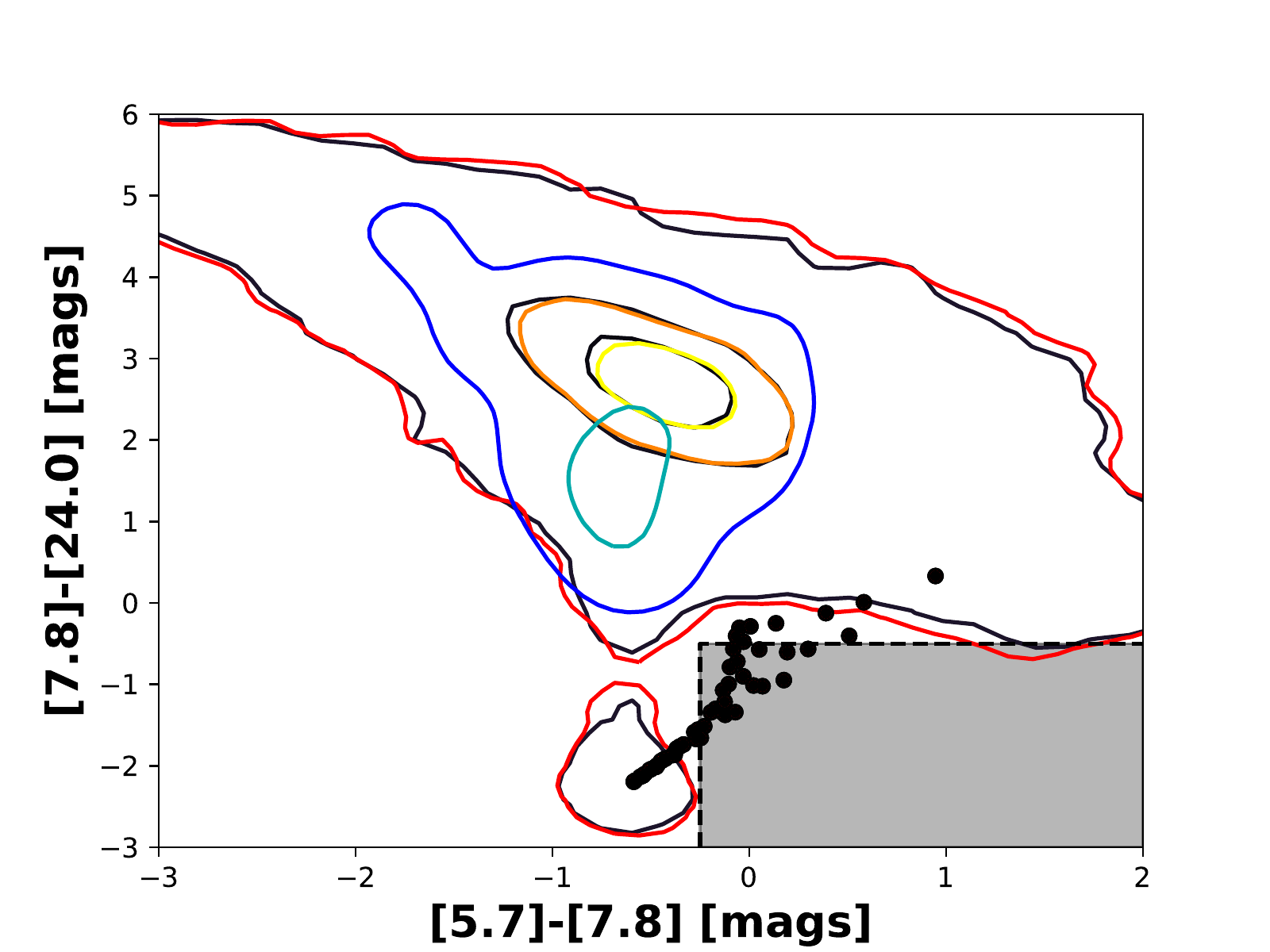}
\caption{Suggested MIR selection criteria, represented by the grey areas, for \textit{WISE} (top row) and COSMOS with IRAC and MIPS 24 micron band (bottom three panels). For \textit{WISE}, the red contours are for spectroscopically confirmed AGN, the blue contours are stars and the black contours are all photometric point sources. For COSMOS, the red contours are galaxies, the blue contours are stars (i.e. point sources) and the black contours are all sources. The contour values are logarithmically spaced from 35\% to 98\% inclusion of sources. 
The black points represent our $z=1$ model SEDs alone, while the red points are the model SEDs with a 1 Gyr old stellar population added.}
\label{fig:MIR_sel}
\end{figure*}

\begin{figure*}
\center
\includegraphics[trim={0.2cm 0 1.5cm 0},clip,width = 0.48\textwidth]{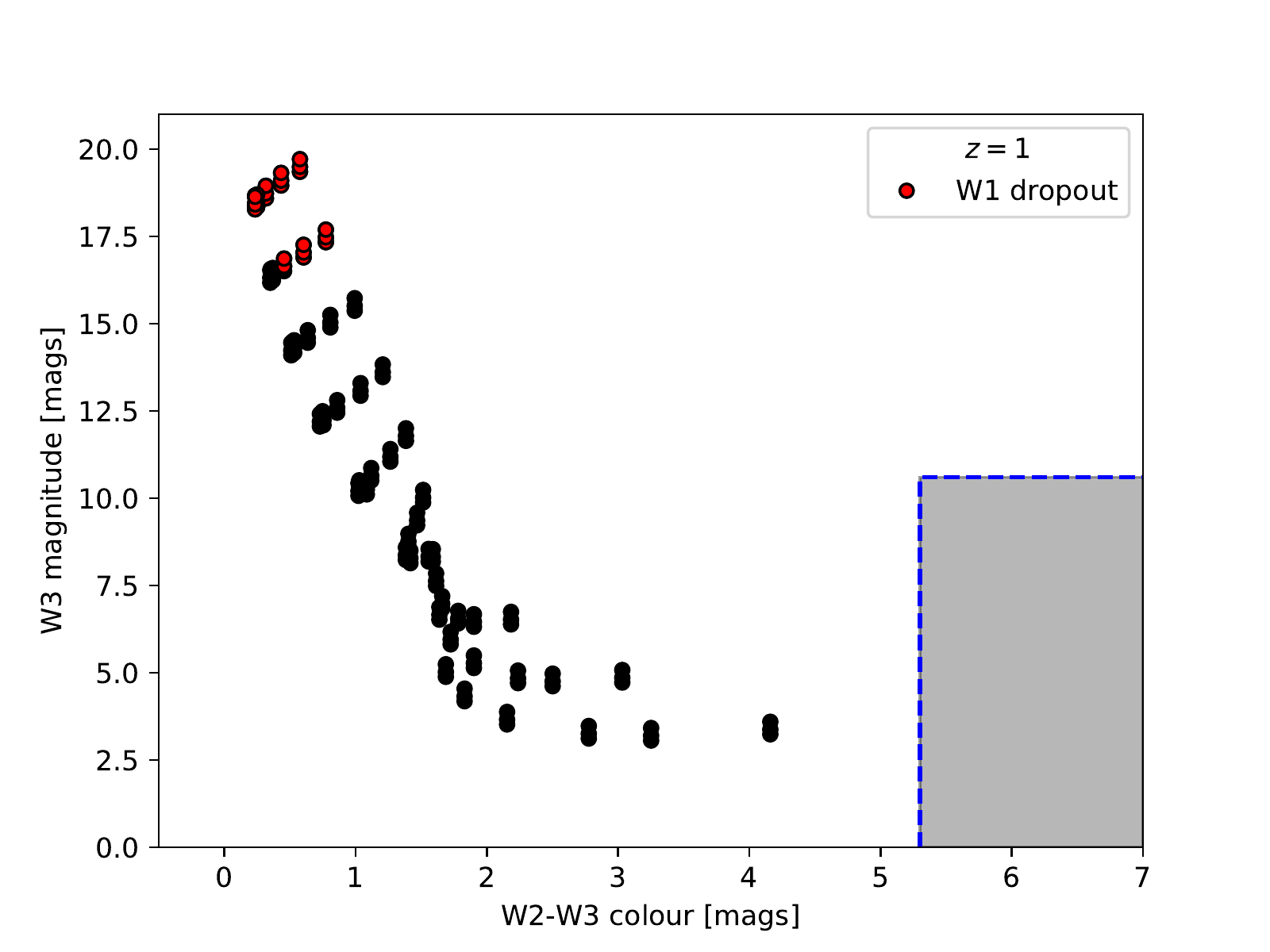}
\includegraphics[trim={0.2cm 0 1.5cm 0},clip,width = 0.48\textwidth]{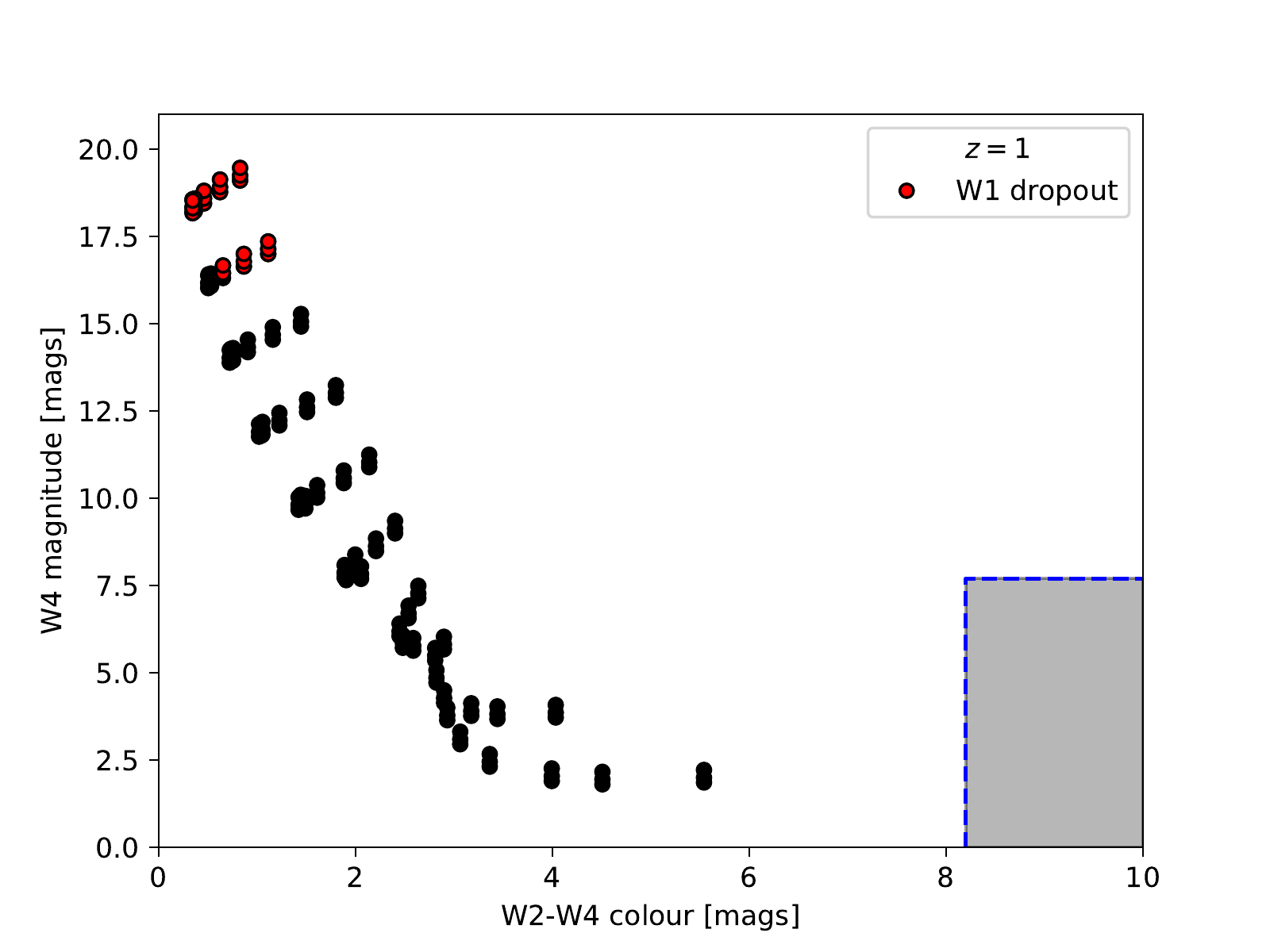}
\caption{The super-critical model colours at $z=1$ plotted along with the Hot DOGs selection criteria as presented in \citet{Assef.etal:15}. The red points are those models that satisfy the $W1 > 17.4$ dropout criteria. W1 dropouts that fall one of the two shaded regions are selected as Hot DOGS. We note that although this plot is for $z=1$, none of our models would be selected as Hot DOGs regardless of redshift.}
\label{fig:DOGS}
\end{figure*}

\subsection{Mid Infra-red Colours}
\label{sec:MIR}

\subsubsection{Mid Infra-Red Selection Criteria}
\label{sec:MIR_sel}

The MIR colour cuts are shown in Figure \ref{fig:MIR_sel}, with contours that follow the same meaning as in previous panels.
Here, the SDSS sources are cross-matched with the \textit{WISE} AllWISE Source Catalog \citep{Cutri.etal:13}, while the COSMOS sources come from the \textit{Spitzer} COSMOS (S-COSMOS) survey \citep{Sander.etal:07}. 
Contrary to the optical and NIR colour spaces, the highest-\mbh\ and $\dot{m}$ (`reddest') models are {\it not} located systematically away from the colours of known sources.
Specifically, the MIR colours of our model SEDs overlap with part of the stellar contour for \textit{WISE}, and several of the contours for COSMOS. 
However, the bluer models with intermediate and even low $\dot{m}$ and \mbh\ are in areas of colour spaces relatively free of known sources, especially following the \textit{WISE} contours in the top two panels of Figure \ref{fig:MIR_sel}. This suggests that MIR colours could be used to seek and identify a different part of the super-Eddington population than in optical and NIR space. As for optical colours however, the flux limits of the survey would play an important role, especially for the bluer models that may be faint in the MIR. The intermediately red models are expected to be detected in \textit{WISE} and (and/or \spitzer/IRAC-like) bands, though will naturally be much dimmer than the reddest models which are unfortunately not selected by the proposed MIR colour cuts. 
These cuts are given by $(W1-W2>0.4)~\wedge~(W2-W3<2.0)~\wedge~(W3-W4>0.7)$ for \textit{WISE}, and $([3.6\mu]-[4.5\mu] >1.0)~\wedge~([5.7\mu]-[7.8\mu]>-0.25)~\wedge~([7.8\mu]-[24.0\mu]<-0.5)$ for COSMOS.

However, MIR also has difficulties beyond potential flux limits, especially with the \spitzer/IRAC colours (bottom panels of Figure \ref{fig:MIR_sel}). Notably, the first two panels with solely IRAC colours reveal almost all of our model SEDs to be found well within COSMOS source contours, to the extent where no cuts can be defined at all for the $[4.5\mu] - [5.7\mu]$ colour. 
The bottom-right panel in Figure \ref{fig:MIR_sel} allows some selection of intermediate $\dot{m}$ and \mbh\ models due to the addition of the MIPS 24$\mu m$ band. 
Further extension into the far IR could potentially allow the bluer class of models to be more reliably selected, as they would certainly be outliers from most photometric sources detected in these bands. The issue of flux limits and whether these bluer models would still be subsequently detected in redder bands would still have to be carefully considered. 
Finally, Figure \ref{fig:MIR_sel} shows that adding host emission to the models has very little effect on the MIR colours. 
Thus, within the simple assumptions of our models, the MIR regime truly probes the emission of the model SED itself, and in principle selection criteria in these colour spaces should be resistant to superfluous extra emission from stellar components.
We note, however, that in reality the MIR may be contaminated by emission from (circumnuclear) dusty gas \cite[e.g.,][]{Mor.etal:09}. This may be the case even when the SMBH-related emission is UV-poor, and the MIR emission is instead related to star formation throughout the host \cite[see, e.g.,][]{Elbaz.etal:07}).

\subsubsection{MIR Luminous Extragalactic Sources and Super-Eddington SEDs}
\label{sec:DOGS}

As some of the model SEDs extend into extremely red colours, it is interesting to consider what sort of known objects they may resemble based (solely) on their NIR and MIR data. 
Such objects include Ultra Luminous Infrared Galaxies (ULIRGs), sub-mm galaxies (SMGs) and Dust Obscured Galaxies (DOGs), at intermediate-to-high redshifts. 
The DOGs differ from the other two classes of IR-dominated galaxies in that they are also extremely red in MIR bands such as the \WISE\ bands, to the point where they are undetected not only in optical but also in NIR. 
Initial colour criteria for identifying DOGs were presented in \citet{Dey.etal:08} as colour $R-24\mu m > 14$, essentially requiring non-detection in the optical regime. 
Much work has since been conducted on DOGs, notably with the introduction of colour criteria defining W1 dropouts that are undetected in the \WISE\ 3.4 \mic\ band \cite[i.e. $W1 > 17.4$ mags;][]{Eisenhardt.etal:12}. The population of galaxies at $z \gtrsim 1$ fitting this criteria have been found to have very high dust temperatures, much greater than those of ULIRGs or SMGs, thus being named `Hot DOGs' \citep{Wu.etal:12}. 
Follow-up, multi-wavelength studies suggest that these systems are powered by rare, extremely luminous ($\Lbol \gtrsim 10^{13}\,\Lsun$), though heavily obscured AGN (visual extinction $A_{\rm V} {\sim} 50$ and/or line-of-sight hydrogen column densities $N_{\rm H} \gtrsim 10^{24}\,{\rm cm}^{-2}$; see \citealt{Eisenhardt.etal:12,Wu.etal:12,Jones.etal:14,Stern.etal:14,Assef.etal:15}).

We next compare our model SEDs with the basic properties (or indeed defininig criteria) of Hot DOGs. 
To do so, we consider the Hot DOG colours defined in \citet{Assef.etal:15}, which expand on the $W1$-dropout criterion:  
 $(W4 < 7.7 \wedge\ W2-W4 > 8.2)$ or $(W3<10.6 \wedge\ W2-W3>5.3)$. 
As shown in Figure \ref{fig:DOGS}, very few of our $z=1$ super-Eddington AGN models would qualify as $W1$-dropouts, and none make it into the Hot DOG selection area. 
At a redshift of $z=2$, only the reddest model with an unrealistically high mass $\mbh = 10^{11}\,\Msol$ and $\dot{m} = 100$ manages to get into the $W4$ Hot DOGs selection area. 
This model is however not a $W1$-dropout; indeed only the bluer models with low \mbh\ and $\dot{m}$ qualify as $W1$ dropouts as the SEDs are still emitting mostly in the optical bands (see left panel in Figure \ref{fig:all_filters}). 
This is not entirely surprising as our model SEDs are simply multi-temperature blackbodies, and thus the decrease in emission between  $22\mu$ and $3.4\mu$ cannot be as sharp as required by the Hot DOGs selection criteria. 

Adding host emission to the super-Eddington models produces a negligible effect, as host emission is very weak in the MIR (see Figure \ref{fig:MIR_sel}). The difference in colours between our super-Eddington models and Hot DOGs is also physically motivated by the fact that the model SEDs do not include any dust related effects, while Hot DOGs are heavily dust obscured by circumnuclear gas. Furthermore, the model SEDs are already lacking much UV and optical emission relative to standard sub-Eddington AGN. If host emission is added to the models, some UV emission is expected which could then be affected by dust. However, it is unlikely that the UV host emission would give rise to a MIR luminosity of $\nu L_{\nu} \sim 6 \times 10^{46} erg s^{-1}$ at $6\mu m$ \citep{Stern.etal:14}.
Thus adding dust to our simplistic super-Eddington models is expected to have a limited effect, since there will be very little UV emission to reprocess as IR emission. 

We conclude that while Hot DOGs may very well be luminous, heavily obscured AGN tracing fast SMBH growth, they are clearly separate from the super-Eddington AGN we study in this work.

\section{Conclusions}
\label{sec:conclusion}

We use a grid of super-Eddington accretion disc SEDs generated by using a simple geometrically thin, optically thick and {\it truncated} disc model, mimicking the effects of photon trapping in the inner part of the disc.
Our grid covers a wide range in several key parameters, including black hole mass \mbh, Eddington ratio $\dot{m}$, and redshift (Table \ref{tab:models}). %
Notably, we cover $1 \leq \dot{m} \leq 100$ for large SMBHs ($\mbh\gtrsim10^8\,\Msol$), at significant redshifts ($z\sim1-2$).
Emission from the stellar population in the host galaxies of such systems is also considered.
Our super-Eddington model SEDs allow us to study the prospects of identifying such systems in large AGN surveys.

\noindent
Our main findings are as follows:

\begin{itemize}

\item Many of our model SEDs resemble stellar sources, and overlap with the `stellar locus' in optical colour-colour space. As a result, the spectroscopic targeting algorithm of the SDSS would have ignored such sources, if they appear in optical imaging. 
Indeed, about 3/4 of the models that lie within the SDSS spectroscopic flux limits are found to be rejected due to proximity to the stellar locus.

\item A vast majority of the super-Eddington models are significantly redder than the known SDSS quasars (and than the stellar locus). They essentially lack the excess (rest-frame) UV emission that motivates the bulk of quasar selection in SDSS and other large surveys of optically-bright quasars. 
Our model SEDs reach colours as extreme as $u-g \simeq 30$ for $z=1$ (Figure \ref{fig:colours}).

\item Increasing \mbh\ and $\dot{m}$ has the degenerate effect of reddening the colours of the model SEDs (Figure \ref{fig:SDSS_tracks}).

\item Many of our model SEDs are photometrically and spectroscopically similar to foreground (Milky Way) dwarf stars (Figure \ref{fig:SED_dwarf}), and these are expected to be an important source of contamination and/or confusion when trying to identify super-Eddington SMBHs such as those described by our models.

\item We devise photometric colour criteria in optical to MIR colour-colour spaces to broadly select super-Eddington AGN populations, for both shallow and wide, and narrow and deep field surveys, which may be used independently or to cross-check potential super-Eddington sources with each other. 

\end{itemize}

While our collection of model SEDs does not effectively span the SDSS quasar population, we emphasize that this simple model does not consider all the complex features of real AGN, but instead focusses on the photon trapping effect. Furthermore, the SDSS quasars are almost all UV bright, unobscured sources and -- importantly -- are sub-Eddington, and thus should indeed be different to the models to some extent. We also note that our models with relatively low Eddington ratios fall well within the quasar population contours.
Although features such as UV line emission are expected to be of little effect on a model utilising photon trapping, other realistic factors such as dust obscuration and AGN-driven outflows would likely play a non-negligible role. 
In terms of completeness, they are expected to reduce the rest-frame UV emission and thus further decrease the probability of SDSS-like selection criteria to identify them. 
Our results may thus be taken as conservative upper limits on the selection probability.
These effects aside, our analysis shows that the 
SDSS may be missing a significant population of fast-growing super-Eddington SMBHs at intermediate redshifts. 

In order to reliably detect and identify these objects, new selection criteria must be devised. We suggest in this work a range of colour cuts extending from optical to MIR for both wide-field and deep, narrow-field surveys (Figure \ref{fig:all_filters} and Table \ref{tab:colour_cuts}). 
The colour cuts suggested here seek to accurately select super-Eddington AGN while avoiding known, non super-Eddington sources. 
This also leads to many models being omitted due to their colours being indistinguishable from known non super-Eddington sources. 
Host emission for the redder, more extreme models contributes heavily to this effect, yielding galaxy-like colours in the optical. 
As discussed in Section \ref{sec:selection}, combinations of criteria across wavebands should not only allow interesting objects to be more reliably selected, but may also increase the number of objects selected. 
We also compare our models in the MIR to Hot DOGs as these have been suggested to be highly obscured, bright AGN. 
However, we find that these objects are even redder than our models, with significant dust effects, and thus confusion between the two should be avoidable (Figure \ref{fig:DOGS}).

Importantly, we note that this current work only tests the completeness of SDSS with regards to colour selection. Other methods, such as radio matching with FIRST, should also be considered in order to make a sweeping statement about the representation of super-Eddington populations. 
We note that the objects successfully selected as quasar candidates by the colour selection may be misidentified during follow-up spectroscopic classification due to the weakness of AGN-driven lines and the potential lack of an X-ray component -- both effects driven by the lack of `seed' UV continuum emission from the inner parts of the (truncated) disc. 
suggested by our photon-trapping models. 
Finally, all the above are based of the truncated thin AD model for super-Eddington accretion, and thus are dependent on the assumptions made in the model. Complex AGN features such as outflows, feedback, emission and absorption lines, and the effects of (circumnuclear) dust are not considered. 

This work thus provides the first hints on how to search for the yet-to-be-seen population of luminous super-Eddington SMBHs, at intermediate redshift, if such intriguing systems indeed exist. 

\section*{Acknowledgements}
We thank the anonymous referee for their constructive comments, which helped us to improve the presentation of our work. We thank A.\ Weigel and L.\ Sartori for useful discussions 
B.T. acknowledges support from the Israel Science Foundation (grant No. 1849/19).



\bibliographystyle{mnras}
\bibliography{biblio} 



\newpage

\appendix

\section{Host Emission Variation with Age}
\label{sec:host_var}

In this study, we examine the effect of host emission on the super-Eddington models' colours by adding the emission from a SSP with a fixed age of 1 Gyr. This choice of age is somewhat arbitrary and in order to examine the effects of a much younger and older stellar population, we also take SSPs of ages 200 Myr and 4 Gyr respectively. The combined AD model with host emission SEDs for these ages are plotted below in Figures \cref{fig:200MYrSSP_SED,fig:4GYrSSP_SED} in the same way as for the 1 Gyr old SSP in Figure \ref{fig:SSP_SED}. From Figures \cref{fig:200MYrSSP_SED,fig:4GYrSSP_SED}, the evolution of the stellar emission with age is very clear. The more massive blue stars responsible for much of the stellar UV emission die out rapidly,  
decreasing not only the optical and UV emission but also the overall luminosity of the host. The smaller stars responsible for red optical and NIR emission however, remain present. Thus the the host emission will always come to dominate the optical emission for the reddest models that emit almost exclusively in the MIR as shown in the rightmost panels of Figures \cref{fig:200MYrSSP_SED,fig:4GYrSSP_SED}. On the other hand, a host galaxy with young SSPs will always have an effect on the total emission of the models, whether by enhancing the optical and UV emission or even the IR emission (see leftmost panel of Figure \ref{fig:200MYrSSP_SED}). In terms of model selection by the SDSS quasar algorithm, the 4 Gyr old SSP host would likely make even less difference than the 1 Gyr old host presented in the body of this work compared to a pure AD SED. For the reddest models, host emission would still dominate the optical wavelengths, leading to the models being with SDSS spectroscopic flux limits and hence observable, but with galaxy like colours and thus within range of the stellar locus. 

\begin{figure*}
\center
\includegraphics[trim={0.2cm 0 1.5cm 0},clip,width = 0.32\textwidth]{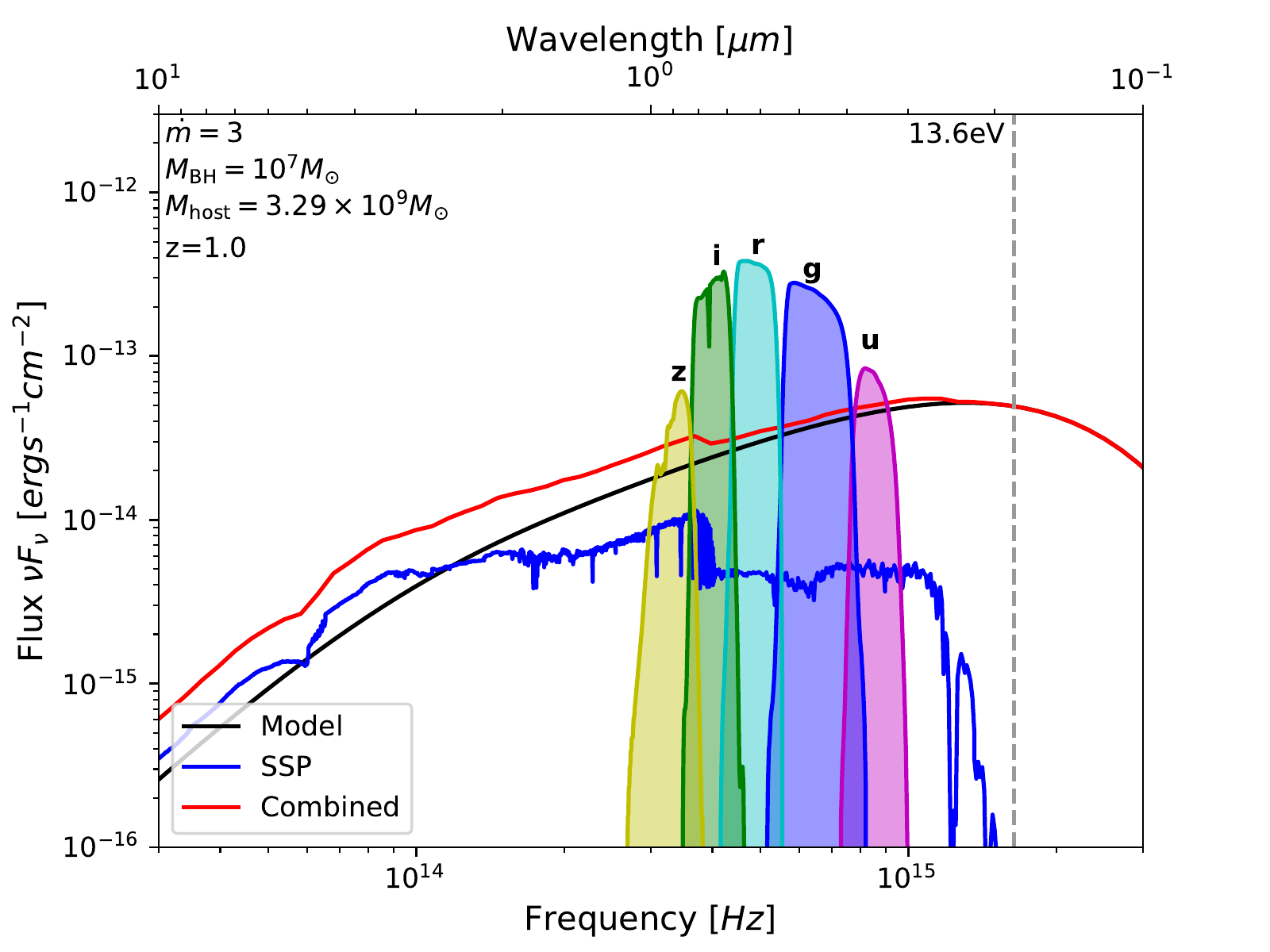}
\includegraphics[trim={0.2cm 0 1.5cm 0},clip,width = 0.32\textwidth]{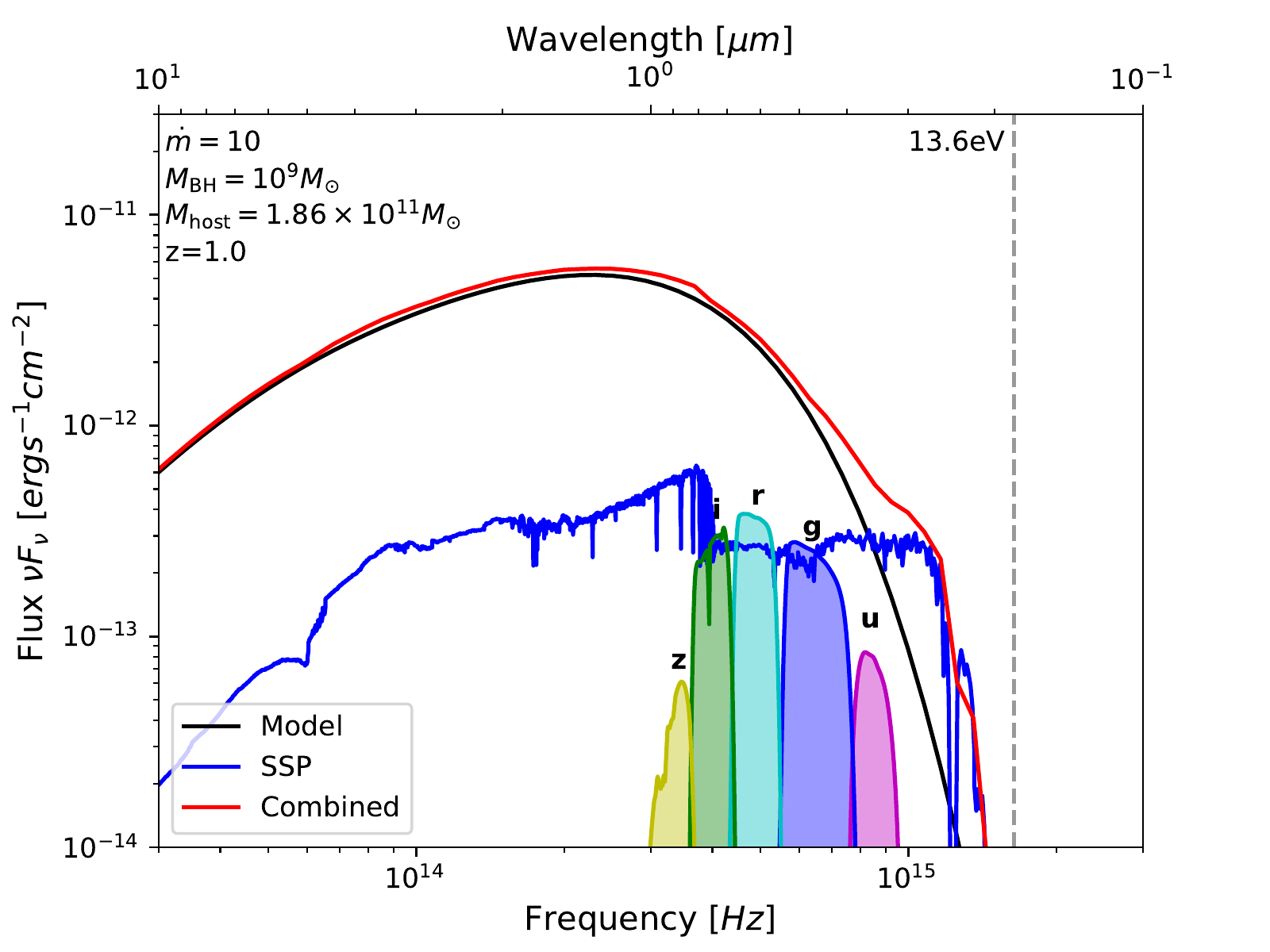}
\includegraphics[trim={0.2cm 0 1.5cm 0},clip,width = 0.32\textwidth]{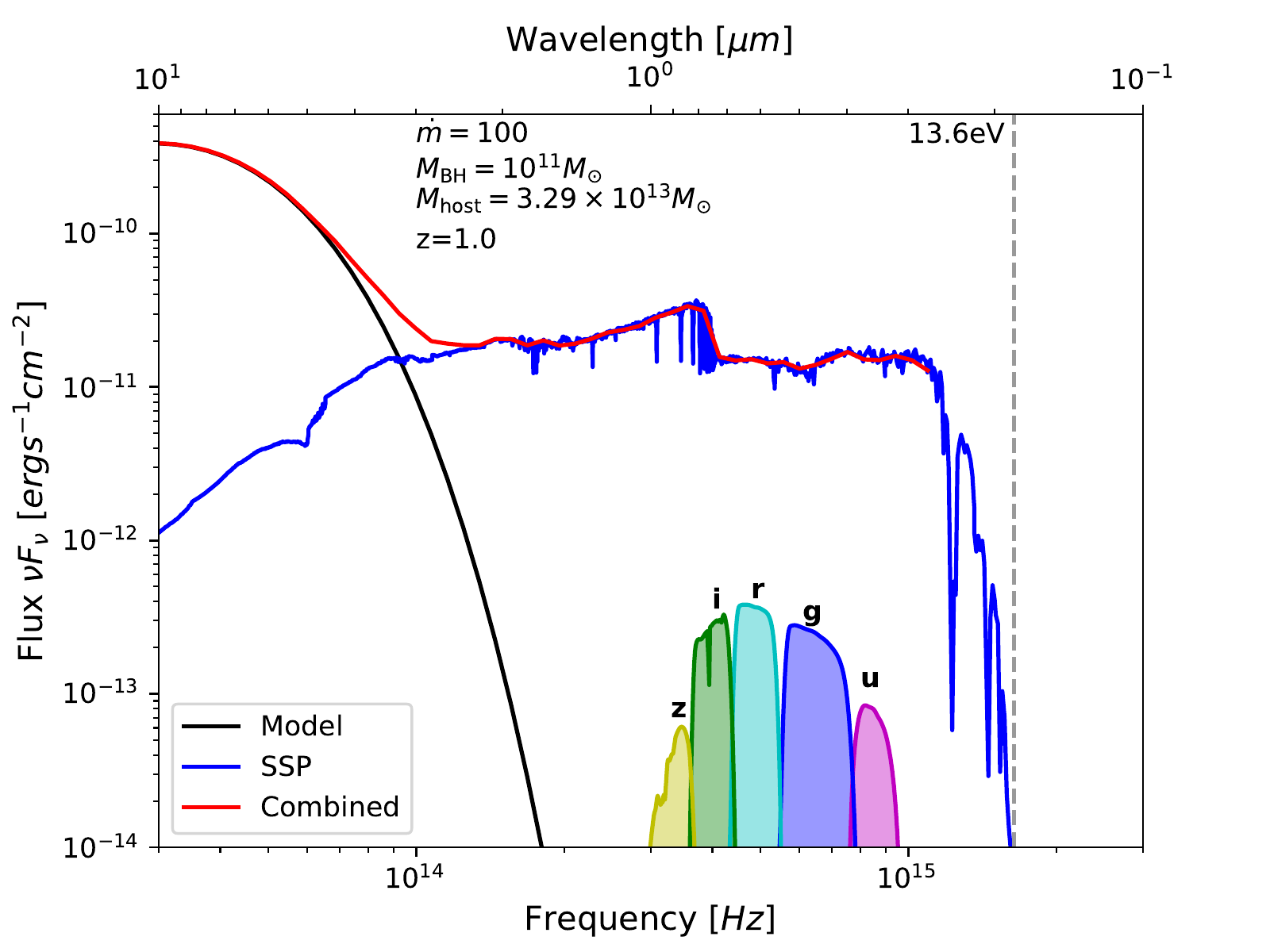}
\caption{The effect of adding the emission from a 200 Myr old host galaxy to the super-critical model SEDs. 
The leftmost panel is chosen to represent a typically bluer model with low \mbh\ and $\dot{m}$. The middle panel represents the "average" SED model with intermediate values, while the rightmost panel represents the reddest models with highest values. The host emission dominates both the optical and near emission for the reddest model, as the AD emission only comes into effect at $\lambda > 2\mu m$. The other models are not as highly affected by the host emission, though the bluest model gain slightly more red emission at $\lambda \gtrsim 1 \mu m$. The model in the middle panel on the other hand has enhanced near UV emission thanks to the blue massive stars still present in the young stellar population.}
\label{fig:200MYrSSP_SED}
\end{figure*}

\begin{figure*}
\center
\includegraphics[trim={0.2cm 0 1.5cm 0},clip,width = 0.32\textwidth]{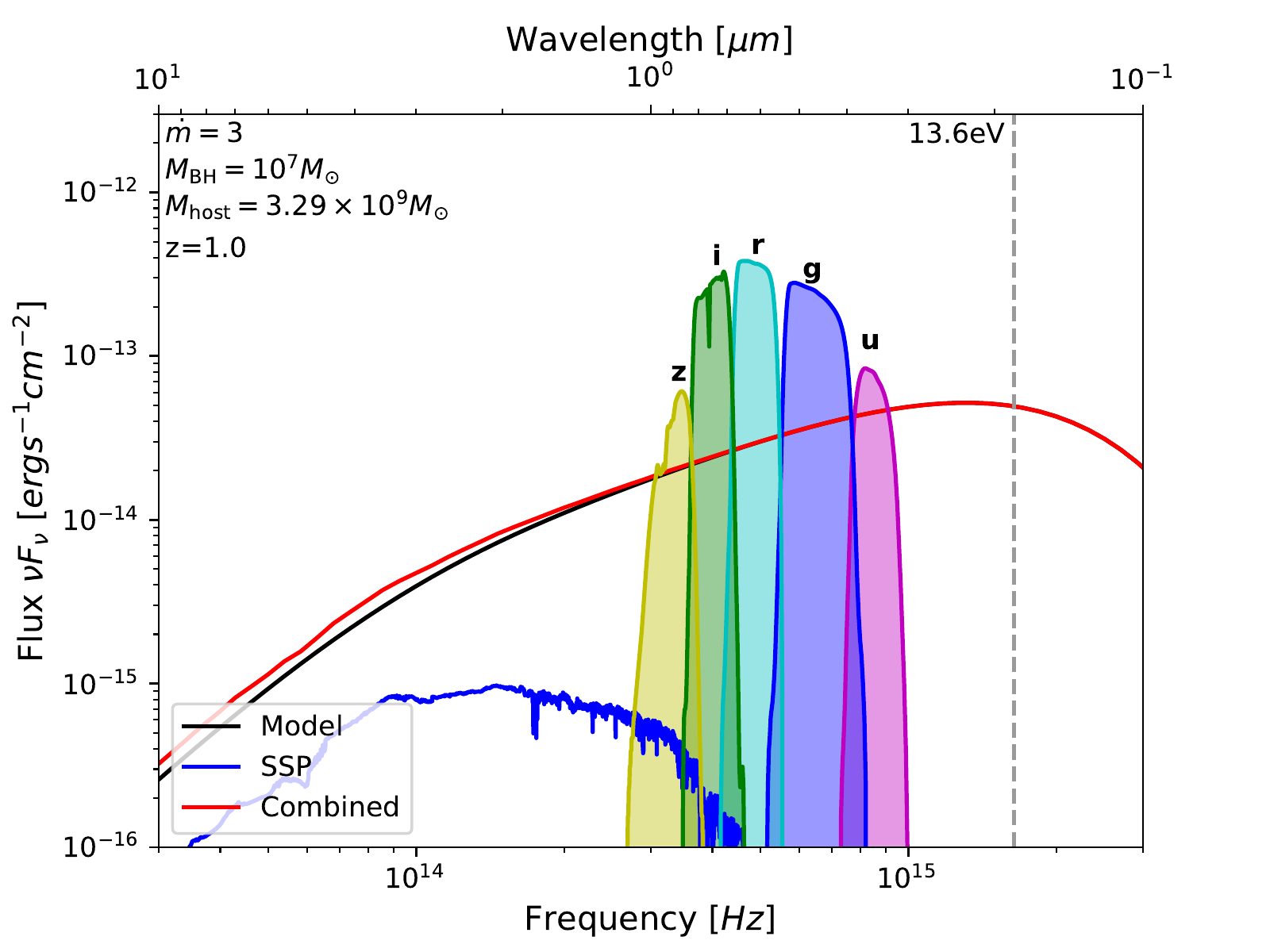}
\includegraphics[trim={0.2cm 0 1.5cm 0},clip,width = 0.32\textwidth]{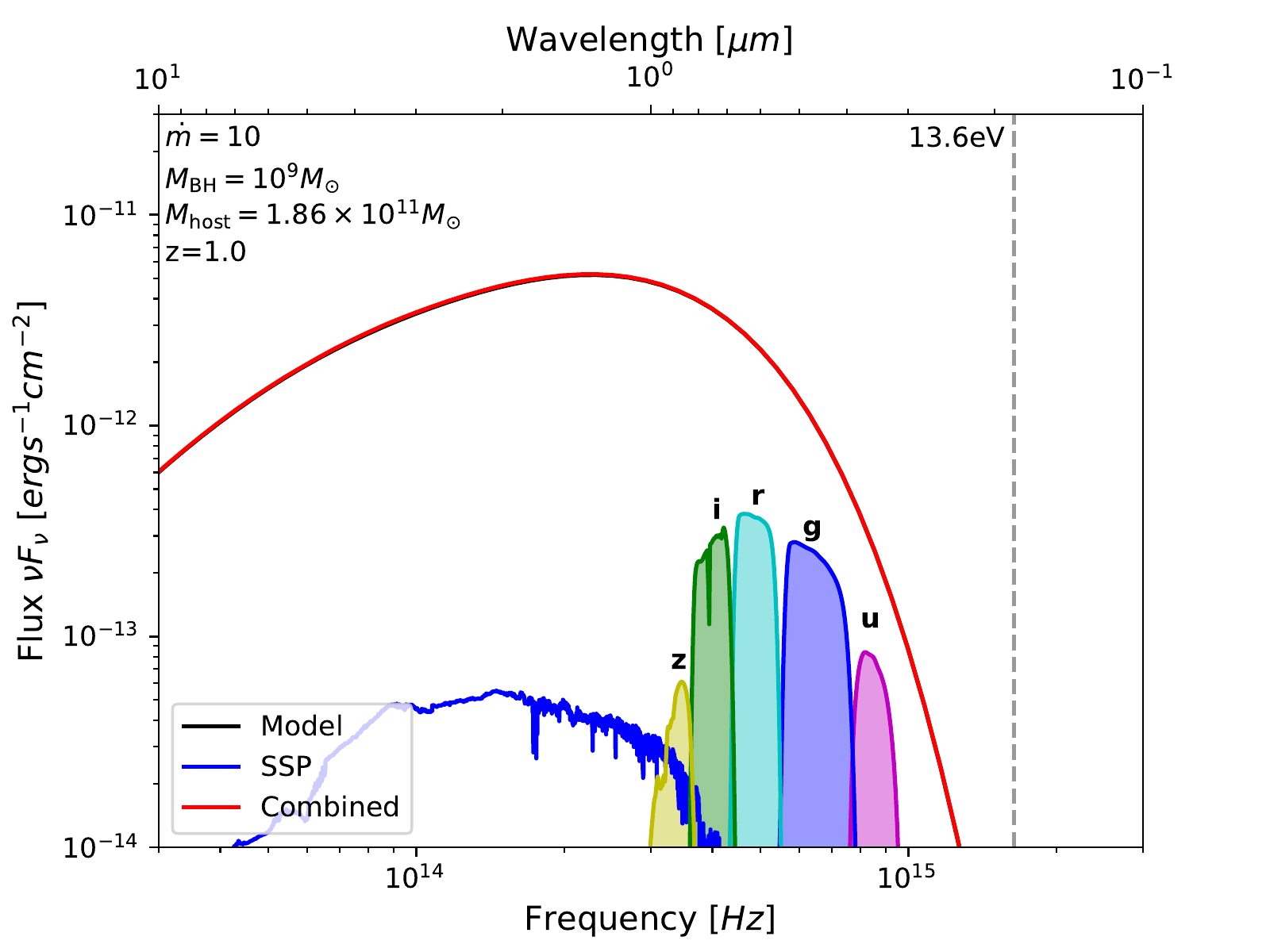}
\includegraphics[trim={0.2cm 0 1.5cm 0},clip,width = 0.32\textwidth]{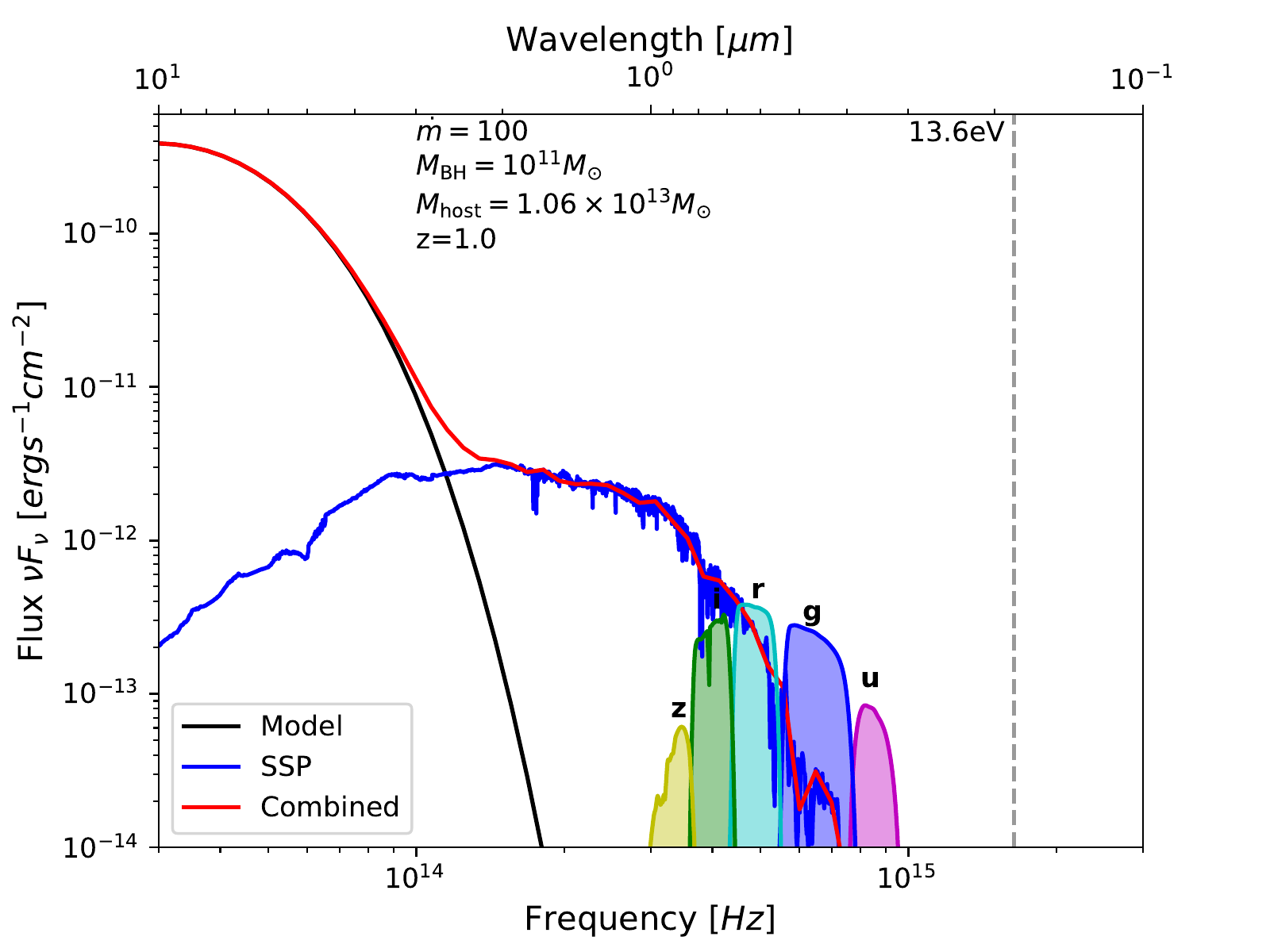}
\caption{Same as Figure~\ref{fig:200MYrSSP_SED}, but with a 4 Gyr old host galaxy SSP. 
The leftmost panel is chosen to represent a typically bluer model with low \mbh\ and $\dot{m}$. The middle panel represents the `average' SED model with intermediate values, while the rightmost panel represents the reddest models with highest values. The host emission dominates the optical emission for the reddest model, as the AD emission only comes into effect at $\lambda > 2$ \mic. However, the other models are not highly affected by the host emission.}
\label{fig:4GYrSSP_SED}
\end{figure*}

The other models represented by the leftmost and middle panels of Figure \ref{fig:4GYrSSP_SED}, have no change in optical emission due to the 4GYr host, and so the SDSS selection statistics would not change at all for these models. As for the 200MYr host, the same reasoning applies with regards to these extremely red models; the algorithm would reject them as most likely being contaminants due to host emission dominating the optical regime and thus having galaxy-like colours. A young host could however increase the number of blue models, represented by the leftmost panel of Figure \ref{fig:200MYrSSP_SED}, being selected by the algorithm, as it enhances \textit{i}-band flux. Many of these bluer models were firmly within the UV excess inclusion zone, but were beneath SDSS flux limits and so were not selected (see Figures \cref{fig:colours,fig:sel_grids_flux}. The addition of \textit{i}-band flux without changing the $u-g$ colour would increase the chances of these models being selected. The models with intermediate colours represented by the central panel of Figure \ref{fig:200MYrSSP_SED} would most likely still be rejected, as the added \textit{g} and \textit{u}-band emission would not be enough to take these models sufficiently away from the stellar locus. Thus, younger hosts are slightly better for selecting super-Eddington AGN in terms of the SDSS selection algorithm compared to hosts of increasing age, although this effect is not extremely significant.

\newpage

\section{Sub-Eddington Model Colours}
\label{sec:subEdd}

In this Appendix, we verify that the colours of sub-Eddington versions of our AD model SEDs correspond to those of real, colour-selected SDSS quasars at $1 \lesssim z \lesssim 2$, which are indeed known to accrete at sub-Eddington rates \cite[e.g.,][]{Trakhtenbrot.Netzer:12,Kelly.Shen:13}.

\begin{figure*}
\center
\includegraphics[trim={0.6cm 0 1.5cm 0},clip,width = 0.48\textwidth]{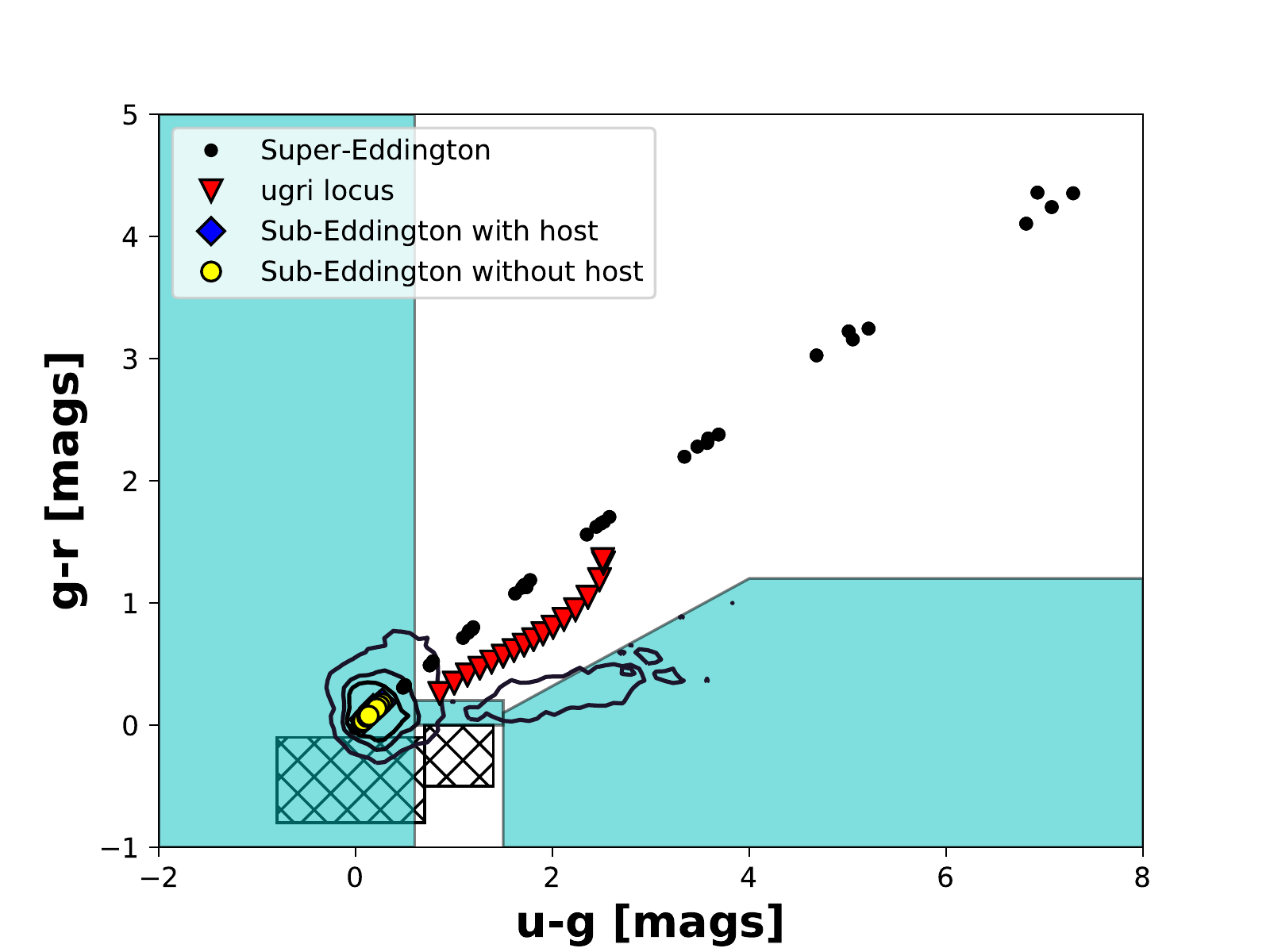}
\includegraphics[trim={0.6cm 0 1.5cm 0},clip,width = 0.48\textwidth]{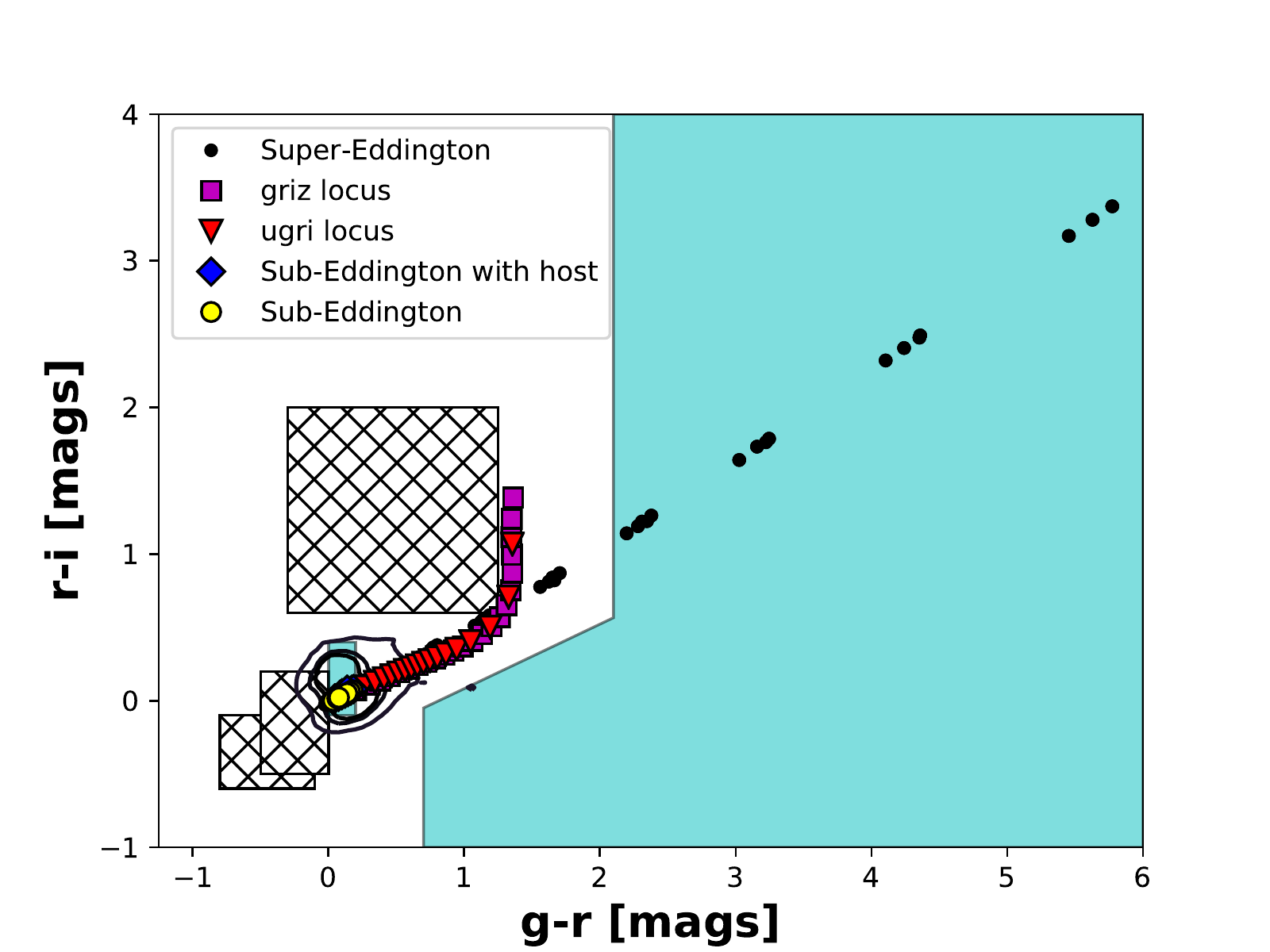}
\includegraphics[trim={0.3cm 0 1.5cm 0},clip,width = 0.48\textwidth]{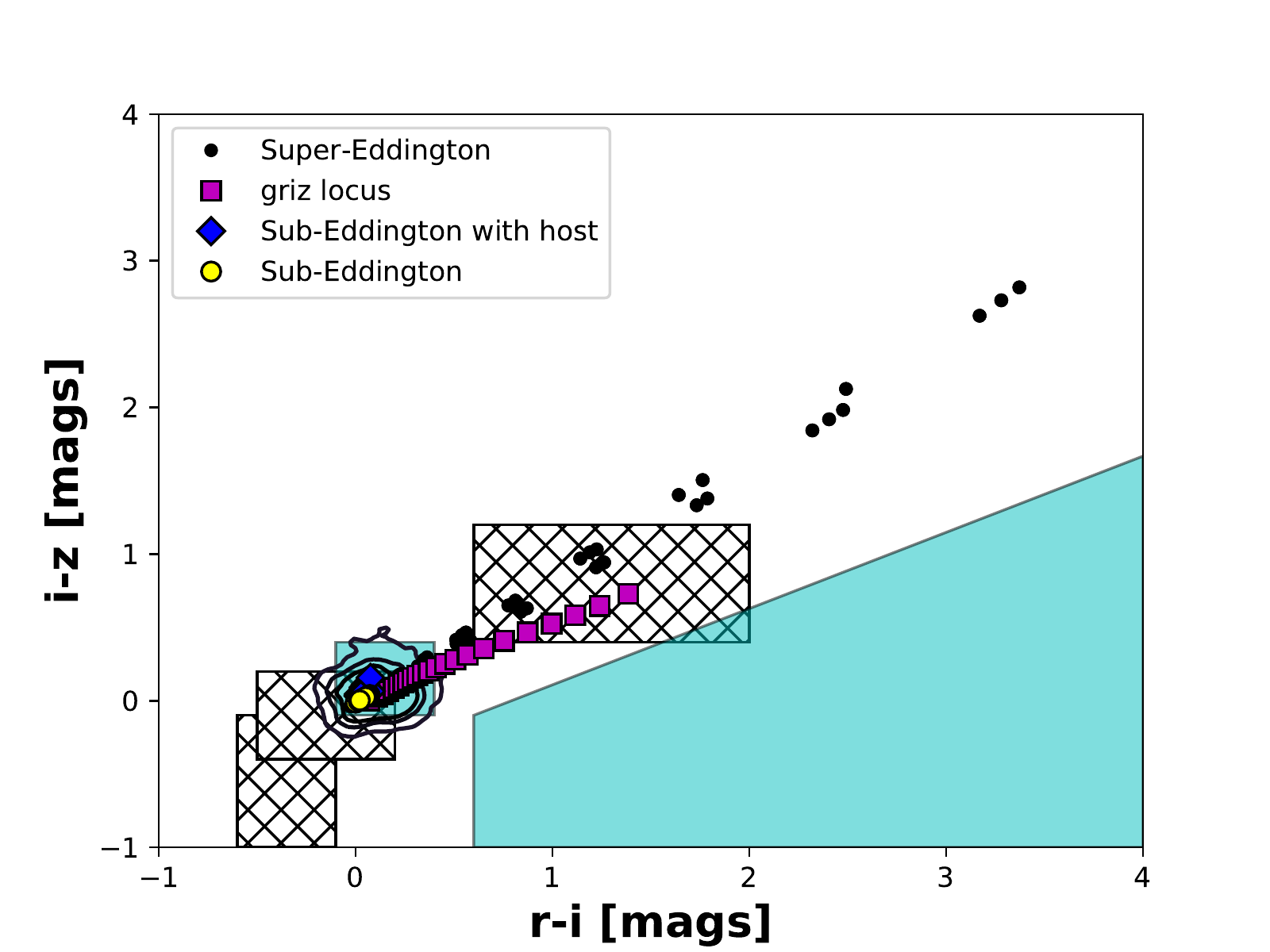}
\caption{The pure AD models at $z=1$, also with host emission considered, along with the SDSS exclusion and inclusion zones. The sub-Eddington models here have $\dot{m} = 0.05 - 0.9$. The quasar contours here are from the SDSS DR7 quasar catalogue \citep{Schneider.etal:10}, and are a subsample of the total catalogue which has been selected by the quasar colour selection algorithm (though not necessarily solely). We note that almost all of the sub-Eddington models are in the UV excess inclusion zone (\textit{ugr} colour space) and that host emission has very little effect on the colours.}
\label{fig:SDSS_subEdd}
\end{figure*}

In Figure~\ref{fig:SDSS_subEdd}, we compare the observed SDSS quasar population with the {\it sub}-Eddington regime of our simplistic thin-disc models, i.e. with little or no photon trapping.
The yellow points trace standard thin disc models with Eddington ratios of $\dot{m} = 0.5,0.1,0.2$, and then truncated AD models for ratios $\dot{m} = 0.3 - 1.0$, increasing in steps of 0.1. 
The latter choice is motivated by the expectation that standard, purely thin disc (Shakura-Sunyaev) models would fail for $\dot{m} \gtrsim 0.3$. 
All the models in Figure \ref{fig:SDSS_subEdd} are at a redshift of $z=1$. Host emission is also added to the sub-Eddington discs (blue squares), however this appears to have very little effect on the colours. 
The quasar contours in Figure \ref{fig:SDSS_subEdd} are the same as in Figure \ref{fig:colours}. 
The use of sub-Eddington SEDs is done to test the accuracy of our basic model in a sub-Eddington regime, in the context of the original SDSS colour selection algorithm.

We note that the sub-Eddington models all lie firmly within the central contour of the SDSS DR7 quasar sample, and are for the vast majority selected by the quasar selection algorithm, as they lie in the UV excess inclusion zone. The few sub-Eddington models that are not selected are the faintest models at $z=2$. The almost complete selection of the sub-Eddington models is to some extent expected, as the UV excess zone was specifically designed to target AGN with significant UV emission arising from the AD, and the sub-Eddington models experience little to no photon trapping. Furthermore, this is consistent with the results of the DR7 survey, which found that UV excess selected quasars constituted 89\% of the colour-selected objects, and 63\% of the total DR7 Catalogue \citep{Schneider.etal:10}. As in Figure \ref{fig:colours}, we see the contours with redder $u-g$ colours in \textit{ugr} colour space remains uninhabited by our sub-Eddington models, for the same reasons as mentioned above for low $\dot{m} \leq 3$ super-Eddington models. Thus, it is unsurprising that our model SEDs do not span the entire DR7 colour selected quasar contours, but reassuring that our basic sub-Eddington models follow the same behaviour as the majority of the colour selected sample in the DR7 catalogue.

\bsp	
\label{lastpage}
\end{document}
